\documentclass[11pt,letterpaper]{article}
\usepackage{latexsym}

\makeatletter
                                                                               
\def\icite{\@ifnextchar [{\@tempswatrue\@citey}{\@tempswafalse\@citey[]}}
\def\@citex[#1]#2{%
\if@filesw \immediate \write \@auxout {\string \citation {#2}}\fi
\@tempcntb\m@ne \let\@h@ld\relax \def\@citea{}%
\@cite{%
  \@for \@citeb:=#2\do {%
    \@ifundefined {b@\@citeb}%
      {\@h@ld\@citea\@tempcntb\m@ne{\bf ?}%
      \@warning {Citation `\@citeb ' on page \thepage \space undefined}}%
      {\@tempcnta\@tempcntb \advance\@tempcnta\@ne%
      \@tempcntb\number\csname b@\@citeb \endcsname \relax%
      \ifnum\@tempcnta=\@tempcntb 
        \ifx\@h@ld\relax%
          \edef \@h@ld{\@citea\csname b@\@citeb\endcsname}%
        \else%
          \edef\@h@ld{\ifmmode{-}\else--\fi\csname b@\@citeb\endcsname}%
        \fi%
      \else
        \@h@ld\@citea\csname b@\@citeb \endcsname%
        \let\@h@ld\relax%
      \fi}%
    \def\@citea{,\penalty\@highpenalty\,}%
  }\@h@ld
}{#1}}
\def\@citey[#1]#2{%
\if@filesw \immediate \write \@auxout {\string \citation {#2}}\fi
\@tempcntb\m@ne \let\@h@ld\relax \def\@citea{}%
\@icite{%
  \@for \@citeb:=#2\do {%
    \@ifundefined {b@\@citeb}%
      {\@h@ld\@citea\@tempcntb\m@ne{\bf ?}%
      \@warning {Citation `\@citeb ' on page \thepage \space undefined}}%
      {\@tempcnta\@tempcntb \advance\@tempcnta\@ne%
      \@tempcntb\number\csname b@\@citeb \endcsname \relax%
      \ifnum\@tempcnta=\@tempcntb 
        \ifx\@h@ld\relax%
          \edef \@h@ld{\@citea\csname b@\@citeb\endcsname}%
        \else%
          \edef\@h@ld{\ifmmode{-}\else--\fi\csname b@\@citeb\endcsname}%
        \fi%
      \else
        \@h@ld\@citea\csname b@\@citeb \endcsname%
        \let\@h@ld\relax%
      \fi}%
    \def\@citea{,\penalty\@highpenalty\,}%
  }\@h@ld
}{#1}}
\def\@cite#1#2{{$^{#1}$\if@tempswa , #2\fi }}
\def\@icite#1#2{{$#1$\if@tempswa , #2\fi }}

\gdef\@publabel{\hfil}
\gdef\@pubdate{\null}
\gdef\@pubnumber{\null}
\gdef\@author{\null}
\gdef\@title{\null}
\gdef\@abstract{\null}
\long\def\pubdate#1{\gdef\@pubdate{#1}}
\long\def\pubnumber#1{\gdef\@pubnumber{#1}}
\long\def\publabel#1{\gdef\@publabel{#1}}
\long\def\author#1{\gdef\@author{#1}}
\long\def\title#1{\gdef\@title{#1}}
\long\def\abstract#1{\gdef\@abstract{#1}}
\def\titlerelax{
\let\maketitle\relax
\let\settitleparameters\relax
\let\consolidatetitle\relax
\let\inittitlepage\relax
\let\finishtitlepage\relax
\let\titlepagecontents\relax
\let\multithanks\relax
\let\titlebaselines\relax
\let\@makepub\relax
\let\@maketitle\relax
\let\@makeauthor\relax
\let\@makeabstract\relax
\let\@maketitlenote\relax
\let\thanks\relax
\let\titlerelax\relax}

\def\titleclean
{\gdef\@titlenote{}
\gdef\@abstract{}
\gdef\@author{}
\gdef\@title{}
\gdef\@pubdate{}\gdef\@pubnumber{}\gdef\@publabel{}
\gdef\@dpublabel{}
}
\def\@makepub{\vbox to \z@{\hbox to \textwidth{\hfill
\@publabel \hfill
\llap{\parbox[t]{0.25\textwidth}{\raggedleft\@pubnumber}}}%
\vss}}
\def\@maketitle{\vskip 60pt \begin{center}
 {\LARGE \@title \par}
 \end{center}}
\def\@makeauthor{{%
\def\and{\smallskip {\normalsize \rm and\smallskip }}
\def\And{\medskip {\normalsize \rm and\\}\medskip}
\long\def\address##1{{\def\and{\\and\\}\medskip
				{\small \it \\##1\\}
}}
{\centering
 \vskip 3em
 \large \lineskip .75em
 \@author}
 \par}}
\def\@makedate{\vskip 1.5em
 {\raggedright \small \noindent\@pubdate \par}}
\def\@makeabstract{\vskip 1.5em
{\small
\begin{center}
{\bf ABSTRACT\vspace{-.5em}\vspace{0pt}}
\end{center}
\quotation \@abstract \endquotation}}
\def\maketitle{\titlepage
\let\footnotesize\small \setcounter{page}{0}
\def\thefootnote{\arabic{footnote}}
\@makepub
\vfil
\@maketitle
\@makeauthor
\vfil
\@makeabstract
\@thanks
\vfil
\@makedate
\if@restonecol\twocolumn \else \eject \fi
\titlerelax \titleclean
\def\thefootnote{\alph{footnote}}
\setcounter{footnote}{0}
}

\ifcase\@ptsize
 \font\tenmsa=msam10
 \font\sevenmsa=msam7
 \font\fivemsa=msam5
 \font\tenmsb=msbm10
 \font\sevenmsb=msbm7
 \font\fivemsb=msbm5
\or
 \font\tenmsa=msam10 scaled \magstephalf
 \font\sevenmsa=msam8
 \font\fivemsa=msam6
 \font\tenmsb=msbm10 scaled \magstephalf
 \font\sevenmsb=msbm8
 \font\fivemsb=msbm6
\or
 \font\tenmsa=msam10 scaled \magstep1
 \font\sevenmsa=msam8
 \font\fivemsa=msam6
 \font\tenmsb=msbm10 scaled \magstep1
 \font\sevenmsb=msbm8
 \font\fivemsb=msbm6
\fi

\newfam\msafam
\newfam\msbfam
\textfont\msafam=\tenmsa  \scriptfont\msafam=\sevenmsa
  \scriptscriptfont\msafam=\fivemsa
\textfont\msbfam=\tenmsb  \scriptfont\msbfam=\sevenmsb
  \scriptscriptfont\msbfam=\fivemsb

\def\hexnumber@#1{\ifnum#1<10 \number#1\else
 \ifnum#1=10 A\else\ifnum#1=11 B\else\ifnum#1=12 C\else
 \ifnum#1=13 D\else\ifnum#1=14 E\else\ifnum#1=15 F\fi\fi\fi\fi\fi\fi\fi}

\def\msa@{\hexnumber@\msafam}
\def\msb@{\hexnumber@\msbfam}
\mathchardef\boxdot="2\msa@00
\mathchardef\boxplus="2\msa@01
\mathchardef\boxtimes="2\msa@02
\mathchardef\square="0\msa@03
\mathchardef\blacksquare="0\msa@04
\mathchardef\centerdot="2\msa@05
\mathchardef\lozenge="0\msa@06
\mathchardef\blacklozenge="0\msa@07
\mathchardef\circlearrowright="3\msa@08
\mathchardef\circlearrowleft="3\msa@09
\mathchardef\rightleftharpoons="3\msa@0A
\mathchardef\leftrightharpoons="3\msa@0B
\mathchardef\boxminus="2\msa@0C
\mathchardef\Vdash="3\msa@0D
\mathchardef\Vvdash="3\msa@0E
\mathchardef\vDash="3\msa@0F
\mathchardef\twoheadrightarrow="3\msa@10
\mathchardef\twoheadleftarrow="3\msa@11
\mathchardef\leftleftarrows="3\msa@12
\mathchardef\rightrightarrows="3\msa@13
\mathchardef\upuparrows="3\msa@14
\mathchardef\downdownarrows="3\msa@15
\mathchardef\upharpoonright="3\msa@16

\mathchardef\downharpoonright="3\msa@17
\mathchardef\upharpoonleft="3\msa@18
\mathchardef\downharpoonleft="3\msa@19
\mathchardef\rightarrowtail="3\msa@1A
\mathchardef\leftarrowtail="3\msa@1B
\mathchardef\leftrightarrows="3\msa@1C
\mathchardef\rightleftarrows="3\msa@1D
\mathchardef\Lsh="3\msa@1E
\mathchardef\Rsh="3\msa@1F
\mathchardef\rightsquigarrow="3\msa@20
\mathchardef\leftrightsquigarrow="3\msa@21
\mathchardef\looparrowleft="3\msa@22
\mathchardef\looparrowright="3\msa@23
\mathchardef\circeq="3\msa@24
\mathchardef\succsim="3\msa@25
\mathchardef\gtrsim="3\msa@26
\mathchardef\gtrapprox="3\msa@27
\mathchardef\multimap="3\msa@28
\mathchardef\therefore="3\msa@29
\mathchardef\because="3\msa@2A
\mathchardef\doteqdot="3\msa@2B

\mathchardef\triangleq="3\msa@2C
\mathchardef\precsim="3\msa@2D
\mathchardef\lesssim="3\msa@2E
\mathchardef\lessapprox="3\msa@2F
\mathchardef\eqslantless="3\msa@30
\mathchardef\eqslantgtr="3\msa@31
\mathchardef\curlyeqprec="3\msa@32
\mathchardef\curlyeqsucc="3\msa@33
\mathchardef\preccurlyeq="3\msa@34
\mathchardef\leqq="3\msa@35
\mathchardef\leqslant="3\msa@36
\mathchardef\lessgtr="3\msa@37
\mathchardef\backprime="0\msa@38
\mathchardef\risingdotseq="3\msa@3A
\mathchardef\fallingdotseq="3\msa@3B
\mathchardef\succcurlyeq="3\msa@3C
\mathchardef\geqq="3\msa@3D
\mathchardef\geqslant="3\msa@3E
\mathchardef\gtrless="3\msa@3F
\mathchardef\sqsubset="3\msa@40
\mathchardef\sqsupset="3\msa@41
\mathchardef\vartriangleright="3\msa@42
\mathchardef\vartriangleleft="3\msa@43
\mathchardef\trianglerighteq="3\msa@44
\mathchardef\trianglelefteq="3\msa@45
\mathchardef\bigstar="0\msa@46
\mathchardef\between="3\msa@47
\mathchardef\blacktriangledown="0\msa@48
\mathchardef\blacktriangleright="3\msa@49
\mathchardef\blacktriangleleft="3\msa@4A
\mathchardef\vartriangle="3\msa@4D
\mathchardef\blacktriangle="0\msa@4E
\mathchardef\triangledown="0\msa@4F
\mathchardef\eqcirc="3\msa@50
\mathchardef\lesseqgtr="3\msa@51
\mathchardef\gtreqless="3\msa@52
\mathchardef\lesseqqgtr="3\msa@53
\mathchardef\gtreqqless="3\msa@54
\mathchardef\Rrightarrow="3\msa@56
\mathchardef\Lleftarrow="3\msa@57
\mathchardef\veebar="2\msa@59
\mathchardef\barwedge="2\msa@5A
\mathchardef\doublebarwedge="2\msa@5B
\mathchardef\angle="0\msa@5C
\mathchardef\measuredangle="0\msa@5D
\mathchardef\sphericalangle="0\msa@5E
\mathchardef\varpropto="3\msa@5F
\mathchardef\smallsmile="3\msa@60
\mathchardef\smallfrown="3\msa@61
\mathchardef\Subset="3\msa@62
\mathchardef\Supset="3\msa@63
\mathchardef\Cup="2\msa@64

\mathchardef\Cap="2\msa@65

\mathchardef\curlywedge="2\msa@66
\mathchardef\curlyvee="2\msa@67
\mathchardef\leftthreetimes="2\msa@68
\mathchardef\rightthreetimes="2\msa@69
\mathchardef\subseteqq="3\msa@6A
\mathchardef\supseteqq="3\msa@6B
\mathchardef\bumpeq="3\msa@6C
\mathchardef\Bumpeq="3\msa@6D
\mathchardef\lll="3\msa@6E

\mathchardef\ggg="3\msa@6F

\mathchardef\circledS="0\msa@73
\mathchardef\pitchfork="3\msa@74
\mathchardef\dotplus="2\msa@75
\mathchardef\backsim="3\msa@76
\mathchardef\backsimeq="3\msa@77
\mathchardef\complement="0\msa@7B
\mathchardef\intercal="2\msa@7C
\mathchardef\circledcirc="2\msa@7D
\mathchardef\circledast="2\msa@7E
\mathchardef\circleddash="2\msa@7F
\def\ulcorner{\delimiter"4\msa@70\msa@70 }
\def\urcorner{\delimiter"5\msa@71\msa@71 }
\def\llcorner{\delimiter"4\msa@78\msa@78 }
\def\lrcorner{\delimiter"5\msa@79\msa@79 }
\def\yen{\mathhexbox\msa@55 }
\def\checkmark{\mathhexbox\msa@58 }
\def\circledR{\mathhexbox\msa@72 }
\def\maltese{\mathhexbox\msa@7A }
\mathchardef\lvertneqq="3\msb@00
\mathchardef\gvertneqq="3\msb@01
\mathchardef\nleq="3\msb@02
\mathchardef\ngeq="3\msb@03
\mathchardef\nless="3\msb@04
\mathchardef\ngtr="3\msb@05
\mathchardef\nprec="3\msb@06
\mathchardef\nsucc="3\msb@07
\mathchardef\lneqq="3\msb@08
\mathchardef\gneqq="3\msb@09
\mathchardef\nleqslant="3\msb@0A
\mathchardef\ngeqslant="3\msb@0B
\mathchardef\lneq="3\msb@0C
\mathchardef\gneq="3\msb@0D
\mathchardef\npreceq="3\msb@0E
\mathchardef\nsucceq="3\msb@0F
\mathchardef\precnsim="3\msb@10
\mathchardef\succnsim="3\msb@11
\mathchardef\lnsim="3\msb@12
\mathchardef\gnsim="3\msb@13
\mathchardef\nleqq="3\msb@14
\mathchardef\ngeqq="3\msb@15
\mathchardef\precneqq="3\msb@16
\mathchardef\succneqq="3\msb@17
\mathchardef\precnapprox="3\msb@18
\mathchardef\succnapprox="3\msb@19
\mathchardef\lnapprox="3\msb@1A
\mathchardef\gnapprox="3\msb@1B
\mathchardef\nsim="3\msb@1C
\mathchardef\napprox="3\msb@1D
\mathchardef\varsubsetneq="3\msb@20
\mathchardef\varsupsetneq="3\msb@21
\mathchardef\nsubseteqq="3\msb@22
\mathchardef\nsupseteqq="3\msb@23
\mathchardef\subsetneqq="3\msb@24
\mathchardef\supsetneqq="3\msb@25
\mathchardef\varsubsetneqq="3\msb@26
\mathchardef\varsupsetneqq="3\msb@27
\mathchardef\subsetneq="3\msb@28
\mathchardef\supsetneq="3\msb@29
\mathchardef\nsubseteq="3\msb@2A
\mathchardef\nsupseteq="3\msb@2B
\mathchardef\nparallel="3\msb@2C
\mathchardef\nmid="3\msb@2D
\mathchardef\nshortmid="3\msb@2E
\mathchardef\nshortparallel="3\msb@2F
\mathchardef\nvdash="3\msb@30
\mathchardef\nVdash="3\msb@31
\mathchardef\nvDash="3\msb@32
\mathchardef\nVDash="3\msb@33
\mathchardef\ntrianglerighteq="3\msb@34
\mathchardef\ntrianglelefteq="3\msb@35
\mathchardef\ntriangleleft="3\msb@36
\mathchardef\ntriangleright="3\msb@37
\mathchardef\nleftarrow="3\msb@38
\mathchardef\nrightarrow="3\msb@39
\mathchardef\nLeftarrow="3\msb@3A
\mathchardef\nRightarrow="3\msb@3B
\mathchardef\nLeftrightarrow="3\msb@3C
\mathchardef\nleftrightarrow="3\msb@3D
\mathchardef\divideontimes="2\msb@3E
\mathchardef\varnothing="0\msb@3F
\mathchardef\nexists="0\msb@40
\mathchardef\mho="0\msb@66
\mathchardef\thorn="0\msb@67
\mathchardef\beth="0\msb@69
\mathchardef\gimel="0\msb@6A
\mathchardef\daleth="0\msb@6B
\mathchardef\lessdot="3\msb@6C
\mathchardef\gtrdot="3\msb@6D
\mathchardef\ltimes="2\msb@6E
\mathchardef\rtimes="2\msb@6F
\mathchardef\shortmid="3\msb@70
\mathchardef\shortparallel="3\msb@71
\mathchardef\smallsetminus="2\msb@72
\mathchardef\thicksim="3\msb@73
\mathchardef\thickapprox="3\msb@74
\mathchardef\approxeq="3\msb@75
\mathchardef\succapprox="3\msb@76
\mathchardef\precapprox="3\msb@77
\mathchardef\curvearrowleft="3\msb@78
\mathchardef\curvearrowright="3\msb@79
\mathchardef\digamma="0\msb@7A
\mathchardef\varkappa="0\msb@7B
\mathchardef\hslash="0\msb@7D
\mathchardef\hbar="0\msb@7E
\mathchardef\backepsilon="3\msb@7F
\def\Bbb{\ifmmode\let\next\Bbb@\else
 \def\next{\errmessage{Use \string\Bbb\space only in math mode}}\fi\next}
\def\Bbb@#1{{\Bbb@@{#1}}}
\def\Bbb@@#1{\fam\msbfam#1}

\def\bk {{\hskip 0.2 cm}}

\def\acknowledgements{\@startsection{section}{4}
{\z@}{-3.5ex plus -1ex minus -.2ex}{2.3ex plus .2ex}{\normalsize\bf}
{Acknowledgements}}

\newcommand{\tab}[1]{{\sc Tab.}\,{\sf #1}}	  
\newcommand{\eq}[1]{{\sc Eq.}\,{\sf (#1)}}	  
\newcommand{\eqs}[1]{{\sc Eqs.}\,{\sf (#1)}}	  
\newcommand{\refoth}[1]{{\sf #1}}  	  	  
\newcommand{\eqoth}[1]{{\sf (#1)}}  	  	  

\def\bbbz {\Bbb{Z}}		  

\def\bbbn {\Bbb{N}}   	          
\def\bbbc {\Bbb{C}}

\newtheorem{definition}{Definition}[section]
\newtheorem{theorem}[definition]{Theorem}
\newtheorem{lemma}[definition]{Lemma}
\newtheorem{proposition}[definition]{Proposition}
\newcounter{defs}[section]

\newcommand{\be}{\begin{equation}}
\newcommand{\ee}{\end{equation}}
\newcommand{\bea}{\begin{eqnarray}}
\newcommand{\eea}{\end{eqnarray}}

\newcommand{\bdf}{\stepcounter{defs}\begin{definition}}
\newcommand{\edf}{\end{definition}}
\newcommand{\bth}{\stepcounter{defs}\begin{theorem}}
\newcommand{\eth}{\end{theorem}}
\newcommand{\blm}{\stepcounter{defs}\begin{lemma}}
\newcommand{\elm}{\end{lemma}}
\newcommand{\bpr}{\stepcounter{defs}\begin{proposition}}
\newcommand{\epr}{\end{proposition}}
\newcommand{\bprf}{Proof: }
\newcommand{\eprf}{\hfill $\Box$ \\}
\newcounter{pics}
\newcommand{\bpic}[4]{\begin{center}\begin{picture}(#1,#2)(#3,#4)
\refstepcounter{pics}}
\renewcommand{\thepics}{{\sf\roman{pics}}}
\newcommand{\epic}[1]{\end{picture}\\
{\small {\sc Fig.} \thepics \bk #1} \end{center}}
\newcommand{\epicspl}{\end{picture}\\		
\addtocounter{pics}{-1}\end{center}}		

\renewcommand{\thefootnote}{\rm{\alph{footnote}}}
\newcounter{tabs}
\newcommand{\btab}[1]{\refstepcounter{tabs}\begin{center}
\begin{tabular}{#1}}
\renewcommand{\thetabs}{{\sf\alph{tabs}}}
\newcommand{\etab}[1]{\end{tabular}\\[1.5ex]
{\small {\sc Tab.} \thetabs \bk #1} \end{center}}

\def\noi {\noindent}
\newcommand{\nn}{\nonumber}

\newcommand{\ket}[1]{\left| {#1} \right\rangle}	
\newcommand{\spn}[1]{{\rm span}\{{#1}\}}	
\newcommand{\vm}[1]{{\langle #1 \rangle}}	

\def\pmb#1{\setbox0=\hbox{#1}%
 \kern-.025em\copy0\kern-\wd0
 \kern.05em\copy0\kern-\wd0
 \kern-.025em\raise.0433em\box0 }

\def\cQ{{\cal Q}}
\def\cG{{\cal G}}
\def\cL{{\cal L}}
\def\cH{{\cal H}}

\makeatother

\def\ta{{\sf T}_2}       
\def\vm{{\cal V}}        
\def\vir{{\sf V}}        
\def\salg{{\cal A}}      
\def\ordering{{\cal O}}  
\def\bsvm{{\cal B}}      
\def\cset{{\cal C}}      
\def\sset{{\cal S}}      
\def\lset{{\sf L}}      
\def\hset{{\sf H}}      
\def\gset{{\sf G}}      
\def\qset{{\sf Q}}      
\newcommand{\level}[1]{{{|}#1{|}}}	
\newcommand{\levelt}[1]{{{|}#1{|}_{L}}}	
\newcommand{\charget}[1]{{{|}#1{|}_{H}}}	
\newcommand{\length}[1]{{{\|}#1{\|}}}	
\newcommand{\osm}[1]{{{<}_{{}_{#1}}}}           

\title{Singular dimensions of the $N=2$ superconformal algebras. I}
\author{Matthias D\"{o}rrzapf\thanks{matthias@feynman.harvard.edu}
and Beatriz Gato-Rivera\thanks{bgato@pinar1.csic.es}$^{,3}$
\address{$^1$Lyman Laboratory of Physics\\
Harvard University,
Cambridge, MA 02138, USA \\[.3cm]
$^2$Instituto de Matem\'aticas y F\'\i sica Fundamental, CSIC,\\
Serrano 123, Madrid 28006, Spain \\[.3cm]
$^3$NIKHEF-H, Kruislaan 409, NL-1098 SJ Amsterdam, The Netherlands}}
\pubdate{July 1998}
\pubnumber{HUTP-98/A021 \\ IMAFF-FM-98/06 \\ NIKHEF-98-022\\
hep-th/9807234}

\abstract{Verma modules of superconfomal algebras can have singular vector
spaces with dimensions greater than $1$. Following a method developed for the 
Virasoro algebra by Kent, we introduce the concept of adapted orderings on
superconformal algebras. We prove several general results on the ordering
kernels associated to the adapted orderings and show that the size of an
ordering kernel implies an upper limit for the dimension of a singular vector 
space. We apply this method to the topological $N=2$ algebra and obtain the
maximal dimensions of the singular vector spaces in the topological Verma
modules: 0, 1, 2 or 3 depending on the type of Verma module and the type of
singular vector. As a consequence we prove the conjecture of Gato-Rivera and
Rosado on the possible existing types of topological singular vectors (4 in
chiral Verma modules and 29 in complete Verma modules). Interestingly, we have 
found two-dimensional spaces of singular vectors at level 1. Finally, by using
the topological twists and the spectral flows, we also obtain the maximal
dimensions of the singular vector spaces for the Neveu-Schwarz $N=2$ algebra
(0, 1 or 2) and for the Ramond $N=2$ algebra (0, 1, 2 or 3).}

\begin{document}

\maketitle



\section{Introduction}

More than two decades ago, superconformal algebras were first constructed
independently and almost at the same time
by Kac\cite{kac3} and by Ademollo et al.\cite{ademollo}.
 Whilst Kac\cite{kac3} derived them for mathematical purposes
along with his classification of
Lie super algebras, Ademollo et al.\cite{ademollo}
constructed the superconformal algebras for physical purposes
in order to define supersymmetric strings.
Since then the study of superconformal algebras
has made much progress on both mathematics and physics. On the mathematical
side Kac and van de Leuer\cite{kac6} and Cheng and Kac\cite{kac4}
have classified all possible superconformal algebras
and Kac recently has proved that their
classification is complete (see footnote in Ref. \icite{kac5}).
As far as the physics side is concerned,
superconformal models are gaining increasing
importance. Many areas of physics make use of
superconformal symmetries but the importance is above all due to
the fact that superconformal
algebras supply the underlying symmetries of Superstring Theory.

The classification of the irreducible highest weight
representations of the superconformal
algebras is of interest to both, mathematicians and physicists.
After more than two decades, only the simpler superconformal
highest weight representations have been fully understood.
Namely, only the representations of $N=1$ are completely
classified and proven\cite{ast1,ast2}. For $N=2$ remarkable efforts
have been taken by several research
groups\cite{adrian,Dobrev,Kir,thesis,paper5,beatriz2}.
Already the $N=2$ superconformal algebras contain several surprising
features regarding their representation theory, most of them related to
the rank $3$ of the algebras, making them more difficult to study than
the N=1 superconformal algebras. The rank of the
superconformal algebras keeps growing with $N$ and therefore
even more difficulties can be expected for higher $N$.

The standard procedure of finding all possible irreducible highest
weight representations starts off with defining
freely generated modules over a highest weight vector, denoted as
{\it Verma modules}. A Verma module is in general
not irreducible, but the corresponding irreducible representation
is obtained as the quotient space of the Verma module divided by
all its proper submodules. Therefore, the
task of finding irreducible highest weight representations can be
reduced to the classification of all submodules of a Verma module.
Obviously, every proper submodule needs to have at least
one highest weight vector different from the highest weight vector of
the Verma module. These vectors are usually called singular vectors
of the Verma module. Conversely, a module generated on
such a singular vector defines a submodule of the Verma module.
Thus, singular vectors play a crucial r\^ole in finding submodules
of Verma modules.
However, the set of singular vectors may not generate all the
submodules. The quotient space of a Verma module divided by the submodules
generated by all singular vectors may still be reducible and may hence
contain further submodules that again contain singular vectors. But
this time they are singular vectors of the quotient space,
known as subsingular vectors of the Verma module.
Repeating this division procedure successively would ultimately lead to
an irreducible quotient space.

On the Verma modules one introduces a
hermitian contravariant form. The vanishing of the corresponding
determinant indicates the existence of a singular vector.
Therefore, a crucial step towards analysing irreducible highest weight
representations is to compute the inner product determinant. This
has been done for $N=1$\cite{kac2,meur,nam},
$N=2$\cite{adrian,nam,kato,beatriz1,DGR3}, $N=3$\cite{adrian3}, and
$N=4$\cite{adrian4,matsuda}. Once the
determinant vanishes we can conclude the existence of a singular
vector $\Psi_l$ at a certain {\it level} $l$, although
there may still be other singular vectors at higher levels even
outside the submodule generated by $\Psi_l$, the so called {\it isolated
singular vectors}. Thus the determinant
may not give all singular vectors neither does it give the dimension
of the space of singular vectors at a given level $l$,
since at levels where the determinant predicts one singular vector,
of a given type, there could in fact be more than
one linearly independent singular vectors, as it happens for the
$N=2$ superconformal algebras\cite{paper2,beatriz2}. Therefore,
the construction of specific singular vectors at levels given by
the determinant formula may not be enough. One needs in addition
information about the dimension of the space of singular vectors,
apart from the (possible) existence of isolated singular vectors .

The purpose of this paper is to give a simple procedure that derives
necessary conditions on the space of dimensions of singular vectors
of the N=2 superconformal algebras. This
will result in an upper limit for the dimension of the spaces of singular
vectors at a given level. For most
weight spaces of a Verma module
these upper limits on the dimensions will be trivial and
we obtain a rigorous proof that there cannot exist any
singular vectors for these weights.
For some weights, however, we will find necessary conditions that allow
one-dimensional singular vector spaces, as is the case for the Virasoro
algebra, or even higher dimensional spaces. The method shown in
this paper for the superconformal algebras originates from the method
used by Kent\cite{adrian1} for the Virasoro algebra\footnote{Besides the
later application to the Neveu-Schwarz $N=2$ algebra in
Ref. \icite{paper2}, only one further application is known
to us which has been achieved by Bajnok\cite{zoltan2}
for the $WA_2$ algebra.}.
Kent analytically continued the Virasoro Verma modules to generalised
Verma modules. In these generalised Verma modules he
constructed generalised singular vector expressions in terms of analytically
continued Virasoro operators. Then he
proved that if a generalised singular vector
exists at level $0$ in a generalised Verma module,
then it is proportional to the highest weight vector.
And consequently, if a generalised singular vector
exists at a given level in a generalised Verma module,
then it is unique up to proportionality.
This uniqueness can therefore be used in order
to show that the generalised singular vector expressions
for the analytically continued modules are actually singular vectors
of the Virasoro Verma module, whenever the Virasoro Verma module has
a singular vector.
As every Virasoro singular vector is at the same time a generalised
singular vector, this implies that Virasoro singular vectors also
have to be unique up to proportionality.

In this paper we focus on the uniqueness proof of Kent and show that
similar ideas can be applied directly to the superconformal
algebras. Our procedure does not require any analytical continuation
of the algebra, however, and therefore gives us a powerful method
that can easily be applied to a vast number of algebras without
the need of constructing singular vectors. We shall define the
underlying idea as {\it the concept of adapted orderings}.
For pedagogical reasons we will first apply Kent's ordering
directly to the Virasoro Verma modules. Then we will present adapted
orderings for the topological $N=2$ superconformal algebra, which is the
most interesting $N=2$ algebra for current research in this field. The
results obtained will be translated finally to the Neveu-Schwarz and to
the Ramond $N=2$ algebras. In a future publication we will further
apply these ideas to the twisted $N=2$ superconformal algebra.

The paper is structured as follows. In section \refoth{\ref{sec:Virasoro}}
we explain the concept of adapted orderings for the case of
the Virasoro algebra, which will also serve to illustrate Kent's proof
in our setting. In section \refoth{\ref{sec:SCA}},
we prove some general results on adapted orderings for superconformal
algebras, which justify the use of this method. In section
\refoth{\ref{sec:Nis2}} we review some basic results concerning the
topological $N=2$ superconformal algebra. Section \refoth{\ref{sec:aoG}}
introduces adapted orderings on generic Verma modules of the topological
$N=2$ superconformal algebra (those built on $G_0$-closed or $Q_0$-closed
highest weight vectors). This procedure is extended to {\it chiral}
Verma modules in section \refoth{\ref{sec:aoGQ}} and to
{\it no-label} Verma modules in section \refoth{\ref{sec:aonl}}.
Section \refoth{\ref{sec:dimensions}} summarises the implications
of the adapted orderings on the dimensions of the singular vector
spaces for the corresponding topological Verma modules.
Section \refoth{\ref{sec:NSR}} translates these results to the singular
vector spaces of the Neveu-Schwarz and the Ramond $N=2$
superconformal algebras.
Section \refoth{\ref{sec:conclusions}}
is devoted to conclusions and prospects. The proof of theorem
\refoth{\ref{th:kernelG}} fills several pages and readers that
are not interested in the details of this proof can simply continue
with theorem \refoth{\ref{th:kernelQ}}. In this case, the preliminary
remarks to theorems \refoth{\ref{th:kernelGQ}}
and \refoth{\ref{th:kernelnl}} should also be skipped.
Nevertheless, the main idea of the concept can easily be understood
from the introductory example of the Virasoro Verma modules in section
\refoth{\ref{sec:Virasoro}}.


\section{Virasoro algebra}
\label{sec:Virasoro}

It is a well-known fact that at a given level of
a Verma module of the Virasoro algebra there can only be one singular
vector which is unique up to proportionality. This is an immediate
consequence of the proof of the Virasoro embedding diagrams
by Feigin and Fuchs\cite{ff1}. Using an analytically
continued algebra of the Virasoro algebra,
Kent constructed in Ref. \icite{adrian1} all Virasoro
singular vectors in terms of products of analytically continued operators.
Although similar methods had already been used earlier on Verma
modules over Kac-Moody algebras\cite{malikov},
the construction by Kent not only shows the existence of
analytically continued singular vectors for any complex level
but also their uniqueness\footnote{The exact proof of Kent
showed that generalised Virasoro
singular vectors at level $0$ are scalar multiples of the identity.}.
This issue is our main interest in this paper. We shall therefore
concentrate on the part of Kent's proof that shows the uniqueness
of Virasoro singular vectors rather than the existence
of analytically continued singular vectors.
It turns out that the extension of the Virasoro algebra to
an analytically continued algebra, although needed for the part of Kent's
proof showing the existence claim, is however not necessary
for the uniqueness claim on which we will focus in this paper.
We will first motivate and define our concept of {\it adapted
orderings} for the Virasoro algebra
and will then prove some first results for the implications
of adapted orderings on singular vectors. Following
Kent\cite{adrian1} we will then introduce an ordering on the basis of a
Virasoro Verma module and describe it in our framework.
If we assume that a singular vector
exists at a fixed level, then this total ordering will show that this
singular vector has to be unique up to proportionality.

The Virasoro algebra $\vir$ is generated
by the operators $L_m$ with $m\in{\bf {Z}}$
and the {\it central extension} $C$ satisfying the commutation relations
\bea
\left[L_m,L_n\right] = (m-n) L_{m+n} +\frac{C}{12} (m^3-m) \delta_{m+n,0}
\,, & \left[C,L_m\right] = 0 \,, & m,n\in\bbbz \,.
\eea
$\vir$ can be written in its {\it triangular decomposition}
$\vir=\vir^-\oplus \vir^0 \oplus \vir^+$, with
$\vir^+=\spn{L_m:m\in{\bf {N}}}$, the
{\it positive Virasoro operators}, and $\vir^-=\spn{L_{-m}:m\in\bbbn}$,
the {\it negative Virasoro operators}. The {\it Cartan subalgebra}
is given by $\vir^0=\spn{L_0,C}$. For elements $Y$ of $\vir$
that are eigenvectors
of $L_0$ with respect to the adjoint representation we call the
$L_0$-eigenvalue the {\it level} of $Y$ and denote it by\footnote{Note
that positive generators $L_m$ have negative level $\level{L_m}= -m$.
Therefore, any positive operators $\Gamma \in \vir^+$ have a
negative level $\level{\Gamma}$.} $\level{Y}$:
$[L_0,Y]=\level{Y}Y$. The same shall be used for the universal
enveloping algebra $U(\vir)$. In particular,
elements of $U(\vir)$ of the form
$Y=L_{-p_I}\ldots L_{-p_1}$, $p_q\in\bbbz$ for $q=1,\ldots, I$,
$I\in\bbbn$,
are at level $\level{Y}=\sum_{q=1}^{I}p_q$ and we furthermore define
them to be of {\it length} $\length{Y}=I$. Finally,
for the identity operator we set $\length{1}=
\level{1}=0$. For convenience we define the graded class of subsets of
operators in $U(\vir)$ at positive level:
\bea
\sset_m &=& \{S=L_{-m_I} \ldots L_{-m_1}:\; \level{S}=m\,;\;  m_I\geq\ldots
m_1\geq 2\,; \; m_1,\ldots,m_I,I\in\bbbn\} \,,
\eea
for $m\in\bbbn$,
$\sset_0 = \{1\}$,
and also
\bea
\cset_n &=& \{X=S_m L_{-1}^{n-m}: \, S_m\in\sset_m, \, m\in\bbbn_0, \,
m\leq n\} \,,
\eea
for $n\in\bbbn_0$, which will serve to construct a basis
for Virasoro {\it Verma modules} later on.

We consider representations of $\vir$ for which the Cartan
subalgebra $\vir^0$ is diagonal. Furthermore, $C$ commutes with
all operators of $\vir$ and can hence be taken to be constant $c\in\bbbc$
(in an irreducible representation). A representation with
$L_0$-eigenvalues bounded from below contains a vector
with $L_0$-eigenvalue $\Delta$ which
is annihilated by $\vir^+$, a
{\it highest weight vector} $\ket{\Delta,c}$:
\bea
\vir^+ \ket{\Delta,c}=0\,, & L_0\ket{\Delta,c}=\Delta\ket{\Delta,c}\,, &
C\ket{\Delta,c}=c\ket{\Delta,c}\,. \label{eq:hwc}
\eea
The {\it Verma module} $\vm_{\Delta,c}$ is the left-module
$\vm_{\Delta,c}=U(\vir)\otimes_{\vir^0\oplus \vir^+}\ket{\Delta,c}$.
For $\vm_{\Delta,c}$
we choose the standard basis $\bsvm^{\Delta,c}$ as:
\bea
\bsvm^{\Delta,c} &=&
\{S_mL_{-1}^n\ket{\Delta,c}:S_m\in\sset_m\,, m,n\in\bbbn_0\} \,.
\label{eq:bbasis}
\eea
$\vm_{\Delta,c}$ and $\bsvm^{\Delta,c}$ are $L_0$-graded in a natural way.
The corresponding $L_0$-eigenvalue is called the {\it conformal weight}
and the $L_0$-eigenvalue relative to $\Delta$ is
the {\it level}. Let us introduce
\bea
\bsvm^{\Delta,c}_k &=&
\{X_k\ket{\Delta,c}:X_k\in\cset_k\} \;,\;\;
k\in\bbbn_0
\,. \label{eq:bbasisk}
\eea
Thus, $\bsvm^{\Delta,c}_k$ has conformal weight $k$ and
$\spn{\bsvm^{\Delta,c}_k}$
is the grade space of $\vm_{\Delta,c}$ at level $k$. For
$x\in\spn{\bsvm^{\Delta,c}_k}$ we again denote the level
by $\level{x}=k$.

Verma modules may not be irreducible. In order to obtain physically
relevant irreducible highest weight representations one thus needs to trace
back the proper submodules of $\vm_{\Delta,c}$ and divide them out. This
finally leads to the notion of singular vectors as any proper submodule
of $\vm_{\Delta,c}$ needs to contain a vector $\Psi_l$ that is not
proportional to the highest weight vector $\ket{\Delta,c}$ but still
satisfies the {\it highest weight vector conditions}\footnote{Note
that the trivial vector $0$ satisfies the highest weight conditions
\eq{\ref{eq:hwc}} at any level $l$. However, $0$ is proportional to
$\ket{\Delta,c}$ and is therefore not a singular vector.}
with conformal weight\footnote{An eigenvector
of $L_0$ with eigenvalue $\Delta+l$ in the Verma module $\vm_{\Delta,c}$
is usually labeled by the level $l$.}
$\Delta+l$ for some $l\in\bbbn_0$:
\bea
\vir^+\Psi_l=0 \,, & L_0\Psi_l=(\Delta+l)\Psi_l \,, & C\Psi_l=c\Psi_l
\label{eq:hwc2}
\eea
$l$ is the level of $\Psi_l$, denoted by
$\level{\Psi_l}$.
An eigenvector $\Psi_l$ of $L_0$ at level $l$ in
$\vm_{\Delta,c}$, in particular a singular vector,
can thus be written using the basis \eqoth{\ref{eq:bbasisk}}:
\bea
\Psi_l &=& \sum_{m=0}^{l}
\sum_{S_m\in \sset_m} c_{S_m} \, S_m
L^{l-m}_{-1} \ket{\Delta,c} \,, \label{eq:psil}
\eea
with coefficients $c_{S_m}\in\bbbc$. The basis decomposition
\eqoth{\ref{eq:psil}} of an $L_0$-eigenvector in $\vm_{\Delta,c}$
will be denoted the {\it normal form} of $\Psi_l$, where
$S_mL_{-1}^{l-m}\in\cset_l$ and $c_{S_m}$ will be referred to
as the {\it terms} and {\it coefficients} of $\Psi_l$, respectively.
A {\it non-trivial term} $Y\in\cset_l$ of $\Psi_l$ refers to a term
$Y$ in \eq{\ref{eq:psil}} with non-trivial coefficient $c_{Y}$.

Let $\ordering$ denote a total ordering on
$\cset_l$ with global minimum.
Thus $\Psi_l$ in \eq{\ref{eq:psil}} needs to contain an
$\ordering$-{\it smallest} $X_0\in\cset_l$
with $c_{X_0}\neq 0$ and $c_Y=0$ for
all $Y\in\cset_l$ with $Y\osm{\ordering} X_0$ and $Y\neq X_0$.
Let us assume that the ordering $\ordering$ exists with global
minimum $L_{-1}^l\in\cset_l$ and is such that for any
term $X\in\cset_l$, $X\neq L_{-1}^l$ of a vector $\Psi_l$ at level
$l\in\bbbn_0$, the action of at least one positive
operator $\Gamma$ of $\vir^+$ on $X\ket{\Delta,c}$ contains a
non-trivial basis term of $\bsvm^{\Delta,c}_{l'}$, with $l'=l+|\Gamma|$,
that cannot be obtained from any other term of $\Psi_l$
$\ordering$-larger than $X$. However, for a singular vector $\Psi_l$
the action of any positive operator of $\vir^+$ needs to vanish
and thus we find immediately that any given non-trivial singular
vector $\Psi_l$ contains in its normal form
the non-trivial term $L_{-1}^l\in\cset_l$.
Otherwise there must exist a smallest non-trivial term $X_0$
of $\Psi_l$ different from $L_{-1}^l$. However, by assumption
we can find a positive operator $\Gamma$ annihilating $\Psi_l$
but also creating a term that can only be generated from
the term $X_0$ and from no other terms of $\Psi_l$. Thus the coefficient
$c_{X_0}$ of $X_0$ is trivial and therefore $\Psi_l$ is trivial.
This motivates our definition of {\it adapted orderings} for Virasoro
Verma modules:

\bdf \label{def:viradapt}
A total ordering $\ordering$ on $\cset_l$ $(l\in\bbbn_0)$ with global
minimum is called adapted to the
subset $\cset^A_l\subset\cset_l$ in the Verma module $\vm_{\ket{\Delta,c}}$
if for any element $X_0\in\cset^A_l$ at least one positive operator
$\Gamma\in{\vir^+}$ exists for which
\bea
\Gamma{\ }X_0\ket{\Delta,c} &=&
\sum_{m=0}^{l+\level{\Gamma}}
\sum_{S_m\in \sset_m} c_{S_m}^{\Gamma X_0}  \,
S_m L^{l+\level{\Gamma}-m}_{-1} \ket{\Delta,c}
\label{eq:viradapt1}
\eea
contains a non-trivial term $\tilde{X}\in\cset_{l+\level{\Gamma}}$
(i.e. $c_{\tilde{X}}^{\Gamma X_0}\neq0$) such
that for all $Y\in\cset_l$ with $X_0 \osm{\ordering} Y$ and $X_0\neq Y$
the coefficient $c_{\tilde{X}}^{\Gamma Y}$ in
\bea
\Gamma{\ }Y\ket{\Delta,c} &=&
\sum_{m=0}^{l+\level{\Gamma}}
\sum_{S_m\in \sset_m} c_{S_m}^{\Gamma Y}  \,
S_m L^{l+\level{\Gamma}-m}_{-1} \ket{\Delta,c}
\eea
is trivial: $c_{\tilde{X}}^{\Gamma Y}=0$. The complement of $\cset^A_l$,
 ${\ }\cset^K_l=\cset_l\setminus\cset^A_l$, is the kernel with respect
to the ordering $\ordering$ in the Verma module $\vm_{\Delta,c}$.
\edf

Obviously, any total ordering on $\cset_l$
is always adapted to the subset $\emptyset\subset\cset_l$ with
ordering kernel $\cset_l^K=\cset_l$, what does not give much information.
For our purposes we need to find suitable ordering restrictions
in order to obtain the smallest possible ordering kernels (which is not
a straightforward task). In the Virasoro case
we will give an ordering such that the ordering kernel
for each $l\in\bbbn$ has just one element: $L_{-1}^l$. As indicated
in our motivation, it is then fairly simple to show that a singular
vector at level $l$ needs to have a non-trivial coefficient
for the term $L_{-1}^l$ in its normal form.
If two singular vectors have the same coefficient for this term,
then their difference is either trivial or a singular vector
with trivial $L_{-1}^l$ term. Again the latter is not allowed and
hence all singular vectors at level $l$
are unique up to proportionality.
This will be summarised in the following theorem:

\bth \label{th:virkernel}
Let $\ordering$ denote an adapted ordering in $\cset^A_l$ at level
$l\in\bbbn$ with a kernel $\cset^K_l$ consisting of just one term $K$
for a given Verma module $\vm_{\Delta,c}$. If two vectors
$\Psi_l^1$ and $\Psi_l^2$ at level $l$ in $\vm_{\Delta,c}$,
both satisfying the highest weight conditions \eq{\ref{eq:hwc2}},
have $c_{K}^1=c_{K}^2$, then
\bea
\Psi_l^1 &\equiv & \Psi_l^2 \,.
\eea
\eth
\bprf
Let us consider $\tilde{\Psi}_l=\Psi_l^1-\Psi_l^2$. The normal form
of $\tilde{\Psi}_l$ does not contain the term $K$
as $c_{K}^1=c_{K}^2$. As $\cset_l$ is a totally ordered set
with respect to $\ordering$,
the non-trivial terms of $\tilde{\Psi}_l$,
provided $\tilde{\Psi}_l$ is non-trivial,
need to have an $\ordering$-minimum $X_0\in\cset_l$.
By construction, the coefficient $\tilde{c}_{X_0}$
of $X_0$ in $\tilde{\Psi}_l$ is non-trivial, hence, $X_0$ is
contained in $\cset^A_l$. As $\ordering$ is adapted to $\cset^A_l$
we can find a positive generator $\Gamma \in\vir^+$ such that
$\Gamma X_0\ket{\Delta,c}$ contains a non-trivial term that
cannot be created from any other term of $\tilde{\Psi}_l$ which is
$\ordering$-larger than $X_0$. But $X_0$ was chosen to be the
$\ordering$-minimum of the non-trivial terms of $\tilde{\Psi}_l$.
Therefore, $\Gamma X_0\ket{\Delta,c}$ contains a non-trivial term
that cannot be created from any other term of $\tilde{\Psi}_l$. The
coefficient of this term is obviously given by $a\tilde{c}_{X_0}$ with
a non-trivial complex number $a$. Like $\Psi_l^1$ and $\Psi_l^2$,
$\tilde{\Psi}_l$ is also annihilated by any positive generator, in
particular by $\Gamma$. It follows that $\tilde{c}_{X_0}=0$ which leads
to contradiction. Thus, the set of non-trivial terms of $\tilde{\Psi}_l$
is empty and therefore $\tilde{\Psi}_l=0$. This results in
$\Psi_l^1=\Psi_l^2$.
\eprf

Equipped with definition \refoth{\ref{def:viradapt}}
and theorem \refoth{\ref{th:virkernel}}
we can now easily prove the well-known\cite{adrian1,ff1}
uniqueness of Virasoro singular vectors. Let us first review the
total ordering on $\cset_l$ defined by Kent\cite{adrian1}.
Whilst Kent used the following ordering to show that
in his generalised Virasoro Verma modules vectors at level
$0$ satisfying the highest weight conditions are actually
proportional to the highest weight vector, we will
use theorem \refoth{\ref{th:virkernel}}
to show that, furthermore, already the ordering implies
that all Virasoro singular vectors are unique at their levels up to
proportionality.

\bdf \label{def:virorder}
On the set $\cset_l$ of Virasoro operators
we introduce the total ordering $\ordering_{\vir}$ for $l\in\bbbn$.
For two elements
$X_1,X_2\in\cset_l$, $X_1\neq X_2$,
with $X_i=L_{-m^i_{I_i}}\ldots L_{-m^i_1}L_{-1}^{n^i}$,
$n^i=l-m^i_{I_i}\ldots-m^i_1$,
or $X_i=L_{-1}^l$,
$i=1,2$ we define
\bea
X_1 \osm{\ordering_{\vir}}X_2 & {\rm if} & n^1>n^2\,. \label{eq:virord1}
\eea
If, however, $n^1=n^2$ we compute the index
$j_0=\min\{j:m^1_j-m^2_j\neq 0,j=1,\ldots,\min(I_1,I_2)\}$. We then
define
\bea
X_1 \osm{\ordering_{\vir}}X_2 & {\rm if} & m^1_{j_0}<m^2_{j_0}\,.
\eea
For $X_1=X_2$ we set $\, X_1 \osm{\ordering_{\vir}}X_2$ and
$\, X_2\osm{\ordering_{\vir}}X_1$.
\edf
In order to show that definition \refoth{\ref{def:virorder}} is
well-defined, we need to prove that the set of indices $J=
\{j:m^1_j-m^2_j\neq 0,j=1,\ldots,\min(I_1,I_2)\}$ is non-trivial for
the cases that $n^1=n^2$ but $X_1\neq X_2$
and thus the minimum $j_0$ is well-defined.
Indeed, trivial $J$ either implies that at least one of $X_1$ or
$X_2$ is equal to $L_{-1}^l$, or that the positive numbers $m^i_j$ agree for
$i=1$ and $i=2$ for all $j$ from $1$ to $\min(I_1,I_2)$.
In the first case, however, as $L_{-1}^l$ is the only element of $\cset_l$
with l operators $L_{-1}$ and as we assumed $n^1=n^2$ we find
that both $X_1$ and $X_2$ must be $L_{-1}^l$. In the second case, let us
assume that $I_1<I_2$, then obviously $n^1=n^2$ and $\level{X_1}=\level{X_2}
=l$
imply $\sum_{j=1}^{I_1} m_j^1 = \sum_{j=1}^{I_1} m_j^2 +
\sum_{j=I_1+1}^{I_2} m_j^2 =  \sum_{j=1}^{I_1} m_j^1 +
\sum_{j=I_1+1}^{I_2} m_j^2$ and thus
$\sum_{j=I_1+1}^{I_2} m_j^2 =0$. As all the numbers $m_j^2$ are strictly
positive we obtain $I_1=I_2$ and thus again $X_1=X_2$.

The index $j_0$ is therefore defined for all pairs $X_1,X_2$ with $n^1=n^2$
($X_1\neq X_2$).
$j_0$ describes the first index, read from the
right to the left, for which the generators in $X_1$ and $X_2$
($L_{-1}$ excluded) are
different. For example $L_{-3}L_{-2}L_{-2}L_{-1}^3$ is
$\ordering_{\vir}$-smaller than $L_{-5}L_{-2}L_{-1}^3$ with
index $j_0=2$.
Before proceeding we ought to remark that $L_{-1}^l\in\cset_l$ is obviously
the global $\ordering_{\vir}$-minimum in $\cset_l$.
The following theorem combines
our results so far and shows the significance of $\ordering_{\vir}$.

\bth \label{th:viradap}
The ordering $\ordering_{\vir}$ is adapted to $\cset^A_l=
\cset_l\setminus\{L_{-1}^l\}$ for each level $l\in\bbbn$ and
for all Verma modules $\vm_{\Delta,c}$. The ordering kernel is
given by the single element set $\cset^K_l=\{L_{-1}^l\}$.
\eth
\bprf
The idea of the proof is a generalisation of Kent's proof
in Ref. \icite{adrian1}.
Let us consider $X_0=L_{-m_I}\ldots L_{-m_1}L_{-1}^{n_0}\in\cset^A_l$,
$n_0=l-m_I\ldots-m_1$,
$m_I\geq\ldots\geq m_1\geq 2$ for $I\in\bbbn$. We then construct
a vector $\Psi_{l}=X_0\ket{\Delta,c}$ at level $l$ in the Verma
module $\vm_{\Delta,c}$. We apply the positive
operator $L_{m_1-1}$ to $\Psi_{l}$ and write the result
in its normal form
\bea
L_{m_1-1}\Psi_{l} &=& \sum_{m=0}^{l-m_1+1}
\sum_{S_m\in \sset_m} c_{S_m} S_m L^{l-m_1+1-m}_{-1} \ket{\Delta,c} \,,
\label{eq:viradappf}
\eea
following \eq{\ref{eq:viradapt1}}.
\eq{\ref{eq:viradappf}} contains a non-trivial contribution
of $\tilde{S}=L_{-m_I}\ldots L_{-m_2}$, simply by commuting
$L_{m_1-1}$ with $L_{-m_1}$ in $X_0$ which creates another operator
$L_{-1}$
but lets $L_{-m_I}\ldots L_{-m_2}$ unchanged and thus creates the term
$\tilde{X}=L_{-m_I}\ldots L_{-m_2}L_{-1}^{n_0+1}$.
In the case $m_1=m_2=\ldots =m_j$ we simply obtain multiple copies
of this term. However, for any other
term $Y=L_{-m^Y_{J}}\ldots L_{-m^Y_1}L_{-1}^{n^Y}$ with $Y\in\cset_l$
($n^Y=l-m^Y_{J}\ldots-m^Y_1$, $J\in\bbbn_0$)
producing the term $\tilde{X}$ under the action of $L_{m_1-1}$, either
the term $Y$ needs to have already at least one $L_{-1}$ more than
$X_0$, and would consequently
be $\ordering$-smaller than $X_0$ due
to \eq{\ref{eq:virord1}}, or $L_{m_1-1}$
needs to create $L_{-1}$ by commuting through
$L_{-m^Y_{J}}\ldots L_{-m^Y_1}$.
The latter, however, is only possible if $m^Y_1<m_1$.
Otherwise the commutation relations would
not allow $L_{-1}$ being created from $L_{m^Y_1-1}$ and for
$m_1^Y=m_1$ we would ultimately find $X_0=Y$, as both terms need
to create $\tilde{X}$. Hence $m_1^Y<m_1$
and therefore one finds $Y\osm{\ordering_{\vir}} X_0$. Consequently,
there is no term $\ordering_{\vir}$-bigger than $X_0$ producing
the term $\tilde{X}$ under the action of the positive generator $L_{m_1-1}$.
\eprf

Theorem \refoth{\ref{th:viradap}} implies as an immediate
consequence the following theorem about the
uniqueness of Virasoro singular vectors.

\bth \label{th:viruni}

If the Virasoro Verma module
$\vm_{\Delta,c}$ contains a singular vector $\Psi_l$ at level $l$,
$l\in\bbbn$, then $\Psi_l$ is unique up to
proportionality. The coefficient of the term $L_{-1}^l\in\cset_l$
in the normal form of $\Psi_l$, i.e. the coefficient
of $L_{-1}^{l}\ket{\Delta,c}$, is non-trivial.
\eth
\bprf
We first show that the $L_{-1}^{l}\ket{\Delta,c}$ component in
the normal form of $\Psi_l$ is non-trivial. Let us assume
this component is trivial. The trivial vector $0$ also satisfies
the highest weight conditions \eq{\ref{eq:hwc2}} for any level $l$
and has trivial $L_{-1}^{l}\ket{\Delta,c}$ component.
According to theorem \refoth{\ref{th:viradap}},
$\{L_{-1}^l\}$ is an ordering kernel for the ordering $\ordering_{\vir}$ on
$\cset_l$. Therefore, from theorem \refoth{\ref{th:virkernel}} we
know that $\Psi_l=0$ and therefore $\Psi_l$ is not a singular vector,
which is a contradiction. Hence, we obtain that the component
of $L_{-1}^{l}\ket{\Delta,c}$ in $\Psi_l$ has to be non-trivial.
Let us now assume $\Psi_l^{\prime}$ is another singular vector
in $\vm_{\Delta,c}$ at the same level $l$ as $\Psi_l$.
We know that the coefficients
$c_{L_{-1}^l}^{\prime}$ and $c_{L_{-1}^l}$ of $\Psi_l^{\prime}$ and
$\Psi_l$ respectively are both non-trivial. Therefore
$c_{L_{-1}^l}^{\prime}\Psi_l$ and
$c_{L_{-1}^l}\Psi_l^{\prime}$ are two singular
vectors at the same level which agree in their $L_{-1}^l$ coefficient
and according to theorem \refoth{\ref{th:virkernel}} are identical.
Thus, $\Psi_l$ and $\Psi_l^{\prime}$ are proportional.
\eprf

Feigin and Fuchs\cite{ff1} have proven for which
Verma modules these unique Virasoro singular vectors do exist.



\section{Superconformal algebras and adapted orderings}
\label{sec:SCA}

A superconformal algebra is a Lie super algebra that contains
the Virasoro algebra as a subalgebra. Therefore superconformal
algebras are also known as {\it super extensions of the Virasoro
algebra}.
Thanks to Kac\cite{kac3,kac5}, Cheng and Kac\cite{kac4}, and
Kac and van de Leur\cite{kac6} all superconformal algebras
are known by now.
Let $\salg$ denote a superconformal algebra and let
$U(\salg)$ be the universal enveloping algebra of $\salg$.
We want that the
{\it energy operator} $L_0$ of the Virasoro subalgebra
is contained in our choice of the Cartan subalgebra ${\cal H}_{\salg}$
of $\salg$. $\salg$ thus
decomposes in $L_0$-grade spaces which we want to group according
to the sign of the grade:
$\salg = \salg^- \oplus \salg^0 \oplus \salg^-$ where the $L_0$-grades
of $\salg^-$, $\salg^0$, or $\salg^+$ are positive, zero, or
negative respectively\footnote{Note that for historical reasons
the elements of $\salg^-$ have positive $L_0$-grade.}.
Consequently, $U(\salg)$ also decomposes in
$L_0$-grade spaces: $U(\salg)= U(\salg)^- \oplus U(\salg)^0 \oplus
U(\salg)^+$. Obviously,
the Cartan subalgebra ${\cal H}_{\salg}$
is contained in $\salg^0$ but does not
need to be identical to $\salg^0$.
The $L_0$-grade is just one component of the roots
$\mu\in{\cal H}_{\salg}^{\ast}$ (the dual space of
${\cal H}_{\salg}$) of $\salg$. For simplicity,
let us fix a basis for ${\cal H}_{\salg}$ that contains $L_0$:
$\{L_0,H^2,\ldots,H^r\}$,
and hence let us denote the roots as $(\Delta,\mu)$ where
$\Delta$ indicates the $L_0$ component and $\mu=(\mu_2,\ldots,\mu_r)$
the vector of all other components.

Physicists are mainly interested in {\it positive energy
representations}. One thus defines a {\it highest weight vector}
$\ket{\Delta,\mu}$
as a simultaneous eigenvector of ${\cal H}_{\salg}$ with eigenvalues, the
{\it weights}, $(\Delta,\mu)$
and vanishing $\salg^+$ action: $\salg^+ \ket{\Delta,\mu}=0$.
The $L_0$-weight $\Delta = \mu(L_0)$ is the conformal
weight, which is for convenience always denoted explicitlty in addition
to the other weights $\mu$. Depending on the algebra $\salg$,
physical as well as mathematical applications may require
highest weight vectors that satisfy
additional vanishing conditions with respect to
operators of $\salg^0$ (the {\it zero modes}) with
${\cal H}_{\salg}$ normally excluded. Later on, this shall be
further explained in section \refoth{\ref{sec:Nis2}}
for the topological $N=2$ algebra.

\bdf \label{def:shw}
For a subalgebra ${\cal N}$ of $\salg^0$ that includes the Cartan subalgebra
${\cal H}_{\salg}$ we define a highest weight vector
$\ket{\Delta,\mu}^{\cal N}$
with weight $(\Delta,\mu)$:
\bea
L_0 \ket{\Delta,\mu}^{\cal N} &=& \Delta \ket{\Delta,\mu}^{\cal N} \,,
\label{eq:shwc} \\
H^i \ket{\Delta,\mu}^{\cal N} &=& \mu_i \ket{\Delta,\mu}^{\cal N} \,,
\; i=2,\ldots,r \,,  \\
\salg^+ \ket{\Delta,\mu}^{\cal N} &=& 0 \,, \label{eq:shwc2} \\
\Gamma^0 \ket{\Delta,\mu}^{\cal N} &=& 0 \,, \; \forall \: \Gamma^0 \in{\cal N}
/{\cal H}_{\salg} \,.
\eea
A Verma module is then defined analogously to
the Virasoro case as the left module $\vm^{\cal N}_{\Delta,\mu}=
U(\salg)\otimes_{{\cal N} \oplus \salg^+}\ket{\Delta,\mu}^{\cal N}$ where
we use the representation \eqs{\ref{eq:shwc}}-\eqoth{\ref{eq:shwc2}}
to act with
${\cal N} \oplus \salg^+$ on $\ket{\Delta,\mu}^{\cal N}$.
If ${\cal N}={\cal H}_{\salg}$ we shall simply write
$\vm_{\Delta,\mu}$ and $\ket{\Delta,\mu}$.
\edf

The Verma module $\vm_{\Delta,\mu}^{\cal N}$ is again graded with respect to
${\cal H}_{\salg}$ into weight spaces $\vm_{\Delta,\mu}^{{\cal N},(l,q)}$
with weights $(\Delta+l,\mu+q)$. For convenience
we shall only use the relative weights $(l,q)$ with $q=(q_2,\ldots,q_r)$
whenever we want to refer to a weight. The $L_0$ relative weight $l$
is again called the {\it level}. Also for the universal enveloping
algebra an element $Y$ with well-defined $L_0$-grade $l$ is said to
be at level $\level{Y}=l$, i.e. $[L_0,Y]=\level{Y}Y$.
As for the Virasoro case we shall define a {\it singular vector}
of a Verma module $\vm_{\Delta,\mu}^{\cal N}$ to be a vector which
is not proportional to the highest weight vector but satisfies the
highest weight conditions \eqs{\ref{eq:shwc}}-\eqoth{\ref{eq:shwc2}} with
possibly different weights.

\bdf \label{def:ssvec}
A vector
$\Psi^{{\cal N}^{\prime}}_{l,q}\in\vm_{\Delta,\mu}^{{\cal N},(l,q)}$ is
said to satisfy the highest weight conditions if
\bea
L_0 {\ }\Psi^{{\cal N}^{\prime}}_{l,q} &=& (\Delta+l)
\Psi^{{\cal N}^{\prime}}_{l,q} \,, \label{eq:shwc3} \\
H^i {\ }\Psi^{{\cal N}^{\prime}}_{l,q} &=& (\mu_i+q_i)
\Psi^{{\cal N}^{\prime}}_{l,q} \,,
\; i=2,\ldots,r \,,  \\
U(\salg)^+ {\ }\Psi^{{\cal N}^{\prime}}_{l,q} &=& 0 \,,  \\
\Gamma^0 {\ }\Psi^{{\cal N}^{\prime}}_{l,q} &=& 0 \,, \; \forall \: \Gamma^0
 \in{\cal N}^{\prime}/ U({\cal H}_{\salg}) \,, \label{eq:shwc4}
\eea
for a subalgebra\footnote{Note that this time, for convenience,
we have chosen the
universal enveloping algebra to define the highest weight conditions which
is equivalent to the earlier definition
\eqs{\ref{eq:shwc}}-\eqoth{\ref{eq:shwc2}}.}
${\cal N}^{\prime}$ of $U(\salg)^0$ that contains
$U({\cal H}_{\salg})$ and may or may not be equal to ${\cal N}$.
$\Psi^{{\cal N}^{\prime}}_{l,q}$
is called a singular vector
if in addition $\Psi^{{\cal N}^{\prime}}_{l,q}$
is not proportional to the highest weight
vector $\ket{\Delta,\mu}^{\cal N}$.
\edf

Each weight space $\vm_{\Delta,\mu}^{{\cal N},(l,q)}$ can be generated
by choosing the subset of the root space of the universal enveloping
algebra $U(\salg)$ with root $(l,q)$ that only consists of generators
not taken from ${\cal N}\oplus\salg^+$. From this set we choose elements
such that the vectors generated by acting on the highest weight vector
$\ket{\Delta,\mu}^{\cal N}$ are all linearly independent and thus form a
basis for $\vm_{\Delta,\mu}^{{\cal N},(l,q)}$.
Let $\cset^{\cal N}_{l,q}$ denote such a {\it basis set}, analogous to the
basis set for the Virasoro algebra, to be specified later.
Further, let $\bsvm^{{\cal N}}_{\Delta,\mu}$ denote any basis for
the Verma module $\vm_{\Delta,\mu}^{{\cal N}}$.
The {\it standard basis} for the weight
space $\vm_{\Delta,\mu}^{{\cal N},(l,q)}$
shall be the basis
$\tilde{\bsvm}^{{\cal N},(l,q)}_{\Delta,\mu}$
generated by the sets $\cset^{\cal N}_{l,q}$ on the highest weight
vector and the basis decomposition of $\Psi_{l,q}$ with respect to
$\tilde{\bsvm}^{{\cal N},(l,q)}_{\Delta,\mu}$ is called
{\it its normal form}.
As in the Virasoro case, an element $X$ of $\cset^{\cal N}_{l,q}$
that generates a particular basis vector in
$\tilde{\bsvm}^{{\cal N},l}_{\Delta,\mu}$ is called a {\it term} of
a vector $\Psi_{l,q}\in\vm_{\Delta,\mu}^{{\cal N},(l,q)}$ and the
corresponding coefficient $c_X$ in its basis decomposition is simply
called a {\it coefficient} of $\Psi_{l,q}$. Finally,
we call the term $X$ of $\Psi_{l,q}$ a {\it non-trivial term} if
$c_X\neq 0$. This completes the necessary notation to define adapted
orderings on $\cset^{\cal N}_{l,q}$ just like in the Virasoro case.

\bdf \label{def:adapt}
A total ordering $\ordering$ on $\cset^{\cal N}_{l,q}$ with global
minimum is called adapted to the
subset $\cset^{{\cal N},A}_{l,q}\subset\cset_{l,q}^{\cal N}$
in the Verma module $\vm^{\cal N}_{\Delta,\mu}$
with annihilation operators
${\cal K}\subset U(\salg)^+\oplus U(\salg)^0/U({\cal H}_{\salg})$
if for any element $X_0\in\cset^{{\cal N},A}_{l,q}$ at least one
annihilation operator
$\Gamma \in{\cal K}$ exists for which
\bea
\Gamma \, X_0\ket{\Delta,\mu} &=&
\sum_{X\in \bsvm^{{\cal N}}_{\Delta,\mu}} c_X^{\Gamma X_0} \, X
\label{eq:adapt1}
\eea
contains a non-trivial term
$\tilde{X}\in\bsvm^{{\cal N}}_{\Delta,\mu}$
(i.e. $c_{\tilde{X}}^{\Gamma X_0}\neq0$) such
that for all $Y\in\cset^{\cal N}_{l,q}$
with $X_0 \osm{\ordering} Y$ and $X_0\neq Y$ the coefficient
$c_{\tilde{X}}^{\Gamma Y}$ in
\bea
\Gamma \, Y\ket{\Delta,\mu} &=&
\sum_{X\in \bsvm^{{\cal N}}_{\Delta,\mu}} c_X^{\Gamma Y} \, X
\eea
 is trivial: $c_{\tilde{X}}^{\Gamma Y}=0$.
The complement of $\cset^{{\cal N},A}_{l,q}$,
${\ }\cset^{{\cal N},K}_{l,q}=\cset^{\cal N}_{l,q}\setminus
\cset^{{\cal N},A}_{l,q}$ is the kernel with respect to
the ordering $\ordering$ in the Verma module $\vm^{\cal N}_{\Delta,\mu}$.
Here $\bsvm^{{\cal N}}_{\Delta,\mu}$ represents a basis
that can be chosen suitably for each $X_0$ and may or may not be the
standard basis\footnote{In the $N=2$ case we will
choose for most $X_0$ the standard basis
with only very few but important exceptions.}.
\edf

In the motivation to theorem \refoth{\ref{th:virkernel}} we assumed the
existence of an ordering with the smallest kernel consisting of one
element only. For the $N=2$ algebras we will find ordering kernels which
contain more than one element and also ordering kernels that are trivial.
We saw in theorem \refoth{\ref{th:viruni}} that if the ordering kernel
has only one element (the global minimum $L_{-1}^l$ in the case of
the Virasoro algebra) then
any singular vector needs to have a non-trivial coefficient for
this element. The following theorems reveal
what can be implied if the ordering kernel
consists of more than one element or of none at all.
\bth \label{th:kernel}
Let $\ordering$ denote an adapted ordering on $\cset^{{\cal N},A}_{l,q}$
at weight $(l,q)$ with kernel $\cset^{{\cal N},K}_{l,q}$
for a given Verma module $\vm^{\cal N}_{\Delta,\mu}$ and annihilation
operators ${\cal K}$. If two vectors $\Psi^{{\cal N}^{\prime},1}_{l,q}$
and $\Psi^{{\cal N}^{\prime},2}_{l,q}$ at the same level $l$ and weight
$q$, satisfying the highest weight conditions
\eqs{\ref{eq:shwc3}}-\eqoth{\ref{eq:shwc4}}
with ${\cal N}^{\prime}={\cal K}$,
have $c_{X}^1=c_{X}^2$ for all $X\in\cset^{{\cal N},K}_{l,q}$, then
\bea
\Psi^{{\cal N}^{\prime},1}_{l,q} &\equiv &
\Psi^{{\cal N}^{\prime},2}_{l,q}\,.
\eea
\eth
\bprf
Let us consider $\tilde{\Psi}_{l,q} = \Psi^{{\cal N}^{\prime},1}_{l,q} -
\Psi^{{\cal N}^{\prime},2}_{l,q}$.
The normal form of $\tilde{\Psi}_{l,q}$
does not contain any terms of the ordering kernel
$\cset^K_{l,q}$, simply because $c_{X}^1=c_{X}^2$ for all
$X\in\cset^{{\cal N},K}_{l,q}$. As $\cset^{\cal N}_{l,q}$
is a totally ordered set with respect to $\ordering$ which has a global
minimum, the non-trivial terms of $\tilde{\Psi}_{l,q}$,
provided $\tilde{\Psi}_{l,q}$ is non-trivial,
need to have an $\ordering$-minimum $X_0\in\cset^{\cal N}_{l,q}$.
By construction, the coefficient $\tilde{c}_{X_0}$
of $X_0$ in $\tilde{\Psi}_{l,q}$ in its normal form
is non-trivial, hence, $X_0$ is also
contained in $\cset^{{\cal N},A}_{l,q}$. As $\ordering$ is adapted
to $\cset^{{\cal N},A}_{l,q}$ one can find an annihilation operator
$\Gamma \in{\cal K}$ such that
$\Gamma X_0\ket{\Delta,\mu}^{\cal N}$ contains a non-trivial
term (for a suitably chosen basis depending on $X_0$)
that cannot be created by any other term of $\tilde{\Psi}_{l,q}$ which is
$\ordering$-larger than $X_0$. But $X_0$ was chosen to be the
$\ordering$-minimum of $\tilde{\Psi}_{l,q}$. Therefore,
$\Gamma X_0\ket{\Delta,\mu}^{\cal N}$ contains a non-trivial term
that cannot be created from any other term of $\tilde{\Psi}_{l,q}$. The
coefficient of this term is obviously given by $a\tilde{c}_{X_0}$ with
a non-trivial complex number $a$. Together with
$\Psi^{{\cal N}^{\prime},1}_{l,q}$ and $\Psi^{{\cal N}^{\prime},2}_{l,q}$,
$\tilde{\Psi}_{l,q}$ is also annihilated by any annihilation operator, in
particular by $\Gamma$. It follows that $\tilde{c}_{X_0}=0$,
contrary to our original assumption.
Thus, the set of non-trivial terms of $\tilde{\Psi}_{l,q}$
is empty and therefore $\tilde{\Psi}_{l,q}=0$. This results in
$\Psi^{{\cal N}^{\prime},1}_{l,q}=\Psi^{{\cal N}^{\prime},2}_{l,q}$.
\eprf

Theorem \refoth{\ref{th:kernel}}
states that if two singular vectors at the same
level and weight agree on the ordering kernel, then they are identical.
The coefficients of a singular vector with respect to the ordering
kernel are therefore sufficient to distinguish singular vectors.
If the ordering kernel is trivial we consequently find $0$ as the only
vector that can satisfy the highest weight conditions.

\bth \label{th:striv}
Let $\ordering$ denote an adapted ordering on
$\cset^{{\cal N},A}_{l,q}$ at weight $(l,q)$ with trivial kernel
$\cset^{{\cal N},K}_{l,q}=\emptyset$
for a given Verma module $\vm^{\cal N}_{\Delta,\mu}$ and annihilation
operators ${\cal K}$. A vector $\Psi^{{\cal N}^{\prime}}_{l,q}$
at level $l$ and weight $q$ satisfying the highest weight conditions
\eqs{\ref{eq:shwc3}}-\eqoth{\ref{eq:shwc4}}
with ${\cal N}^{\prime}={\cal K}$, is therefore
trivial. In particular, this shows that there are no singular vectors.
\eth
\bprf
We again make use of the fact that the trivial vector $0$
satisfies any vanishing conditions for any level $l$ and weight $q$.
As the ordering kernel is trivial the components of the vectors $0$
and $\Psi^{{\cal N}^{\prime}}_{l,q}$
agree on the ordering kernel and using theorem \refoth{\ref{th:kernel}}
we obtain $\Psi^{{\cal N}^{\prime}}_{l,q}=0$.
\eprf

We now know that singular vectors can be classified by their
components on the ordering kernel.
As we shall see if the ordering kernel has $n$ elements,
then the space of singular vectors for this weight is at most
$n$-dimensional. Conversely, one could ask if there
are singular vectors corresponding to all possible combinations of
elements of the ordering kernel. In general this will not be the case,
however, for the Virasoro algebra\cite{adrian1} and
for the Neveu-Schwarz $N=2$ algebra\cite{paper2} it has been shown that
for each element of the ordering kernel there exists a
singular vector
for suitably defined analytically continued
Verma modules. Some of these generalised singular vectors lie in the
embedded original non-continued Verma module and are therefore singular
in the above sense. We finally conclude with the following theorem
summarising all our findings so far.

\bth \label{th:dims}
Let $\ordering$ denote an adapted ordering on $\cset^{{\cal N},A}_{l,q}$
at weight $(l,q)$ with kernel $\cset^{{\cal N},K}_{l,q}$
for a given Verma module $\vm^{\cal N}_{\Delta,\mu}$ and annihilation
operators ${\cal K}$.
If the ordering kernel $\cset^{{\cal N},K}_{l,q}$ has $n$ elements,
then there are at most $n$ linearly independent singular vectors
$\Psi^{{\cal N}^{\prime}}_{l,q}$ in $\vm^{\cal N}_{\Delta,\mu}$ with
weight $(l,q)$ and ${\cal N}^{\prime}={\cal K}$.
\eth
\bprf
Suppose there were more than $n$ linearly independent singular vectors
$\Psi^{{\cal N}^{\prime}}_{l,q}$
in $\vm^{\cal N}_{\Delta,\mu}$ with weight $(l,q)$.
We choose $n+1$ linearly independent singular vectors
among them $\Psi_1$,$\ldots$,$\Psi_{n+1}$. The ordering
kernel $\cset^{{\cal N},K}_{l,q}$ has the $n$ elements
$K_1$,$\ldots$,$K_n$. Let $c_{jk}$ denote the coefficient of
the term $K_j$ in the vector $\Psi_k$ in its standard basis
decomposition. The coefficients $c_{jk}$ thus form
a $n$ by $n+1$ matrix $C$. The homogeneous system
of linear equations $C\lambda=0$ thus has a non-trivial
solution $\lambda^0=(\lambda^0_1,\ldots,\lambda^0_{n+1})^{T}$
for the vector $\lambda$.
We then form the linear combination $\Psi=\sum_{i=1}^{n+1}
\lambda^0_i\Psi_i$. Obviously, the coefficient of $K_j$ in
the vector $\Psi$ in its normal form
is just given by the $j$-th component of the vector $C\lambda$
which is trivial for $j=1,\ldots,n$. Hence, the coefficients of $\Psi$
are trivial on the ordering kernel. On the other hand, $\Psi$ is a linear
combination of singular vectors and therefore also satisfies the highest
weight conditions with ${\cal N}^{\prime}={\cal K}$
just like the trivial vector $0$. Due to theorem \refoth{\ref{th:kernel}}
one immediately finds that $\Psi\equiv 0$
and therefore $\sum_{i=1}^{n+1}\lambda_i\Psi_i=0$. This, however, is a
non-trivial decomposition of $0$ contradicting the assumption
that $\Psi_1$,$\ldots$, $\Psi_{n+1}$ are linearly independent.
\eprf


\section{Topological $N=2$ superconformal Verma modules}
\label{sec:Nis2}

We will now apply the construction developed in the previous
section to the topological $N=2$ superconformal algebra. We first introduce
an adapted ordering on the basis of the $N=2$ Verma modules.
Consequently the size of the ordering kernel will reveal a
maximum for the degrees of
freedom of the singular vectors in the same $N=2$ grade space.
As the representation theory of the $N=2$ superconformal algebras
has different types of Verma modules we
will see that the corresponding ordering kernels also allow
different degrees of freedom.

The topological $N=2$ superconformal algebra $\ta$ is a super Lie algebra
which contains the Virasoro generators $\cL_m$ with trivial central
extension\footnote{Note our slightly different notation for the Virasoro
generators $\cL_{n}$ in the topological $N=2$ case.},
a Heisenberg algebra $\cH_m$ corresponding to the U(1) current, and the
fermionic generators $\cG_m$ and $\cQ_m$, $m\in\bbbz$ corresponding to
two anticommuting fields with conformal weights 2 and 1 respectively.
$\ta$ satisifies the (anti-)commutation relations\cite{DVV}

\be
\begin{array}{ll}
\left[\cL_m,\cL_n\right] = (m-n)\cL_{m+n}\,, &
\left[\cH_m,\cH_n\right] = \frac{C}{3}m\delta_{m+n}\,,\\
\left[\cL_m,\cG_n\right] = (m-n)\cG_{m+n}\,, &
\left[\cH_m,\cG_n\right] = \cG_{m+n}\,,\\
\left[\cL_m,\cQ_n\right] = -n\cQ_{m+n}\,, &
\left[\cH_m,\cQ_n\right] = -\cQ_{m+n}\,,\\
\left[\cL_m,\cH_n\right] = -n\cH_{m+n}+\frac{C}{6}(m^2+m)\delta_{m+n}\,,\\
\left\{\cG_m,\cQ_n\right\} = 2\cL_{m+n}-2n\cH_{m+n}
+\frac{C}{3}(m^2+m)\delta_{m+n}\,,\\
\left\{\cG_m,\cG_n\right\} = \left\{\cQ_m,\cQ_n\right\} = 0\,, &
 m,~n\in\bbbz\,.\label{topalgebra}
\end{array}
\ee

The central term $C$ commutes with all other operators and can therefore
be fixed again as $c\in\bbbc$.
${\cal H}_{\ta}=\spn{\cL_0,\cH_0, C}$
defines a commuting subalgebra
of $\ta$, which can therefore be
diagonalised simultaneously. Generators with positive
index span the set of {\it positive operators} $\ta^+$ of $\ta$
and likewise generators with negative index
span the set of {\it negative operators} $\ta^-$ of $\ta$:
\bea
\ta^{+} &=& \spn{\cL_{m},\cH_{m},\cG_{n},\cQ_{n}: m,n\in\bbbn} \,, \\
\ta^{-} &=& \spn{\cL_{-m},\cH_{-m},\cG_{-n},\cQ_{-n}: m,n\in\bbbn} \,.
\eea
\noi The {\it zero modes} are spanned by $\ta^0=\spn{\cL_0,\cH_0,C,\cG_0,
\cQ_0}$ such that the generators $\{\cG_0,\cQ_0\}$ classify the
different choices of Verma modules. $\cQ_0$ has the properties of a
BRST-charge\cite{DVV} so that the energy-momentum tensor is BRST-exact:
$\cL_m=1/2 \, \{\cG_m,\cQ_0\}$.

Using definition \refoth{\ref{def:shw}}
a simultaneous eigenvector $\ket{\Delta,q,c}^{\cal N}$ of
${\cal H}_{\ta}$ with $\cL_0$
eigenvalue $\Delta$, $\cH_0$ eigenvalue $q$, $C$ eigenvalue\footnote{For
simplicity from now on we will supress the eigenvalue of $C$ in
$\ket{\Delta,q,c}^{\cal N}$ and simply write $\ket{\Delta,q}^{\cal N}$.}
$c$, and vanishing
$\ta^{+}$ action is called a {\it highest weight vector}. Each
representation with lower bound for the
eigenvalues of $\cL_0$ needs to contain
a highest weight vector. Additional vanishing conditions ${\cal N}$
are possible only with respect to the operators $\cG_0$ and $\cQ_0$
which may or may not annihilate a highest weight vector.
The different types of annihilation conditions have been
analysed in Ref. \icite{beatriz2} resulting as follows. One can
distinguish $4$ different types of highest weight vectors
$\ket{\Delta,q}^{\cal N}$ labeled by a superscript
$N\in\{G,Q,GQ\}$, or no superscript at all: highest weight vectors
$\ket{\Delta,q}^{GQ}$ annihilated by both $\cG_0$ and $\cQ_0$
({\it chiral})\footnote{Chirality conditions are important for
physics\cite{LVW,DVV,beatriz1}.},
highest weight vectors $\ket{\Delta,q}^{G}$ annihilated by $\cG_0$ but
not by $\cQ_0$  ($\cG_0$-closed), highest weight vectors
$\ket{\Delta,q}^{Q}$ annihilated by $\cQ_0$ but not by $\cG_0$
($\cQ_0$-closed), and finally highest weight vectors  $\ket{\Delta,q}$
that are neither annihilated by $\cG_0$ nor by $\cQ_0$  ({\it no-label}).

Since $2\cL_0=\cG_0\cQ_0+\cQ_0\cG_0$, a chiral vector, annihilated by
both $\cG_0$ and $\cQ_0$, necessarily has vanishing $\cL_0$-eigenvalue.
On the other hand, any highest weight vector $\ket{\Delta,q}$
that is neither annihilated by $\cG_0$
nor by $\cQ_0$ can be decomposed into $\frac{1}{2\Delta}\cG_0\cQ_0
\ket{\Delta,q}
+ \frac{1}{2\Delta}\cQ_0\cG_0 \ket{\Delta,q}$,
provided $\Delta\neq 0$. In this case
the whole representation decomposes into a direct sum of two submodules
one of them containing the $\cG_0$-closed highest weight vector
$\cG_0\cQ_0 \ket{\Delta,q}$ and the other one containing the
$\cQ_0$-closed highest weight vector $\cQ_0\cG_0 \ket{\Delta,q}$.
Therefore, for {\it no-label} highest weight vectors, annihilated
neither by $\cG_0$ nor by $\cQ_0$, we only need to consider the cases
with $\Delta =0$; i.e. the highest weight vectors that cannot be
expressed as linear combinations of $\cG_0$-closed and $\cQ_0$-closed
highest weight vectors. From now on {\it no-label} will refer
exclusively to such highest weight vectors with $\Delta=0$.

\btab{|l|l|l|}
\hline \label{tab:hw}
$\ket{0,q}$ & $\cG_0\ket{0,q}\neq 0$ and
$\cQ_0\ket{0,q}\neq 0$ &  no-label \\
\hline
$\ket{\Delta,q}^G$ & $\cG_0\ket{\Delta,q}=0$ and
$\cQ_0\ket{\Delta,q}\neq 0$ & $\cG_0$-closed \\
\hline
$\ket{\Delta,q}^Q$ & $\cG_0\ket{\Delta,q}\neq 0$ and
$\cQ_0\ket{\Delta,q}=0$ & $\cQ_0$-closed \\
\hline
$\ket{0,q}^{GQ}$ & $\cG_0\ket{0,q}=0$ and
$\cQ_0\ket{0,q}=0$ & chiral \\
\hline
\etab{Topological highest weight vectors.}

\noi
Hence, according to definition \refoth{\ref{def:shw}}, we have the
following $4$ different types of topological Verma modules,
as shown in \tab{\ref{tab:hw}}:

\bea
\label{eq:vms}
\vm_{0,q} &=& U(\ta)\otimes_{{\cal H}_{\ta}\oplus\ta^+}\ket{0,q} \,,\\
\vm_{\Delta,q}^G &=& U(\ta)\otimes_{{\cal H}_{\ta}\oplus\ta^+\oplus
\spn{\cG_0}}\ket{\Delta,q}^G \,,\\
\vm_{\Delta,q}^Q &=& U(\ta)\otimes_{{\cal H}_{\ta}\oplus\ta^+\oplus
\spn{\cQ_0}}\ket{\Delta,q}^Q \,,\\
\vm_{0,q}^{GQ} &=& U(\ta)\otimes_{{\cal H}_{\ta}\oplus\ta^+\oplus
\spn{\cG_0,\cQ_0}}\ket{0,q}^{GQ} \,.
\eea

The Verma modules of types $\vm_{\Delta,q}^G$ and $\vm_{\Delta,q}^Q$,
based on $\cG_0$-closed or $\cQ_0$-closed highest weight vectors, are
called\cite{beatriz2} {\it generic} Verma modules, whereas the Verma
modules of types $\vm_{0,q}$ and $\vm_{0,q}^{GQ}$ are called {\it no-label}
and {\it chiral} Verma modules, respectively, for obvious
reasons\footnote{As explained in Ref. \icite{beatriz2}, the chiral Verma
modules $\vm_{0,q}^{GQ}$, built on chiral highest weight vectors, are not
complete Verma modules because the chirality constraint is not required
(just allowed) by the algebra.}.

For elements $Y$ of $\ta$ which are eigenvectors of ${\cal H}_{\ta}$
with respect
to the adjoint representation we define similarly to the Virasoro case
the {\it level} $\levelt{Y}$ as $[\cL_0,Y]=\levelt{Y}Y$ and in
addition the {\it charge} $\charget{Y}$ as $[\cH_0,Y]=\charget{Y}Y$.
In particular, elements of the form
\bea
Y&=&\cL_{-l_L} \ldots \cL_{-l_1}\cH_{-h_H} \ldots \cH_{-h_1}
\cQ_{-q_Q} \ldots \cQ_{-q_1}\cG_{-g_G} \ldots \cG_{-g_1}
\eea
\noi and any reorderings of $Y$ have level $\levelt{Y}=\sum_{j=1}^{L}l_j
+\sum_{j=1}^{H}h_j+\sum_{j=1}^{Q}q_j+\sum_{j=1}^{G}g_j$ and
charge $\charget{Y}=G-Q$. For these elements we shall also
define their {\it length} $\length{Y}=L+H+G+Q$. Again, we shall set
$\levelt{1}=\charget{1}=\length{1}=0$.
For convenience we define the following sets of negative
operators for $m\in\bbbn$
\bea
\lset_m &=& \{ Y=\cL_{-l_L} \ldots \cL_{-l_1}: \; l_L\geq\ldots\geq l_1\geq 2,
\; \levelt{Y}=m\} \,, \\
\hset_m &=& \{ Y=\cH_{-h_H} \ldots \cH_{-h_1}: \; h_H\geq\ldots\geq h_1\geq 1,
\; \levelt{Y}=m\} \,, \\
\gset_m &=& \{ Y=\cG_{-g_G} \ldots \cG_{-g_1}: \; g_G > \ldots > g_1\geq 2,
\; \levelt{Y}=m\} \,, \\
\qset_m &=& \{ Y=\cQ_{-q_Q} \ldots \cQ_{-q_1}: \; q_Q > \ldots > q_1\geq 2,
\; \levelt{Y}=m\} \,, \\
\lset_0 &=& \hset_0 ~=~ \gset_0 ~=~ \qset_0 ~=~ \{1\} \,.
\eea
We are now able to define a graded basis for the Verma modules
as described in the previous section.
We choose $l\in\bbbn_0$, $n\in\bbbz$ and
define:
\bea
\sset^G_{m,n} &=&
\left\{Y=LHGQ: \; L\in\lset_l, \; H\in\hset_h, \; G\in\gset_g, \; Q\in\qset_q, 
\right. \nn
\\
&& \left. \levelt{Y}=m=l+h+g+q, \;\charget{Y}=n=\charget{G}+\charget{Q},\;
\,l,h,g,q\in\bbbn_0\right\} \,, \\
\sset^Q_{m,n} &=&
\left\{Y=LHQG: \;L\in\lset_l, \;H\in\hset_h, \;Q\in\qset_q, \;G\in\gset_g,
\right. \nn
\\
&& \left. \levelt{Y}=m=l+h+g+q, \;\charget{Y}=n=\charget{Q}+\charget{G},\;
\,l,h,g,q\in\bbbn_0\right\} \,.
\eea
And finally for $m\in\bbbn_0$, $n\in\bbbz$:
\bea
\cset^G_{m,n}
&=& \left\{S_{p,q} \, \cL_{-1}^{m-p-r_1-r_2}
\cG_{-1}^{r_1}\cQ_{-1}^{r_2}\cQ_0^{r_3}: \;S_{p,q}\in\sset^G_{p,q},
\;p\in\bbbn_0,\;r_1,r_2,r_3\in\{0,1\}, \right. \nn \\
&& \left. m-p-r_1-r_2\geq 0, \, n=q+r_1-r_2-r_3 \right\} \,,
\label{eq:csetG} \\
\cset^Q_{m,n}
&=& \left\{S_{p,q} \, \cL_{-1}^{m-p-r_1-r_2}
\cQ_{-1}^{r_1}\cG_{-1}^{r_2}\cG_0^{r_3}: \;S_{p,q}\in\sset^Q_{p,q},
\;p\in\bbbn_0,\;r_1,r_2,r_3\in\{0,1\}, \right. \nn \\
&& \left. m-p-r_1-r_2\geq 0, \, n=q-r_1+r_2+r_3 \right\} \,.
\eea
Thus, a typical element of $\cset^G_{m,n}$ is of the form
\bea
Y &=& \cL_{-l_L}\ldots\cL_{-l_1}\cH_{-h_H}\ldots\cH_{-h_1}
\cG_{-g_G}\ldots\cG_{-g_1}\cQ_{-q_Q}\ldots\cQ_{-q_1}
\cL_{-1}^{m^{\prime}}\cG_{-1}^{r_1}\cQ_{-1}^{r_2}\cQ_0^{r_3} \,,
\eea
$r_1,r_2,r_3\in\{0,1\}$, such that
$\levelt{Y}=m$ and $\charget{Y}=n$.
$S_{p,q}$ of $Y\in\cset_{m,n}^G$ or of $Y\in\cset_{m,n}^Q$ is
called the {\it leading part of} $Y$ and is denoted by $Y^{\ast}$.
Hence, one can define the following standard bases:
\bea
{\cal B}_{\ket{\Delta,q}^G} &=& \left\{ Y\ket{\Delta,q}^G:
\; Y\in\cset^G_{m,n} \; m\in\bbbn_0, \;n\in\bbbz \right\} \,, \nn \\
{\cal B}_{\ket{\Delta,q}^Q} &=& \left\{ Y\ket{\Delta,q}^Q:
\; Y\in\cset^Q_{m,n} \; m\in\bbbn_0, \;n\in\bbbz  \right\} \,,
\label{eq:basis1}
\eea

\noi
obtaining finally $\vm_{\Delta,q}^G =\spn{{\cal B}_{\ket{\Delta,q}^G}}$
and $\vm_{\Delta,q}^Q=\spn{{\cal B}_{\ket{\Delta,q}^Q}}$.
For Verma modules built on no-label and chiral highest weight vectors,
$\vm_{0,q}$ and $\vm_{0,q}^{G,Q}$ both with
$\Delta=0$, one defines in exactly the same way:
\bea
\sset_{m,n} &=&
\left\{Y=LHGQ: \;L\in\lset_l, \;H\in\hset_h, \;G\in\gset_g, \;Q\in\qset_q,
\right. \nn
\\
&& \left. \levelt{Y}=m=l+h+g+q, \;\charget{Y}=n=\charget{G}+\charget{Q},
\;l,h,g,q\in\bbbn_0\right\} \, \\
\cset_{m,n}
&=& \left\{S_{p,q} \, \cL_{-1}^{m-p-r_1-r_2}
\cG_{-1}^{r_1}\cQ_{-1}^{r_2}\cG_0^{r_4}\cQ_0^{r_3}: \;S_{p,q}\in\sset_{p,q},
\;p\in\bbbn_0,\;r_1,r_2,r_3,r_4\in\{0,1\}, \right. \nn \\
&& \left. m-p-r_1-r_2\geq 0, \, n=q+r_1-r_2+r_4-r_3 \right\} \,, \\
\cset^{GQ}_{m,n}
&=& \left\{S_{p,q} \, \cL_{-1}^{m-p-r_1-r_2}
\cG_{-1}^{r_1}\cQ_{-1}^{r_2}: \;S_{p,q}\in\sset_{p,q},
\;p\in\bbbn_0,\;r_1,r_2\in\{0,1\}, \nn \right. \\
&& \left. m-p-r_1-r_2\geq 0, \, n=q+r_1-r_2 \right\} \,.
\eea
$S_{p,q}$ of $Y\in\cset_{m,n}$ or of $Y\in\cset_{m,n}^{GQ}$ will be
also called the {\it leading part} $Y^{\ast}$ {\it of} $Y$.
One obtains the following standard bases for the modules
$\vm_{0,q}$ and $\vm_{0,q}^{G,Q}$:
\bea
{\cal B}_{\ket{0,q}} &=& \left\{ Y\ket{0,q}:
\; Y\in\cset_{m,n}, \; m\in\bbbn_0, \;n\in\bbbz \right\} \,, \nn \\
{\cal B}_{\ket{0,q}^{GQ}} &=& \left\{ Y\ket{0,q}^{GQ}:
\; Y\in\cset^{GQ}_{m,n}, \; m\in\bbbn_0, \;n\in\bbbz  \right\} \,.
\label{eq:basis2}
\eea

The bases \eq{\ref{eq:basis1}} and \eq{\ref{eq:basis2}} are naturally
$\bbbn_0\times\bbbz$ graded with respect to their ${\cal H}_{\ta}$
eigenvalues relative to the eigenvalues $(\Delta,q)$ of the highest
weight vector. For an eigenvector $\Psi_{l,p}$ of ${\cal H}_{\ta}$
in $\vm_{\Delta,q}^N$ the $\cL_0$-eigenvalue is $\Delta+l$
and the $\cH_0$-eigenvalue is $q+p$ with $l\in\bbbn_0$ and $p\in\bbbz$.
We define the {\it level} $\levelt{\Psi_{l,p}}=l$ and
{\it charge} $\charget{\Psi_{l,p}}=p$.

Like for the Virasoro case, we shall use $\cset^G_{m,n}$, $\cset^Q_{m,n}$,
$\cset_{m,n}$ and $\cset^{GQ}_{m,n}$
in order to define the {\it normal form} of an eigenvector $\Psi_{l,p}$
of ${\cal H}_{\ta}$. It is defined to be the basis decomposition with
respect to the corresponding standard bases \eq{\ref{eq:basis1}} and
\eq{\ref{eq:basis2}}. Again, we call the operators
$X\in\cset_{l,p}^{G}$, or $X\in\cset_{l,p}^{Q}$, or $X\in\cset_{l,p}$, or
$X\in\cset_{l,p}^{GQ}$ simply the {\it terms} of
$\Psi_{l,p}$ and the coefficients $c_X$ its {\it coefficients}.

We introduce topological singular vectors according to definition
\refoth{\ref{def:ssvec}} as ${\cal H}_{\ta}$ eigenvectors that are not
proportional to the highest weight vector but are annihilated by
$\ta^+$ and may also satisfy additional vanishing conditions with
respect to the operators $\cG_0$ and $\cQ_0$. Therefore one also
distinguishes singular vectors $\Psi_{l,p}$, $\Psi^G_{l,p}$, $\Psi^Q_{l,p}$
and $\Psi^{GQ}_{l,p}$ carrying the superscript $G$ and/or $Q$ depending on
whether the singular vector is annihilated by $\cG_0$ and/or
$\cQ_0$. Obviously one obtains similar restrictions on the eigenvalues
of $\Psi_{l,p}^{{\cal N}^{\prime}} \in\vm^{\cal N}_{\Delta,q}$
as for the highest weight vectors, as shown in \tab{\ref{tab:svecs}}.

\btab{|l|l|l|}
\hline \label{tab:svecs}
$\Psi_{l,p}$ & $\cG_0\Psi_{l,p}\neq 0$ and
$\cQ_0\Psi_{l,p}\neq 0$; $l+\Delta=0$ & {\it no-label} \\
\hline
$\Psi_{l,p}^G$ & $\cG_0\Psi_{l,p}=0$ and
$\cQ_0\Psi_{l,p}\neq 0$ & $\cG_0$-closed \\
\hline
$\Psi_{l,p}^Q$ & $\cG_0\Psi_{l,p}\neq 0$ and
$\cQ_0\Psi_{l,p}=0$ & $\cQ_0$-closed \\
\hline
$\Psi_{l,p}^{GQ}$ & $\cG_0\Psi_{l,p}=0$ and
$\cQ_0\Psi_{l,p}=0$; $l+\Delta=0$& chiral \\
\hline
\etab{Topological singular vectors.}

As there are $4$ types of topological Verma modules and $4$ types of
topological singular vectors one might think of $16$ different
combinations of singular vectors in Verma modules. However, as will be
explained later, no-label and chiral singular vectors do not exist
neither in no-label Verma modules nor in chiral Verma modules (with
one exception: chiral singular vectors at level 0 in no-label
Verma modules). Most of these types
of singular vectors are connected via the $N=2$
{\it topological spectral flow mappings}\cite{BJI3,beatriz3,beatriz2}
which have been analysed in detail in Refs. \icite{beatriz3,beatriz2}.


\section{Adapted orderings on generic Verma modules $\vm_{\Delta,q}^G$ and
$\vm_{\Delta,q}^Q$}
\label{sec:aoG}

We will now introduce total orderings $\ordering_G$ and
$\ordering_Q$ on $\cset^G_{m,n}$ and $\cset^Q_{m,n}$ respectively.
For convenience, however, we shall first give an ordering
on the sets $\lset_m$, $\hset_m$, $\gset_m$ and $\qset_m$.

\bdf \label{def:lorder}
Let $Y$ denote either $\lset$, $\hset$, $\gset$, or $\qset$,
(but the same throughout this definition)
and take two elements $X_i\in Y_{m^i}$ for $m^i\in\bbbn_0$,
$i=1,2$, such that $X_i=Z^i_{-m_{\length{X_i}}^i}
\ldots Z^i_{-m^i_1}$, $\levelt{X_i}=m^i$ or
$X_i=1$, $i=1,2$, with $Z^i_{-m^i_j}$ being an operator of the type
$\cL_{-m^i_j}$, $\cH_{-m^i_j}$, $\cG_{-m^i_j}$, or $\cQ_{-m^i_j}$
depending on
whether $Y$ denotes $\lset$, $\hset$, $\gset$ or $\qset$ respectively.
For $X_1\neq X_2$
we compute the index\footnote{For subsets of $\bbbn$ we define
$\min\emptyset=0$.}
$j_0=\min\{j:m^1_j-m^2_j\neq 0,j=1,\ldots,\min(\length{X_1},\length{X_2})\}$.
$j_0$ is, if non-trivial, the index for which the level of the operators
in $X_1$ and $X_2$ first
disagree when read from the right to the left.
For $j_0>0$ we then define
\bea
X_1\osm{Y}X_2 & {\rm if} & m^1_{j_0}<m^2_{j_0} \,.
\eea
If, however, $j_0=0$, we set
\bea
X_1\osm{Y}X_2 & {\rm if} & \length{X_1}>\length{X_2} \,.
\eea
For $X_1=X_2$ we set $\, X_1\osm{Y}X_2$ and $\, X_2\osm{Y}X_1$.
\edf

\noi
Some examples of definition \refoth{\ref{def:lorder}} are:
\bea
\cL_{-4}\cL_{-3}\cL_{-2} & \osm{L} & \cL_{-4}\cL_{-2} \,, \nn \\
\cH_{-4}\cH_{-3}\cH_{-2} & \osm{H} & \cH_{-3}\cH_{-2} \,, \nn \\
\cQ_{-2} & \osm{Q} & 1 \,.
\eea

\noi
We can now define an ordering on $\cset^G_{m,n}$ which
will turn out to be adapted with a very small kernel.
\bdf \label{def:cord}
On the set $\cset^G_{m,n}$ we introduce the total ordering $\ordering_G$.
For two elements $X_1,X_2\in\cset^G_{m,n}$, $X_1\neq X_2$
with $X_i=L^i H^i G^i Q^i \cL_{-1}^{k^i}\cG_{-1}^{r^i_1}\cQ_{-1}^{r^i_2}
\cQ_0^{r^i_3}$, $L^i\in\lset_{l_i}$,
$H^i\in\hset_{h_i}$,$G^i\in\gset_{g_i}$,$Q^i\in\qset_{q_i}$ for
some $l_i,h_i,g_i,q_i,k^i,r_1^i,r_2^i,r_3^i\in\bbbn_0$, $i=1,2$ such that
$m\in\bbbn_0$, $n\in\bbbz$ we define
\bea
X_1\osm{\ordering_G}X_2 & {\rm if} & k^1>k^2 \,. \label{eq:cord1}
\eea
For $k^1=k^2$ we set
\bea
X_1\osm{\ordering_G}X_2 & {\rm if} & r^1_1+r^1_2>r^2_1+r^2_2 \,.
\eea
If $r^1_1+r^1_2=r^2_1+r^2_2$, then we set
\bea
X_1\osm{\ordering_G}X_2 & {\rm if} & Q^1\osm{Q} Q^2 \,.
\label{eq:qord}
\eea
In the case where also $Q^1=Q^2$ we define
\bea
X_1\osm{\ordering_G}X_2 & {\rm if} & G^1\osm{G} G^2 \,.
\eea
If even $G^1=G^2$ we then define
\bea
X_1\osm{\ordering_G}X_2 & {\rm if} & L^1\osm{L} L^2 \,.
\eea
If further $L^1=L^2$ we set
\bea
X_1\osm{\ordering_G}X_2 & {\rm if} & H^1\osm{H} H^2 \,, \label{eq:cord2}
\eea
which finally has to give an answer. For $X_1=X_2$ we define
$\, X_1\osm{\ordering_G}X_2$ and $\, X_2\osm{\ordering_G}X_1$.
\edf

Definition \refoth{\ref{def:cord}} is well-defined since one obtains
an answer for any pair $X_1, X_2\in\cset^G_{m,n}$, $X_1\neq X_2$
after going through \eqs{\ref{eq:cord1}}-\eqoth{\ref{eq:cord2}},
and hence the ordering $\ordering_{G}$ proves to be a total ordering
on $\cset^G_{m,n}$. Namely, if \eqs{\ref{eq:cord1}}-\eqoth{\ref{eq:cord2}}
do not give an answer on the ordering of $X_1$ and $X_2$, then
obviously $X_1$ and $X_2$ are of the form
$X_i=LHGQ\cL_{-1}^k\cG_{-1}^{r_1^i}\cQ_{-1}^{r_2^i}\cQ_0^{r_3^i}$,
with common $L$, $H$, $G$, $Q$, $k$ and also $r_1^1+r_2^1=r_1^2+r_2^2$.
The fact that both $X_1$ and $X_2$ has charge $n$ implies
$r_1^1-r_2^1-r_3^1=r_1^2-r_2^2-r_3^2$, and using $r_1^1+r_2^1=r_1^2+r_2^2$
one obtains $2 r_2^1+r_3^1=2 r_2^2+r_3^2$. But this
equation has solutions from $\{0,1\}$ only for $r_2^1=r_2^2$ and
consequently also $r_3^1=r_3^2$ and hence $X_1=X_2$.

Obviously the $\ordering_G$-smallest element of $\cset^G_{m,0}$
is $\cL_{-1}^m$ followed by $\cL_{-1}^{m-1}\cG_{-1}\cQ_0$ whilst
the $\ordering_G$-smallest element of $\cset^G_{m,-1}$
is $\cL_{-1}^{m}\cQ_{0}$ followed by $\cL_{-1}^{m-1}\cQ_{-1}$.
Similarly, for $\cset^G_{m,1}$ we find $\cL_{-1}^{m-1}\cG_{-1}$
as $\ordering_G$-smallest element followed by $\cL_{-1}^{m-2}\cG_{-1}$.
We will now show that the ordering $\ordering_G$
is adapted and we will compute the ordering kernels. We do not give a
theoretical proof that these kernels are the smallest possible ordering
kernels. However, we shall refer later to explicit examples of singular
vectors that show that for general values of $\Delta$ and $q$ most of
the ordering kernels presented here cannot be smaller.

\bth \label{th:kernelG}
If the central extension satisfies $c\neq 3$, then
the ordering $\ordering_G$
is adapted to $\cset^G_{m,n}$ for all
Verma modules $\vm^G_{\Delta,q}$ and for all grades
$(m,n)$ with $m\in\bbbn_0$, $n\in\bbbz$.
Ordering kernels are given by the following
tables for all levels\footnote{Note that for levels $m=0$ and
$m=1$ some of the kernel elements obviously do not exist.}
$m$, depending on the set
of annihilation operators and depending on the charge $n$.

\btab{|r|l|}
\hline \label{tab:adkern1}
 $n$ & ordering kernel \\
\hline
$+1$ &  $\{\cL_{-1}^{m-1}\cG_{-1}\}$ \\
\hline
$0$ &  $\{\cL_{-1}^{m},\cH_{-1}\cL_{-1}^{m-1},
\cL_{-1}^{m-1}\cG_{-1}\cQ_0\}$ \\
\hline
$-1$ & $\{\cL_{-1}^{m}\cQ_{0}, \cH_{-1}\cL_{-1}^{m-1}\cQ_{0},
\cL_{-1}^{m-1}\cQ_{-1}\}$ \\
\hline
$-2$ & $\{\cL_{-1}^{m-1}\cQ_{-1}\cQ_0\}$ \\
\hline
\etab{Ordering kernels for $\ordering_G$,
annihilation operators $\ta^+$.}

\btab{|r|l|}
\hline \label{tab:adkern2}
 $n$ & ordering kernel \\
\hline
$+1$ &  $\{\cL_{-1}^{m-1}\cG_{-1}\}$ \\
\hline
$0$ &  $\{\cL_{-1}^{m}, \cL_{-1}^{m-1}\cG_{-1}\cQ_0\}$\\
\hline
$-1$ &  $\{\cL_{-1}^{m}\cQ_{0}\}$ \\
\hline
\etab{Ordering kernels for $\ordering_G$,
annihilation operators $\ta^+$ and $\cG_0$.}

\btab{|r|l|}
\hline \label{tab:adkern3}
 $n$ & ordering kernel \\
\hline
$0$ &  $\{\cL_{-1}^{m}\}$ \\
\hline
$-1$ &  $\{\cL_{-1}^{m}\cQ_0, \cL_{-1}^{m-1}\cQ_{-1}\}$\\
\hline
$-2$ &  $\{\cL_{-1}^{m-1}\cQ_{-1}\cQ_{0}\}$ \\
\hline
\etab{Ordering kernels for $\ordering_G$,
annihilation operators $\ta^+$ and $\cQ_0$.}

\btab{|r|l|}
\hline \label{tab:adkern4}
 $n$ & ordering kernel \\
\hline
$0$ &  $\{\cL_{-1}^{m}\}$ \\
\hline
$-1$ &  $\{\cL_{-1}^{m}\cQ_0\}$\\
\hline
\etab{Ordering kernels for $\ordering_G$,
annihilation operators $\ta^+$, $\cQ_0$, and $\cG_0$.}
Charges that do not appear in the tables have trivial ordering kernels.
\eth
Like in the Virasoro case our strategy will be to find
annihilation operators that are able to produce an additional
$\cL_{-1}$. Hence, we raise the term in question to the class of terms with
one additional $\cL_{-1}$
and try to prove that terms that can also be raised to this class of terms
have to be $\ordering_G$-smaller.
That we need to focus only on operators that create
$\cL_{-1}$ from the leading part of a term is a consequence of the
following theorem which we therefore shall prove before starting
the proof of theorem \refoth{\ref{th:kernelG}}.

\bth \label{th:littletheorem}
Let us assume that there exists an annihilation operator $\Gamma$ that
creates a term $X^{\Gamma}$ with $n+1$ operators $\cL_{-1}$ by acting on
\bea
X_0 &=& X_0^{\ast}\, \cL_{-1}^n\, \cG_{-1}^{r_1}\,
\cQ_{-1}^{r_2}\, \cQ^{r_3}_0 \;\; \in\cset^G_{m_0,n_0}
\eea
\noi with $X_0^{\ast}=L^0\, H^0\, G^0\, Q^0$,
\bea
L^0 & = & \cL_{-m^L_{\length{L^0}}}\ldots\cL_{-m^L_1}\in\lset_l \,, \nn \\
H^0 & = & \cH_{-m^H_{\length{H^0}}}\ldots\cH_{-m^H_1}\in\hset_h \,, \nn \\
G^0 & = & \cG_{-m^G_{\length{G^0}}}\ldots\cG_{-m^G_1}\in\gset_g \,, \nn \\
Q^0 & = & \cQ_{-m^Q_{\length{Q^0}}}\ldots\cQ_{-m^Q_1}\in\qset_q \,, \nn
\eea
$l,h,g,q\in\bbbn_0$ and $r_1,r_2,r_3\in\{0,1\}$. Let us further assume
that this additional $\cL_{-1}$ is created by commuting $\Gamma$ through
$X_0^{\ast}$. Then any other term $Y\in\cset^G_{m_0,n_0}$ with
$X_0\osm{\ordering_G}Y$ for
which the action of $\Gamma$ also produces the term $X^{\Gamma}$
will also create one additional $\cL_{-1}$ and this by commuting
$\Gamma$ through the leading part $Y^{\ast}$ of $Y$.
\eth
\indent
Theorem \refoth{\ref{th:littletheorem}} thus tells us that any
$Y\in\cset^G_{m_0,n_0}$ that does not
generate an additional $\cL_{-1}$ from $Y^{\ast}$ in the
described sense is automatically $\ordering_G$-smaller
than $X_0$ and is therefore irrelevant for the adapted ordering.

\noi
\bprf
Let us show that if $\Gamma$ does not create one additional
$\cL_{-1}$ by commuting through $Y^{\ast}$ but still satisfies that it
also creates the term $X^{\Gamma}$ acting on $Y$,
then $Y\osm{\ordering_G} X_0$. If $Y$ has already $n+1$
or even more operators $\cL_{-1}$, then $Y$ is obviously
$\ordering_G$-smaller
than $X_0$. Thus the action of $\Gamma$ on $Y$ needs to create at least
one $\cL_{-1}$. However, the action of one operator $\Gamma$ can create at
most one $\cL_{-1}$. This is an imediate consequence of the
commutation relations \eq{\ref{topalgebra}}:
the action of one operator $\Gamma$ can take
several operators in $X_0$ away but it can at most create only
one new operator. We can therefore concentrate on terms $Y$
that have exactly $n$ operators $\cL_{-1}$. The additional
$\cL_{-1}$ is either created from $Y^{\ast}$ or from
$\cG_{-1}^{r_1^Y}\cQ_{-1}^{r_2^Y}\cQ_0^{r_3^Y}$.
For the latter case there are a few possibilities to create $\cL_{-1}$:
only under the action of $\cG_0$, $\cQ_0$, $\cL_1$, or $\cH_1$, and
depending on the values of $r_1^Y$, $r_2^Y$, and $r_3^Y$. The operator
$\Gamma$ could be one of these operators or the commutation of
$\Gamma$ with $Y^{\ast}$ could produce one of them. (Note that it is
not possible to create one of these operators and to create in
addition a $\cL_{-1}$ from $Y^{\ast}$.) If now
 $\cG_0$, $\cQ_0$, $\cL_1$, or $\cH_1$ creates a $\cL_{-1}$
from $\cG_{-1}^{r_1^Y}\cQ_{-1}^{r_2^Y}\cQ_0^{r_3^Y}$, then we find
in each case that at least one of $r_1^Y$ or $r_2^Y$ changes from $1$ to
$0$ whilst $r_3^Y$ remains unchanged. But $X^{\Gamma}$
has the same $r_i$ as $X_0$ for $i=1,2,3$. One therefore deduces that
$r_1^Y+r_2^Y>r_1+r_2$ and thus $Y\osm{\ordering_G}X_0$. Hence the
additional $\cL_{-1}$ must be created by commuting $\Gamma$ through the
leading part $Y^{\ast}$ in order that $X_0\osm{\ordering_G}Y$.
\eprf

\noi Equipped with theorem \refoth{\ref{th:littletheorem}} we can now
proceed with the proof of theorem \refoth{\ref{th:kernelG}}:

\noi
Let us consider the term
\bea
X_0 & = & L^0\, H^0\, G^0\, Q^0\, \cL_{-1}^n\, \cG_{-1}^{r_1}\,
\cQ_{-1}^{r_2}\, \cQ^{r_3}_0 \;\; \in\cset^G_{m_0,n_0} \,, \label{X0term}
\eea
\noi with $L^0, H^0, G^0$ and $Q^0$ given above. We construct the vector
$\Psi^0=X_0\ket{\Delta,q}^G\in\vm_{\Delta,q}^G$
at level $\levelt{X_0}=m_0$ and charge $\charget{X_0}=n_0$.

Let us first consider the annihilation operators to be
those in $U(\ta)^+$ only.
If $Q^0\neq 1$ we act with $\cG_{m^Q_1-1}\in\ta^+$
on $\Psi^0$ and write the result again
in its normal form. We will thus obtain a non-trivial term in
$\cG_{m^Q_1-1}\Psi^0$ with one additional $\cL_{-1}$:
\bea
X^Q & = & L^0\, H^0\, G^0\, \tilde{Q}^0\, \cL_{-1}^{n+1}\, \cG_{-1}^{r_1}
\, \cQ_{-1}^{r_2}\, \cQ^{r_3}_0 \,, \nn \\
\tilde{Q}^0 & = & \cQ_{-m^Q_q}\ldots \cQ_{-m^Q_2}
\eea
or, if $\length{Q^0}=1$, $\tilde{Q}^0=1$ simply
by commuting $\cG_{m^Q_1-1}$ with $\cQ_{-m^Q_1}$ which
produces the additional operator $\cL_{-1}$. Any other term
$Y\in\cset^G_{m_0,n_0}$
also producing $X^Q$ under the action of $\cG_{m^Q_1-1}\in\ta^+$
and being $\ordering_G$-bigger than $X_0$ also needs to create
one $\cL_{-1}$ by commuting $\cG_{m_1^Q-1}$ with $Y^{\ast}$ due
to theorem \refoth{\ref{th:littletheorem}}.
We can therefore focus on terms
$Y=L^YH^YG^YQ^Y\cL_{-1}^{n}\cG_{-1}^{r_1} \cQ_{-1}^{r_2}\cQ^{r_3}_0$.
One finds that by commuting $\cG_{m^Q_1-1}$ with operators in $L^Y$ or
$H^Y$ one can only produce terms of the form $\cG_{m^{\prime}}$
with $m^{\prime}<m^Q_1-1$.
Therefore, in order to create subsequently
the operator $\cL_{-1}$ from $\cG_{m^{\prime}}$ or directly from
$\cG_{m_1^Q-1}$, $Y$ needs to contain an operator
of the form
$\cQ_{-m^{\star}-1}$ that satisfies\footnote{Note that
for $m^{\star}=0,-1$ $\cQ_{-m^{\star}-1}$ would not be in the leading
part $Y^{\ast}$.} $0<m^{\star}+1\leq m^Q_1$. If $m^{\star}+1< m^Q_1$
one finds that $Y$ is $\ordering_G$-smaller than $X_0$
as the equation deciding on the ordering of $X_0$ and $Y$
would in this case be \eq{\ref{eq:qord}}. If $m^{\star}+1 = m^Q_1$, on
the other hand, $Y$ must be necessarily equal to $X_0$.
Note that $\cG_{m^Q_1-1}$ and $\cG_{m^{\prime}}$ simply anticommute with
operators of $G^Y$ and therefore cannot create any $\cL_{-1}$.
Hence there are no terms $Y$ $\ordering_G$-bigger than the terms $X_0$
producing the same terms $X^Q$ under the action of
 $\cG_{m^Q_1-1}\in\ta^+$. We have therefore shown that {\it the ordering}
$\ordering_G$ {\it is adapted on the set of terms} $X_0$ {\it of the form
 given by} \eq{\ref{X0term}} {\it with} $Q^0\neq 1$ {\it for all grades}
$(m_0,n_0)$ {\it and all central terms} $c\in\bbbc$.	

As the terms $X_0$ with $Q^0\neq 1$ are now proven to be adapted,
next we will consider the terms $X_0$ with $Q^0=1$ and $G^0\neq 1$:
\bea
X_0 & = & L^0\, H^0\, G^0\, \cL_{-1}^n\, \cG_{-1}^{r_1} \,
 \cQ_{-1}^{r_2}\, \cQ^{r_3}_0 \, .
\eea
If $G^0\neq 1$ we act with the annihilation operator
$\cQ_{m^G_1-1}$ on $\Psi^0=X_0\ket{\Delta,q}^G$. This produces the term
\bea
X^G & = & L^0\, H^0\, \tilde{G}^0\, \cL_{-1}^{n+1}\, \cG_{-1}^{r_1}
\, \cQ_{-1}^{r_2}\, \cQ^{r_3}_0
\eea
with $\tilde{G}^0=\cG_{-m^G_g}\ldots \cG_{-m^G_2}$ or, if
$\length{G^0}=1$, $\tilde{G}^0=1$.
Again, any other term $Y$ with $X_0\osm{\ordering_G}Y$
also producing $X^G$ under the action
of $\cQ_{m^G_1-1}\in\ta^+$ needs to create one $\cL_{-1}$ by commuting
$\cQ_{m^G_1-1}$ through the leading part $Y^{\ast}$ due to theorem
\refoth{\ref{th:littletheorem}}.
Thus we focus on operators $Y$ of the form
$Y=L^YH^YG^YQ^Y\cL_{-1}^{n}\cG_{-1}^{r_1}\cQ_{-1}^{r_2}\cQ^{r_3}_0$
with $Q^Y=1$ as otherwise $Y\osm{\ordering_G}X_0$.
Commuting $\cQ_{m^G_1-1}$ with operators in $L^Y$ or $H^Y$ can
only create operators of the form $\cQ_{m^{\prime}}$ with
$m^{\prime}<m^G_1-1$. The operators $\cQ_{m^G_1-1}$ and $\cQ_{m^{\prime}}$
can create $\cL_{-1}$ from $G^Y$ only if $G^Y$ contains
$\cG_{-m^{\prime}-1}$ with $m^{\prime} +1\leq m_1^G$ and therefore
$Y$ is again $\ordering_G$-smaller or equal than $X_0$. Commuting
$\cQ_m$ through $G^Y$ can also give rise to operators of the
form $\cL_p$, $\cH_p$ and consequently even to $\cG_p$ with $p<m_1^G-1$.
In order to create $\cL_{-1}$ from $Q^Y$ it would require that
$Q^Y\neq 1$ so that one again finds $Y\osm{\ordering_G}X_0$.
This shows that {\it the ordering} $\ordering_G$
{\it is adapted on the set of terms} $X_0$ {\it of the form given by}
 \eq{\ref{X0term}} {\it with}
$Q^0\neq 1$ {\it or} $G^0\neq 1$ {\it for all grades}
 $(m_0,n_0)$ {\it and all central terms} $c\in\bbbc$.

Next we will consider the terms $X_0$ with $Q^0=1$, $G^0=1$ and
$L^0\neq 1$:
\bea
X_0 & = & L^0\, H^0\, \cL_{-1}^n\, \cG_{-1}^{r_1}\,
\cQ_{-1}^{r_2}\, \cQ^{r_3}_0 \, .
\eea
If $L^0\neq 1$ we act with the annihilation operator $L_{m_1^L-1}\in\ta^+$
on $\Psi^0=X_0\ket{\Delta,q}^G$. This produces a term of the form
\bea
X^L & = & \tilde{L}^0\, H^0\, \cL_{-1}^{n+1}\, \cG_{-1}^{r_1}\,
\cQ_{-1}^{r_2}\, \cQ^{r_3}_0
\eea
with $\tilde{L}^0=\cL_{-m^L_l}\ldots \cL_{-m^L_2}$ or, if
$\length{L^0}=1$, $\tilde{L}^0=1$.
If $m^L_2=m^L_1$
we may simply obtain multiple copies of the same term $X^L$.
Again theorem \refoth{\ref{th:littletheorem}} allows us to
focus on $\ordering_G$-bigger terms $Y$ of the form
$Y=L^YH^Y\cL_{-1}^{n}\cG_{-1}^{r_1}
\cQ_{-1}^{r_2}\cQ^{r_3}_0\in\cset^G_{m_0,n_0}$ (in addition,
$G^Y\neq 1$ or $Q^Y\neq 1$ would lead to $Y\osm{\ordering_G}X_0$).
The operator $\cL_{m_1^L-1}$ commuted with
operators in $H^Y$ cannot create any $\cL_{-1}$ and obviously,
following the arguments of the Virasoro case (proof of theorem
\refoth{\ref{th:viradap}}), terms $Y$ that produce $X^L$
creating $\cL_{-1}$ out of $L^Y$ would again be
$\ordering_G$-smaller than $X_0$. Therefore we can state that
{\it the ordering} $\ordering_G$
{\it is adapted on the set of terms} $X_0$ {\it of the form given by}
 \eq{\ref{X0term}}
{\it with} $Q^0\neq 1$, {\it or} $G^0\neq 1$, {\it or} $L^0\neq 1$
{\it for all grades} $(m_0,n_0)$ {\it and all central terms} $c\in\bbbc$.

We are thus left with terms $X_0$ of the form
\bea
X_0 & = & H^0\, \cL_{-1}^n\, \cG_{-1}^{r_1}\,
\cQ_{-1}^{r_2}\, \cQ^{r_3}_0 \, .
\eea
At this stage it is not
possible to create operators $\cL_{-1}$ by acting directly with positive
operators of $\ta^+$ on $H^0$.
Therefore we cannot use further theorem \refoth{\ref{th:littletheorem}},
which has proven to be very fruitful so far, and a
different strategy must be applied.
Let us first assume that $H^0$ contains operators
other than $\cH_{-1}$ and let $j_0\in\bbbn$ be the smallest
index of $H^0$ such that $m_{j_0}^H\neq 1$.
There are four different cases to study depending on
the values of $r_1$, $r_2$ and $r_3$.
Let us start with the cases where $r_2=0$.
Acting with\footnote{At this stage of the proof we need
the annihilation operators to be
from the universal enveloping algebra. Therefore, definition
\refoth{\ref{def:adapt}} is slighlty modified compared to
definition \refoth{\ref{def:viradapt}}.}
$\cG_{m^H_{j_0}}\cQ_{-1}\in U(\ta)^+$
on $\Psi^0=X_0\ket{\Delta,q}^G$ and writing the result
in its normal form, one obtains a non-trivial term
\bea
X^H & = & \tilde{H}^0\, \cL_{-1}^{n+1}\, \cG_{-1}^{r_1}\,
\cQ^{r_3}_0 \,, \nn \\
\tilde{H}^0 & = & \cH_{-m^H_h}\ldots \cH_{-m^H_{j+1}}\cH_{-1}^{j_0-1}
\eea
or, if $\length{H^0}=j_0$, $\tilde{H}^0=\cH_{-1}^{j_0-1}$ (commuting
$\cQ_{-1}$ with $\cH_{-m^H_{j_0}}$ produces $\cQ_{-m^H_{j_0}-1}$
and subsequently the commutation with $\cG_{m^H_{j_0}}$ produces
$\cL_{-1}$). If $m^H_{j_0+1}=m^H_{j_0}$ one simply obtains
multiple copies of $X^H$. Now we must show that any other term
$Y\in\cset^G_{m_0,n_0}$ also producing $X^H$ under the action of
$\cG_{m^H_{j_0}}\cQ_{-1}$ is $\ordering_G$-smaller than $X_0$.
Just like in theorem \refoth{\ref{th:littletheorem}}, if $Y$ already has
$n+1$ or more operators $\cL_{-1}$, then $Y\osm{\ordering_G}
X_0$. But unlike in theorem \refoth{\ref{th:littletheorem}},
we act now with two operators and could therefore also produce
two new operators. However $\cG_{m^H_{j_0}}\cQ_{-1}$ cannot produce
two $\cL_{-1}$ as there are no operators $\cG_0$ in $Y$. We therefore
take first $Y=H^Y\cL_{-1}^{n}\cG_{-1}^{r_1}
\cQ_{-1}^{r_2}\cQ^{r_3}_0\in\cset^G_{m_0,n_0}$ .
If $\cG_{m^H_{j_0}}\cQ_{-1}$ produces one
$\cL_{-1}$ by commuting through $H^Y$ and leaves $r_1$, $r_2$ and
$r_3$ unchanged one finds $Y\osm{\ordering_G}X_0$,
because then $H^Y$ needs to have more operators
$\cH_{-1}$ than $X_0$, so that $m_{j_0}^Y<m_{j_0}^H$. On the other hand,
$\cQ_{-1}$ acting on $\cG_{-1}^{r^Y_1}\cQ_{-1}^{r^Y_2}\cQ_{0}^{r^Y_3}$
cannot create any $\cL_{-1}$ but it could change $r^Y_2$ from $0$ to
$1$ or $r_1^Y$ from $1$ to $0$. The first case would not produce
$X^H$ as we assumed $r_2=0$ and in addition $\cG_{m^H_{j_0}}$ cannot
create $\cL_{-1}$ from $H^Y$. The latter case can produce $X^H$ but
only for $r_1=0$ and $r_1^Y=r_2^Y=1$
($\cG_2\cQ_{-1}$ creating $\cL_{-1}$ from $\cG_{-1}^{r_1^Y}
\cQ_{-1}^{r_2^Y}\cQ_0^{r_3^Y}$). Thus $r_1^Y+r_2^Y=2>0=r_1+r_2$
resulting in $Y\osm{\ordering_G}X_0$.

Now let us take $r_2=1$ and let us assume $r_1=0$. In this case we proceed
analogously as in the previous case by acting with
$\cQ_{m^H_{j_0}}\cG_{-1}\in U(\ta)^+$. Again one cannot
produce the term $X^H$ from a term $Y$ $\ordering_G$-bigger than $X_0$.
In particular, the only way $\cG_{-1}$ could change the
triple $(r^Y_1,r^Y_2,r^Y_3)$ is by changing $r^Y_1$ from $0$ to $1$,
which is not allowed as $r_1=0$, and in addition $\cL_{-1}$ could not be
created by $\cQ_{m^H_{j_0}}$ acting on $H^Y$.
For the case that both $r_1=r_2=1$ let us first assume that $r_3=0$.
By acting with $\cG_{m^H_{j_0}-1}\cQ_{0}\in U(\ta)^+$ one produces
$\cL_{-1}$ from $\cH_{-m^H_{j_0}}$ in a similar way as before.
$\cQ_0$ can change the triple $(r^Y_1,r^Y_2,r^Y_3)$ in two
ways: it can change $r^Y_3$ from $0$ to $1$ or it could change
$r^Y_1$ from $1$ to $0$. The first case, however, would not lead to
the term $X^H$ as $r_3=0$. The latter case can only lead to $X^H$ if
$\cG_{m^H_{j_0}-1}$ can be converted into $\cG_{-1}$ which requires
$H^Y\osm{H}H^0$. Therefore we find $Y\osm{\ordering_G}X_0$.
Finally, if $r_1=r_2=r_3=1$ we act with $\cQ_{m^H_{j_0}-1}\cG_{0}$. In
this case $r_3^Y$ can only be changed from $1$ to $0$, which does not
lead to $X^H$, and $r_2^Y$ can only be changed from $1$ to $0$,
which requires $\cQ_{m^H_{j_0}-1}$ to be converted into $\cQ_{-1}$ in
order to obtain $X^H$, resulting again in $Y\osm{\ordering_G} X_0$.
We can thus summarise that {\it the ordering}
$\ordering_G$ {\it is adapted on the set of terms} $X_0$ {\it of the form
 given by} \eq{\ref{X0term}}
{\it with} $Q^0\neq 1$, {\it or} $G^0\neq 1$, {\it or} $L^0\neq 1$,
{\it or} $H^0\neq\cH_{-1}^k$ {\it for some} $k\in\bbbn_0$
{\it for all grades} $(m_0,n_0)$ {\it and all central terms} $c\in\bbbc$.

Next let us consider
$X_0 = \cH_{-1}^{m^H}\cL_{-1}^n\cG_{-1}^{r_1}\cQ_{-1}^{r_2}\cQ_0^{r_3}$.
For the ordering we have chosen, $\cH_{1}$ is not capable
to rule out elements of the ordering kernel containing $\cH_{-1}$.
The action of $\cH_1$ rather concerns $\cG_{-1}\cQ_{-1}$ as
this combination of operators is necessarily needed in order to create
$\cL_{-1}$ under the action of $\cH_1$. If we take $r_1=r_2=1$, i.e.
$X_0=\cH_{-1}^{m^H}\cL_{-1}^n\cG_{-1}\cQ_{-1}\cQ_0^{r_3}$,
then the action of $\cH_1$ creates an additional
$\cL_{-1}$ in the only possible way that $\cH_1$ can create $\cL_{-1}$,
producing a term $\cH_{-1}^{m^H}\cL_{-1}^{n+1}\cQ_0^{r_3}$
that cannot be obtained from any other term $Y$
$\ordering_G$-bigger than $X_0$.
{\it Thus elements of the ordering kernel are of the form}
$\cH_{-1}^{m^H}\cL_{-1}^n\cG_{-1}^{r_1}\cQ_{-1}^{r_2}\cQ_0^{r_3}$
{\it with} $r_1+r_2<2$ {\it for all grades and all central terms}
$c\in\bbbc$.

At this stage all restrictions on the ordering kernel arising from
operators in $\ta^+$ which create an additional $\cL_{-1}$ have been
used. One might think that the smallest ordering kernel has been
found. However we will now show that, considering the action
of two annihilation operators at the same time, we can still
reduce the ordering kernel at least for central terms $c\neq 3$. Let us
consider the case
$X_0 = \cH_{-1}^{m^H}\cL_{-1}^n\cG_{-1}^{r_1}\cQ_{-1}^{r_2}\cQ_0^{r_3}$.
with $m^H\neq 0$ and $r_1+r_2<2$. The action of $\cH_{1}\in\ta^+$ on
$\Psi^0=X_0\ket{\Delta,q}^G$ creates, provided $c\neq 0$, a non-trivial
term of the form
$X^H=\cH_{-1}^{m^H-1}\cL_{-1}^{n}\cG_{-1}^{r_1}\cQ_{-1}^{r_2}\cQ_0^{r_3}$
with one $\cH_{-1}$ removed but no new $\cL_{-1}$ created.
Furthermore, as $r_1+r_2<2$ $\cH_1$ cannot create any $\cL_{-1}$.
Now,
depending on $r_1$, $r_2$ and $r_3$ it may be possible to find terms $Y$
$\ordering_G$-bigger than $X_0$ that do not create $\cL_{-1}$ but still
generate $X^H$ under $\cH_1$.
In the case $r_1=1$, $r_2=0$ one finds that
$Y=\cH_{-1}^{m^H-1}\cG_{-2}\cL_{-1}^{n}\cQ_0^{r_3}$ is the only such term.
Thus, $X_0$ and $Y$ both create the same term
$X^H=\cH_{-1}^{m^H-1}\cL_{-1}^{n}\cG_{-1}\cQ_0^{r_3}$
under the action of $\cH_1$ with coefficients
$m^H\frac{c}{3}$ and $1$ respectively\footnote{There are
certainly other terms that are $\ordering_G$-smaller than
$X_0$ and create $X^H$ such as
$\cH_{-1}^{m^H-1}\cL_{-1}^{n}\cG_{-1}\cQ_{-1}$ for $r_2=0$ and $r_3=1$.
But as before, $\ordering_G$-smaller terms are not relevant
due to definition \refoth{\ref{def:adapt}}.},
with $X_0\osm{\ordering_G}Y$.
On the other hand, acting with $\cQ_{1}$ again shows that
$Y$ is the only term $\ordering_G$-bigger than $X_0$, both of them
generating\footnote{Note that we have used the action of $\cQ_{1}$
before to rule out $Y$ in the ordering kernel.} $\tilde{X}^H=
\cH_{-1}^{m^H-1}\cL_{-1}^{n+1}\cQ_0^{r_3}$.
Let us therefore take the combination
$\cH_1 + \cQ_1$ acting on $X_0$  and on $Y$. This results in
\bea
(\cH_1 + \cQ_1) X_0 &=& m^H\frac{c}{3} X^H + 2m^H \tilde{X}^H + ... \,,
\label{eq:H1change} \\
(\cH_1 + \cQ_1) Y &=&  X^H + 2 \tilde{X}^H + ... \,, \nn
\eea
where ``$\ldots$'' denotes terms that are irrelevant
for us\footnote{Note that this is consistent for $c=0$.}. We now alter
the standard basis
of the normal form by defining new basis terms $X^1=m^H\frac{c}{3} X^H
+ 2 m^H \tilde{X}^H$ and $X^2= X^H + 2 \tilde{X}^H$. If this change of basis
is possible the action of $\cH_1 +\cQ_1$ on $X_0$ thus yields a term
$X^1$ that cannot be produced from any other term $Y$ unless
$Y\osm{\ordering_G}X_0$. In order for this basis transformation
to be allowed, the determinant of the transformation coefficients
must be non-trivial: $2m^H \frac{c}{3}-2m^H \neq 0$ with $m^H>0$ and
$c\neq 3$. In the case $r_1=0$, $r_2=1$ we can repeat exactly
the same procedure with
$Y=\cH_{-1}^{m^H-1}\cQ_{-2}\cL_{-1}^{n}\cQ_0^{r_3}$,
$\tilde{X}^H= \cH_{-1}^{m^H-1}\cL_{-1}^{n+1}\cQ_0^{r_3}$ and $\cQ_1$
replaced by $\cG_1$. \eqs{\ref{eq:H1change}} turn in this case into
\bea
(\cH_1 + \cG_1) X_0 &=& m^H\frac{c}{3} X^H - 2m^H \tilde{X}^H + ... \,,
\label{eq:H1changebis} \\
(\cH_1 + \cG_1) Y &=& - X^H + 2 \tilde{X}^H + ... \,, \nn
\eea
and thus result in exactly the same conditions from the determinant:
$m^H>0$ and $c\neq 3$.

Finally in the case $r_1=r_2=0$ we are left with $X_0$ of the form
\bea
X_0 & = & \cH_{-1}^{m^H} \, \cL_{-1}^n \, \cQ_0^{r_3} \,.
\eea
For $m^H\geq 1$ one finds that the action of $\cH_{1}$ on $X_0$ and on
$Y=\cH_{-1}^{m^H-1}\cL_{-1}^{n-1}\cG_{-1}\cQ_{-1}\cQ_0^{r_3}$ produces
a term $X^H=\cH_{-1}^{m^H-1}\cL_{-1}^n\cQ_0^{r_3}$
with coefficients $\frac{c}{3}m^H$ and
$2$ respectively. As $X_0\osm{\ordering_G} Y$ one again needs to find a
suitable second operator creating from both $X_0$ and $Y$ a common
term $\tilde{X}^H$ that cannot be created from any other terms that are
$\ordering_G$-bigger than $X_0$ and $Y$. For $m^H\geq 2$ we act with
$\cH_{1}\cG_0\cQ_0$ on $X_0$ and $Y$. In both cases one obtains a term
$\tilde{X}= \cH_{-1}^{m^H-2}\cL_{-1}^{n+1}\cQ_0^{r_3}$ with coefficients
$2m^H(m^H-1)(\frac{c}{3}-1)$ and $4(m^H-1)(\frac{c}{3}-1)$ respectively.
As above, the change of basis is possible if the determinant of
these coefficients is non-trivial:
\bea
\left|
\begin{array}{cc}
m^H\frac{c}{3} & 2 \\
2m^H(m^H-1)(\frac{c}{3}-1) & 4(m^H-1)(\frac{c}{3}-1)
\end{array}
\right| & = & 4 m^H (m^H-1)(\frac{c}{3}-1)^2 \,,
\eea
and thus result in the conditions: $m^H\geq 2$ and $c\neq 3$.

{\it Therefore the kernel of the ordering} $\ordering_G$, {\it is given by}
\bea
\cset_{m_0,n_0}^{K} &=& \left\{ \cL_{-1}^n\cG_{-1}^{r_1}\cQ_{-1}^{r_2}
\cQ_{0}^{r_3} ,\; \cH_{-1} \cL_{-1}^n\cQ_0^{r_3} : \; r_1+r_2<2 ,\;
m_0=n+r_1+r_2, \; n_0=r_1-r_2-r_3
\right\} \,, \nn
\eea
{\it for all grades} $(m_0,n_0)$ {\it and all central terms} $c\in\bbbc$
{\it with} $c\neq 3$.
 This proves the results shown in \tab{\ref{tab:adkern1}}.

For $\cG_0$-closed vectors $\cG_0$ is also in the set of annihilation
operators. In this case the action of $\cG_0$ on $X_0$
of the form $\cH_{-1}^{m^H}\cL_{-1}^n\cG_{-1}^{r_1}\cQ_{-1}\cQ_0^{r_3}$
produces the term
$\cH_{-1}^{m^H}\cL_{-1}^{n+1}\cG_{-1}^{r_1}\cQ_0^{r_3}$ that cannot
be obtained from any other term $Y$ $\ordering_G$-bigger than $X_0$
(commuting with $\cQ_{-1}$ is the only way to produce $\cL_{-1}$ acting
with $\cG_0$). Thus, the ordering kernel contains no terms with
$\cQ_{-1}$ for all complex values of $c$.

If we now take $X_0=\cH_{-1}\cL_{-1}^n\cQ_0^{r_3}$ we find the (unique)
$\ordering_G$-bigger term
 $Y=\cL_{-1}^{n-1}\cG_{-1} \cQ_{-1}\cQ_0^{r_3}$, both producing the
terms $\cL_{-1}^n\cG_{-1}\cQ_0^{r_3}$ and $\cL_{-1}^n\cQ_0^{r_3}$
under the action of $\cG_0$ and $\cH_{1}$
respectively\footnote{The basis we have to choose in this
case is not $L_0$ graded.}. As a result $\cH_{-1}$ can also be removed
by changing the basis suitably provided $c\neq 3$. The determinant of
the coefficients results in:
\bea
\left|
\begin{array}{cc}
-1 & -2 \\
\frac{c}{3} & 2
\end{array}
\right| & = & 2 (\frac{c}{3}-1) \,,
\eea
which is again non-trivial for $c\neq 3$. This proves the results shown
in \tab{\ref{tab:adkern2}}.

In a completely analogous way one can remove the terms containing
$\cG_{-1}$ or $\cH_{-1}$ in the case of $\cQ_0$-closed singular vectors.
The former can be done by acting with $\cQ_0$ on
$X_0=$$\cH_{-1}^{m^H}\cL_{-1}^n\cG_{-1}\cQ_{-1}^{r_2}\cQ_0^{r_3}$,
creating the term $\cH_{-1}^{m^H}\cL_{-1}^{n+1}
\cQ_{-1}^{r_2}\cQ_0^{r_3}$, whilst the latter is achieved from the
action of $\cQ_0$ and $\cH_1$ on the same $X_0$ and $Y$ as
above. The determinant of the coefficients is again non-trivial
for $c\neq 3$:
\bea
\left|
\begin{array}{cc}
1 & 2 \\
\frac{c}{3} & 2
\end{array}
\right| & = & -2 (\frac{c}{3}-1) \,,
\eea
The results are shown in \tab{\ref{tab:adkern3}}.

Finally, combining our considerations for $\cG_0$-closed
singular vectors and $\cQ_0$-closed singular vectors we obtain that
the ordering kernel for the case of chiral singular
vectors does not contain any operators of the form $\cQ_{-1}$,
$\cG_{-1}$ or $\cH_{-1}$. This proves the results shown in
\tab{\ref{tab:adkern4}} and finally completes the proof of theorem
\refoth{\ref{th:kernelG}}.
\eprf

By replacing the r\^oles of the operators $\cG_{n}$ and $\cQ_{n}$
for all $n\in\bbbz$ we can define analogously an ordering
$\ordering_Q$ on $\cset^Q_{m,n}$ which is adapted for $c\neq 3$
for all levels. The corresponding ordering kernels are as follows.
\bth \label{th:kernelQ}
If the central extension satisfies $c\neq 3$
then the ordering $\ordering_Q$
is adapted to $\cset^Q_{m,n}$ for all
Verma modules $\vm_{\Delta,q}^Q$ and for all grades
$(m,n)$ with $m\in\bbbn_0$, $n\in\bbbz$.
Ordering kernels are given by the following
tables for all levels $m$, depending on the set
of annihilation operators and depending on the charge $n$.

\btab{|r|l|}
\hline \label{tab:adkern5}
 $n$ & ordering kernel \\
\hline
$+2$ &  $\{\cL_{-1}^{m-1}\cG_{-1}\cG_0\}$ \\
\hline
$+1$ &  $\{\cL_{-1}^{m}\cG_0,\cH_{-1}\cL_{-1}^{m-1}\cG_0,
\cL_{-1}^{m-1}\cG_{-1}\}$ \\
\hline
$0$ & $\{\cL_{-1}^{m}, \cH_{-1}\cL_{-1}^{m-1},
\cL_{-1}^{m-1}\cQ_{-1}\cG_0\}$ \\
\hline
$-1$ & $\{\cL_{-1}^{m-1}\cQ_{-1}\}$ \\
\hline
\etab{Ordering kernels for $\ordering_Q$,
annihilation operators $\ta^+$.}

\btab{|r|l|}
\hline \label{tab:adkern6}
 $n$ & ordering kernel \\
\hline
$+1$ &  $\{\cL_{-1}^{m}\cG_{0}\}$ \\
\hline
$0$ &  $\{\cL_{-1}^{m}, \cL_{-1}^{m-1}\cQ_{-1}\cG_0\}$\\
\hline
$-1$ &  $\{\cL_{-1}^{m-1}\cQ_{-1}\}$ \\
\hline
\etab{Ordering kernels for $\ordering_Q$,
annihilation operators $\ta^+$ and $\cQ_0$.}

\btab{|r|l|}
\hline \label{tab:adkern7}
 $n$ & ordering kernel \\
\hline
$+2$ &  $\{\cL_{-1}^{m-1}\cG_{-1}\cG_0\}$ \\
\hline
$+1$ &  $\{\cL_{-1}^{m}\cG_0,\cL_{-1}^{m-1}\cG_{-1}\}$\\
\hline
$0$ &  $\{\cL_{-1}^{m}\}$ \\
\hline
\etab{Ordering kernels for $\ordering_Q$,
annihilation operators $\ta^+$ and $\cG_0$.}

\btab{|r|l|}
\hline \label{tab:adkern8}
 $n$ & ordering kernel \\
\hline
$+1$ &  $\{\cL_{-1}^{m}\cG_0\}$ \\
\hline
$0$ &  $\{\cL_{-1}^{m}\}$\\
\hline
\etab{Ordering kernels for $\ordering_Q$,
annihilation operators $\ta^+$, $\cQ_0$, and $\cG_0$.}
Charges that do not appear in the tables have trivial ordering kernels.
\eth
\bprf
The proof of theorem \refoth{\ref{th:kernelQ}} is completely analogous
to the proof of theorem \refoth{\ref{th:kernelG}}. We just need to swap
the r\^oles of the operators $\cG_n$ and $\cQ_n$ for all $n\in\bbbz$.
\eprf


\section{Adapted orderings on chiral Verma modules $\vm_{0,q}^{GQ}$}
\label{sec:aoGQ}

We saw in section \refoth{\ref{sec:Nis2}} that for both chiral and
no-label highest weight vectors the conformal weight is zero.
This applies to Verma modules as well as to singular vectors.
Thus a chiral singular vector
in the chiral Verma module $\vm_{0,q}^{GQ}$ needs to
have level $0$ and the same is true for no-label singular vectors
in $\vm_{0,q}^{GQ}$. But at level $0$ there are no
singular vectors in $\vm_{0,q}^{GQ}$, as the only
state at level $0$ is the highest weight state $\ket{0,q}^{GQ}$ itself.
For chiral Verma modules $\vm_{0,q}^{GQ}$ we shall therefore only
consider adapted orderings with additional annihilation conditions
corresponding to $\cG_0$ or $\cQ_0$, but not to both.

In section \refoth{\ref{sec:Nis2}} we also introduced the set
$\cset^{GQ}_{m,n}$, \eq{\ref{eq:basis2}},
defining the standard basis $\bsvm_{\ket{0,q}^{GQ}}$ of the chiral
Verma modules $\vm_{0,q}^{GQ}$. $\cset^{GQ}_{m,n}$ can be obtained
by setting $r_3\equiv 0$ in $\cset^{G}_{m,n}$, \eq{\ref{eq:csetG}}.
Therefore, $\ordering_{G}$ is also defined on $\cset^{GQ}_{m,n}$, a subset
of $\cset^{G}_{m,n}$. This suggests that the ordering kernels for
$\ordering_{G}$ on $\cset^{GQ}_{m,n}$ may simply be appropriate subsets
of the ordering kernels of $\ordering_{G}$ on $\cset^{G}_{m,n}$,
given in theorem \refoth{\ref{th:kernelG}}. This can easily be
shown by considering the fact that $r_3$ is never a deciding element of
the ordering $\ordering_G$ in \eqs{\ref{eq:cord1}}-\eqoth{\ref{eq:cord2}}.
Furthermore, during the proof of theorem \refoth{\ref{th:kernelG}} it
happens in each case that the considered term $X^{\Gamma}$,
constructed from $X_0$ under the action of a suitable
annihilation operator $\Gamma$, has always the same
exponent $r_3$ of $\cQ_0$ as $X_0$ itself. Therefore, the whole
proof of theorem \refoth{\ref{th:kernelG}} can also be applied
to $\ordering_{G}$ defined on $\cset^{GQ}_{m,n}$ simply by imposing
$r_3\equiv 0$ in every step. As a result the new ordering
kernels are simply the intersections of $\cset^{GQ}_{m,n}$
with the ordering kernels for $\cset^{G}_{m,n}$. Hence, we have already
proven the following theorem.

\bth \label{th:kernelGQ}
For the set of annihilation operators that contains $\cG_0$ or
$\cQ_0$ but not both and for $c\neq 3$ the ordering $\ordering_G$
is adapted to $\cset^{GQ}_{m,n}$ for all chiral
Verma modules $\vm^{GQ}_{0,q}$ and for all grades
$(m,n)$ with $m\in\bbbn_0$, $n\in\bbbz$. Depending on the set
of annihilation operators and depending on the charge $n$,
ordering kernels are given by the following tables for all levels $m$:

\btab{|r|l|}
\hline
 $n$ & ordering kernel \\
\hline
$+1$ &  $\{\cL_{-1}^{m-1}\cG_{-1}\}$ \\
\hline
$0$ &  $\{\cL_{-1}^{m}\}$\\
\hline
\etab{Ordering kernels for $\ordering_G$ on $\cset^{GQ}_{m,n}$,
annihilation operators $\ta^+$ and $\cG_0$.}

\btab{|r|l|}
\hline
 $n$ & ordering kernel \\
\hline
$0$ &  $\{\cL_{-1}^{m}\}$ \\
\hline
$-1$ &  $\{\cL_{-1}^{m-1}\cQ_{-1}\}$\\
\hline
\etab{Ordering kernels for $\ordering_G$ on $\cset^{GQ}_{m,n}$,
annihilation operators $\ta^+$ and $\cQ_0$.}
Charges that do not appear in the tables have trivial ordering kernels.
\eth


\section{Adapted orderings on no-label Verma modules $\vm_{0,q}$}
\label{sec:aonl}

We will now consider adapted orderings for
no-label Verma modules $\vm_{0,q}$. In section \refoth{\ref{sec:Nis2}},
the standard basis for $\vm_{0,q}$
is defined using $\cset_{m,n}$ of \eq{\ref{eq:basis2}}.
No-label Verma modules have zero conformal weight, like chiral Verma
modules. Consequently chiral singular vectors as well as no-label
singular vectors in $\vm_{0,q}$ can only exist at level $0$.
The space of states in $\vm_{0,q}$ at level $0$ is spanned by
$\{\ket{0,q},\cG_0\ket{0,q},\cQ_0\ket{0,q},\cG_0\cQ_0\ket{0,q}\}$.
Therefore, there are no no-label singular vectors in $\vm_{0,q}$
and there is exactly one chiral singular vector in $\vm_{0,q}$ for
all $q$ and for all central extensions $c$, namely $\cG_0\cQ_0\ket{0,q}$.
Hence, our main interest focuses on the $\cG_0$-closed singular
vectors and the $\cQ_0$-closed singular vectors in $\vm_{0,q}$
and we shall therefore investigate adapted orderings with the
corresponding vanishing conditions. The states $\cG_0\ket{0,q}$
and $\cQ_0\ket{0,q}$ satisfy $\cQ_0\cG_0\ket{0,q}=-\cG_0\cQ_0\ket{0,q}$.
Consequently the norms of these states have opposite signs and can
be set to zero.

Clearly, $\vm_{0,q}^G$ is isomorphic to the quotient module of $\vm_{0,q}$
divided by the submodule generated by the singular vector
$\cG_0\ket{0,q}$, i.e. $\vm_{0,q}^G=\frac{\vm_{0,q}}{\cG_0\ket{0,q}}$
and likewise $\vm_{0,q}^Q=\frac{\vm_{0,q}}{\cQ_0\ket{0,q}}$.
If we consider a singular vector $\Psi$ of $\vm_{0,q}$ then
the canonical projection of $\Psi$ into $\vm_{0,q}^G$
is either trivial or a singular vector in $\vm_{0,q}^G$ and similarly
for $\vm_{0,q}^Q$. The converse, however, is not true, a singular vector
in $\vm_{0,q}^G$ or $\vm_{0,q}^Q$ do not necessarily correspond to a
singular vector in $\vm_{0,q}$, it may only be subsingular in
$\vm_{0,q}$. One may also ask whether all singular vectors in
$\vm_{0,q}$ correspond to singular vectors in either $\vm_{0,q}^G$
or $\vm_{0,q}^Q$, in which case the investigation of no-label
Verma modules would not give us more information than what
we already know, or rather there can also be singular vectors in
$\vm_{0,q}$ that vanish for both canonical projections into the generic
Verma modules $\vm_{0,q}^G$ and $\vm_{0,q}^Q$. This is indeed the case, as
was shown by the explicit examples at level 1 given in Ref. \icite{beatriz2}
(we will come back to this point at the end of next section).

Unlike for chiral Verma modules, the no-label Verma modules are not simply
a subcase of the $\cG_0$-closed Verma modules with respect to the adapted
ordering $\ordering_G$. In fact, rather than $\cset_{m,n}$ being a subset of
$\cset_{m,n}^G$, we find that $\cset_{m,n}^G$ is a subset of $\cset_{m,n}$
and we hence need to extend the ordering $\ordering_G$ suitably.
\bdf \label{def:gqord}
On the set $\cset_{m,n}$ we introduce the total ordering $\ordering_{GQ}$.
For two elements $X_1,X_2\in\cset_{m,n}$, $X_1\neq X_2$
with $X_i=L^i H^i G^i Q^i \cL_{-1}^{k^i}\cG_{-1}^{r^i_1}\cQ_{-1}^{r^i_2}
\cG_0^{r_4^i} \cQ_0^{r^i_3}$, $L^i\in\lset_{l_i}$,
$H^i\in\hset_{h_i}$,$G^i\in\gset_{g_i}$,$Q^i\in\qset_{q_i}$ for
some $l_i,h_i,g_i,q_i,k^i,r_1^i,r_2^i,r_3^i,r_4^i\in\bbbn_0$,
$i=1,2$ such that
$m\in\bbbn_0$, $n\in\bbbz$ we define
\bea
X_1\osm{\ordering_{GQ}}X_2 & {\rm if} & k^1>k^2 \,. \label{eq:gqord1}
\eea
For $k^1=k^2$ we set
\bea
X_1\osm{\ordering_{GQ}}X_2 & {\rm if} & r^1_1+r^1_2>r^2_1+r^2_2 \,.
\eea
If $r^1_1+r^1_2=r^2_1+r^2_2$ we set
\bea
X_1\osm{\ordering_{GQ}}X_2 & {\rm if} & Q^1\osm{Q} Q^2 \,.
\eea
In the case $Q^1=Q^2$ we define
\bea
X_1\osm{\ordering_{GQ}}X_2 & {\rm if} & G^1\osm{G} G^2 \,.
\eea
If also $G^1=G^2$ we then define
\bea
X_1\osm{\ordering_{GQ}}X_2 & {\rm if} & L^1\osm{L} L^2 \,.
\eea
If further $L^1=L^2$ we set
\bea
X_1\osm{\ordering_{GQ}}X_2 & {\rm if} & H^1\osm{H} H^2 \,, \label{eq:gqord2}
\eea
unless $H^1=H^2$ in which case we set
\bea
X_1\osm{\ordering_{GQ}}X_2 & {\rm if} & r_1^1>r_1^2 \,. \label{eq:gqords1}
\eea
If also $r_1^1=r_1^2$ we finally define
\bea
X_1\osm{\ordering_{GQ}}X_2 & {\rm if} & r_3^1+r_4^1>r_3^2+r_4^2
\,, \label{eq:gqords2}
\eea
which necessarily has to give an answer.
For $X_1=X_2$ we define
$\, X_1\osm{\ordering_{GQ}}X_2$ and $\, X_2\osm{\ordering_{GQ}}X_1$.
\edf

If \eqs{\ref{eq:gqord1}}-\eqoth{\ref{eq:gqords2}} do not give an answer
on the ordering of $X_1$ and $X_2$, then $X_1$ and $X_2$ are
of the form $X_i=LHGQ\cL_{-1}^k\cG_{-1}^{r_1}\cQ_{-1}^{r_2}
\cG_0^{r^i_4}\cQ_0^{r^i_3}$, with $r_3^1+r_4^1=r_3^2+r_4^2$. Since
$X_1$ and $X_2$ both have charge $n$ one has
$r_3^1-r_4^1=r_3^2-r_4^2$ and thus $X_1=X_2$.
Hence, definition \refoth{\ref{def:gqord}} is a total
ordering well-defined on $\cset_{m,n}$.

We will now argue that the proof of theorem \refoth{\ref{th:kernelG}}
can easily be modified in such a way that exactly the same restrictions
on the ordering kernels of $\ordering_G$ extend to
the ordering kernels of $\ordering_{GQ}$. As a first step we
see that theorem \refoth{\ref{th:littletheorem}} extends
straightforwardly to $\cset_{m,n}$ simply by replacing
$X_0=X_0^{\ast}\cL_{-1}^n\cG_{-1}^{r_1}
\cQ_{-1}^{r_2}\cQ^{r_3}_0\in\cset^G_{m,n}$ by
$X_0=X_0^{\ast}\cL_{-1}^n\cG_{-1}^{r_1}
\cQ_{-1}^{r_2}\cG_0^{r_4}\cQ^{r_3}_0\in\cset_{m,n}$. Note
that in the proof $r_4^Y$ would behave exactly like $r_3^Y$ which does
not interfere with any arguments. As theorem
\refoth{\ref{th:littletheorem}}
turned out to be the key tool to remove operators of the form $\cL_{-n}$,
$\cG_{-n}$, or $\cQ_{-n}$ from the ordering kernel of $\cset^G_{m,n}$,
we can in exactly the same way already state that {\it the ordering}
$\ordering_{GQ}$ {\it is for all grades}
$(m,n)$ {\it and all central extensions} $c\in\bbbc$
{\it adapted to the set of terms}
\bea
X_0 & = & L^0\, H^0\, G^0\, Q^0\, \cL_{-1}^n\, \cG_{-1}^{r_1}\,
\cQ_{-1}^{r_2}\, \cG_0^{r_4}\, \cQ^{r_3}_0 \;\; \in\cset_{m,n} \,,
\eea
{\it with} $L^0\neq 1$, {\it or} $G^0\neq 1$, {\it or} $Q^0\neq 1$.
We can thus focus on terms $X_0$ of the form
\bea
X_0 & = & \cH_{-m^H_I} \ldots \cH_{-m^H_{j_0}}
\, \cH_{-1}^{j_0 -1}\,  \cL_{-1}^n\, \cG_{-1}^{r_1}\,
\cQ_{-1}^{r_2}\, \cG_0^{r_4}\, \cQ^{r_3}_0 \,,
\eea
with $m^H_{j_0}>1$. In the proof of theorem \refoth{\ref{th:kernelG}}
we dealt with these terms by acting with $\cG_{m^H_{j_0}-1}\cQ_{-1}$
for $r_2=0$. At first, the existence of $\cG_0$ in the no-label case seems
to interfere with this argument. However, the ordering $\ordering_{GQ}$
has been defined in such a way that $\cQ_{-1}$ does never interact with
$\cG_0$ as it would simply be stuck on the left of $\cG_0$. Therefore,
one easily sees that the same arguments as in the proof of theorem
\refoth{\ref{th:kernelG}} hold for $r_2=0$. In the case of
$r_2=1$ and $r_1=0$ the proof even holds without any modification.
For the cases $r_1=r_2=1$, we act with
$\cG_{m^H_{j_0}}\cQ_{0}$ or $\cQ_{m^H_{j_0}}\cG_{0}$ for
$r_3=0$ or $r_3=1$ respectively. In these cases we have to consider the
additional possibility that $r_4^Y$ changes from $1$ to $0$ or from
$0$ to $1$ respectively. However, as $\cG_{m^H_{j_0}}$ or $\cQ_{m^H_{j_0}}$
still needs to create a $\cL_{-1}$ we easily see that any term
$Y$ also satisfying the conditions of proof \refoth{\ref{th:kernelG}}
for this case must contain operators $L^Y$, $G^Y$, $Q^Y$, or $H^Y$ with
$H^Y\osm{H}H^0$ and therefore $Y\osm{GQ} X_0$ as the $r_4$ term decides
very last in the ordering
\eqs{\ref{eq:gqord1}}-\eqoth{\ref{eq:gqords2}}. For the final part of
proof \refoth{\ref{th:kernelG}} where we found restrictions on
$\cH_{-1}^{j_0}$ in $X_0$, we can simply note that $\cG_0$ cannot
interfere with the arguments of the proof for $\cH_1$ (terms that could
interfere to produce $\cG_0$ would be $\ordering_{GQ}$-smaller than $X_0$)
and $\cQ_1$ and $\cG_1$ are used in the proof to
create $\cL_{-1}$ which is neither possible with $\cG_0$. Similar arguments
show that the proof also holds for $r_1=r_2=0$.
It is easy to see that in the $\cG_0$-closed or $\cQ_0$-closed cases all
considerations of proof \refoth{\ref{th:kernelG}} extend to $\cset_{m,n}$.
We have thus shown that the proof of theorem \refoth{\ref{th:kernelG}}
extends to prove the following theorem regarding no-label Verma modules
$\vm_{0,q}$.
\bth \label{th:kernelnl}
For the set of annihilation operators that contains $\cG_0$ or
$\cQ_0$ but not both and for $c\neq 3$ we find that
the ordering $\ordering_{GQ}$ is adapted to $\cset_{m,n}$ for all
Verma modules $\vm_{0,q}$ and for all grades
$(m,n)$ with $m\in\bbbn_0$, $n\in\bbbz$. Depending on the set
of annihilation operators and depending on the charge $n$,
ordering kernels are given by the following tables for all levels $m$:

\btab{|r|l|}
\hline
 $n$ & ordering kernel \\
\hline
$+2$ &  $\{\cL_{-1}^{m-1}\cG_{-1}\cG_0\}$ \\
\hline
$+1$ &  $\{\cL_{-1}^{m}\cG_{0},\cL_{-1}^{m-1}\cG_{-1},
\cL_{-1}^{m-1}\cG_{-1}\cG_{0}\cQ_0\}$ \\
\hline
$0$ &  $\{\cL_{-1}^{m},\cL_{-1}^{m}\cG_{0}\cQ_0,
\cL_{-1}^{m-1}\cG_{-1}\cQ_0\}$\\
\hline
$-1$ &  $\{\cL_{-1}^{m},\cQ_0\}$\\
\hline
\etab{Ordering kernels for $\ordering_{GQ}$ on $\cset_{m,n}$,
annihilation operators $\ta^+$ and $\cG_0$.}

\btab{|r|l|}
\hline
 $n$ & ordering kernel \\
\hline
$+1$ &  $\{\cL_{-1}^{m}\cG_0\}$ \\
\hline
$0$ &  $\{\cL_{-1}^{m},\cL_{-1}^{m}\cG_{0}\cQ_0,
\cL_{-1}^{m-1}\cQ_{-1}\cG_0\}$ \\
\hline
$-1$ &  $\{\cL_{-1}^{m}\cQ_0,\cL_{-1}^{m-1}\cQ_{-1},
\cL_{-1}^{m-1}\cQ_{-1}\cG_0\cQ_0\}$ \\
\hline
$-2$ &  $\{\cL_{-1}^{m-1}\cQ_{-1}\cQ_0\}$\\
\hline
\etab{Ordering kernels for $\ordering_{GQ}$ on $\cset_{m,n}$,
annihilation operators $\ta^+$ and $\cQ_0$.}
Charges that do not appear in the tables have trivial ordering kernels.
\eth


\section{Dimensional analysis}
\label{sec:dimensions}

In previous sections we argued, following Ref. \icite{beatriz2},
that a naive estimate would give $16$ types of singular vectors in
$N=2$ topological Verma modules, depending on whether the highest
weight vector or the singular vector itself satisfy additional
vanishing conditions with respect to the zero modes $\cG_0$ or $\cQ_0$,
each of these types coming with different charges. Three of
these types can be ruled out, however, simply by taking into account
that {\it chiral} highest weight conditions and {\it no-label}
highest weight conditions apply only to states with zero conformal
weight. In chiral Verma modules this rules out chiral singular vectors
as well as no-label singular vectors, whilst in no-label Verma
modules the no-label singular vectors are
ruled out. A fourth type of singular vectors, chiral singular vectors
in no-label Verma modules, turns out to consist of only the level zero
singular vector $\cG_0\cQ_0\ket{0,q}$. In this section we will use
theorem \refoth{\ref{th:dims}}, together with the results for
the ordering kernels of the previous sections, as the main tools
to give upper limits for the dimensions of the remaining $12$ types of
topological singular vectors. For most charges this procedure
will even show that there are no singular vectors corresponding to them.

The dimension of the singular vector spaces in $N=2$ superconformal Verma
modules can be larger than one. This fact was discovered for the
Neveu-Schwarz $N=2$ algebra in Ref. \icite{paper2}. In particular,
sufficient conditions were found (and proved) to guarantee the
existence of two-dimensional spaces of uncharged singular vectors.
Before this had been shown, it was a false common belief that singular
vectors at the same level and with the same charge would always be
linearly dependent. Later some of the results in  Ref. \icite{paper2}
were extended\cite{beatriz2} to the topological $N=2$ algebra. As a
consequence two-dimensional spaces for four different types of topological
singular vectors were shown to exist (those given in
\tab{\ref{tab:dim1}} below). However, as we will discuss, the Neveu-Schwarz
counterpart of most topological singular vectors are not singular vectors
themselves, but either descendants of singular vectors or subsingular
vectors\cite{DGR1,beatriz2,beatriz4}, for which very little is known.
As a consequence, in order to compute the maximal dimensions for the
singular vector spaces of the topological $N=2$ algebra, an independent
method, like the one presented in this paper, was needed.

Let us first proceed with a clear definition of what we mean by
{\it singular vector spaces}.

\bdf
A $\cG_0$-closed singular vector space of the topological Verma module
$\vm_{\Delta,q}^M$ is a subspace of $\vm_{\Delta,q}^M$
of vectors at the same level and with the same charge
for which each non-trivial element is a $\cG_0$-closed singular vector.
$M$ stands for $\cG_0$-closed, $\cQ_0$-closed, chiral, or no-label.
Analogously we define $\cQ_0$-closed singular vector spaces,
chiral singular vector spaces, and no-label singular vector spaces.
\edf

Let us denote by $\Psi_{m,\ket{\Delta,q}^M}^{K,n}$ a singular vector
in the topological Verma module $\vm_{\Delta,q}^M$ at level $m$
and with charge $n$. $K$ denotes the additional vanishing conditions
of the singular vector, whith respect to $\cG_0$ and $\cQ_0$, whilst $M$
denotes the additional vanishing
conditions of the highest weight vector, as introduced in section
\refoth{\ref{sec:Nis2}}. The ordering kernels of theorems
\refoth{\ref{th:kernelG}} and \refoth{\ref{th:kernelQ}} together with
theorem \refoth{\ref{th:dims}} allow us to write down
an upper limit for the dimensions of the singular vector spaces
simply by counting the number of elements of the ordering kernels.

\bth
For singular vectors with additional vanishing conditions in
$\vm_{\Delta,q}^G$ or in $\vm_{\Delta,q}^Q$, $c\neq 3$, we find the
following upper limits for the number of linearly independent
singular vectors at the same level $m\in\bbbn_0$ and with the same
charge $n\in\bbbz$ ($\Delta=-m$ for chiral singular vectors).
\btab{|l|c|c|c|c|c|}
\hline \label{tab:dim1}
 & $n=-2$ & $n=-1$ & $n=0$ & $n=1$ & $n=2$ \\
\hline
$\Psi_{m,\ket{\Delta,q}^G}^{G,n}$ &
$0$ & $1$ & $2$ & $1$ & $0$ \\
\hline
$\Psi_{m,\ket{\Delta,q}^G}^{Q,n}$ &
$1$ & $2$ & $1$ & $0$ & $0$ \\
\hline
$\Psi_{m,\ket{-m,q}^G}^{GQ,n}$ &
$0$ & $1$ & $1$ & $0$ & $0$ \\
\hline
$\Psi_{m,\ket{\Delta,q}^Q}^{Q,n}$ &
$0$ & $1$ & $2$ & $1$ & $0$ \\
\hline
$\Psi_{m,\ket{\Delta,q}^Q}^{G,n}$ &
$0$ & $0$ & $1$ & $2$ & $1$ \\
\hline
$\Psi_{m,\ket{-m,q}^Q}^{GQ,n}$ &
$0$ & $0$ & $1$ & $1$ & $0$ \\
\hline
\etab{Maximal dimensions for singular vectors spaces annihilated by
$\cG_0$ and/or $\cQ_0$ in $\vm_{\Delta,q}^G$ or in $\vm_{\Delta,q}^Q$.}
Singular vectors can only exist if they
contain in their normal form at least one non-trivial term of the
corresponding ordering kernel of theorem \refoth{\ref{th:kernelG}} or
\refoth{\ref{th:kernelQ}}. Charges $n$ that are not given have dimension
$0$ and hence do not allow any singular vectors.
\eth

The ordering kernels for the vanishing conditions $\ta^+$, given in
tables \tab{\ref{tab:adkern1}} and \tab{\ref{tab:adkern5}},
do not include any
conditions requiring the action of $\cG_0$ and $\cQ_0$ not to be trivial.
As a result, the ordering kernels of tables \tab{\ref{tab:adkern1}} and
\tab{\ref{tab:adkern5}} include not only the no-label cases but also
the cases of $\cG_0$-closed singular vectors, $\cQ_0$-closed
singular vectors, and chiral singular vectors. However, for no-label
singular vectors in $\vm_{\Delta,q}^G$ or in $\vm_{\Delta,q}^Q$
we can find in addition the
following restrictions. If $\Psi_{m,\ket{\Delta,q}^M}^{n}$ is
a no-label singular vector, then $\cG_0\Psi_{m,\ket{\Delta,q}^M}^{n}$
must be a singular vector of type $\Psi_{m,\ket{\Delta,q}^M}^{G,n+1}$.
Consequently, the dimension for the space of the no-label
singular vector $\Psi_{m,\ket{\Delta,q}^M}^{n}$ cannot be larger than
the dimension for the space of the $\cG_0$-closed singular
vector\footnote{If $\Psi$ is a no-label singular vector
and $\Xi$ is a $\cG_0$-closed, or $\cQ_0$-closed, or chiral
singular vector, both at the same level and with the same charge,
then $\Psi+\Xi$ is again a no-label singular vector (in the sense
of not being annihilated by $\cG_0$ or $\cQ_0$) which is
linearly idenpendent
of $\Psi$. However, the space spanned by $\Psi$ and $\Psi+\Xi$ is not
considered to be a two-dimensional no-label singular vector space as
it decomposes into a one-dimensional no-label singular vector space
and a one-dimensional $\cG_0$-closed, or $\cQ_0$-closed, or chiral
singular vector space.}
$\Psi_{m,\ket{\Delta,q}^M}^{G,n+1}$.
This can easily be seen as follows. Assume that $\Psi_1$
and $\Psi_2$ are two no-label linearly independent
singular vectors at the same level
and with the same charge, and suppose $\cG_0\Psi_1$
and $\cG_0\Psi_2$ are linearly dependent. Then obviously there
exist numbers $\alpha$, $\beta$ ($\alpha\beta\neq 0$) such that
$\cG_0(\alpha\Psi_1+\beta\Psi_2)=0$ and thus the
$\cG_0$-closed singular vector
$\alpha\Psi_1+\beta\Psi_2$ is contained in the space spanned
by $\Psi_1$ and $\Psi_2$, which is therefore not a no-label singular
vector space. Therefore, linearly independent singular vectors of type
$\Psi_{m,\ket{\Delta,q}^M}^{n}$ imply linearly independent
singular vectors of type $\Psi_{m,\ket{\Delta,q}^M}^{G,n+1}$.
The converse is not true, however, since most $\cG_0$-closed singular
vectors are not generated by the action of $\cG_0$ on a no-label
singular vector (in fact there are many more $\cG_0$-closed
singular vectors than no-label singular vectors, as was shown in Ref.
\icite{beatriz2}). Hence, the dimension for the space of singular vectors
$\Psi_{m,\ket{\Delta,q}^M}^{n}$ is limited
by the dimension for the space of singular vectors
$\Psi_{m,\ket{\Delta,q}^M}^{G,n+1}$. Similarly,
$\cQ_0\Psi_{m,\ket{\Delta,q}^M}^{n}$ is
a singular vector of type $\Psi_{m,\ket{\Delta,q}^M}^{Q,n-1}$.
This again restricts the dimension for
$\Psi_{m,\ket{\Delta,q}^M}^{n}$ to be less or equal to the dimension
of $\Psi_{m,\ket{\Delta,q}^M}^{Q,n-1}$.

\bth
For no-label singular vectors in
$\vm_{\Delta,q}^G$ or in $\vm_{\Delta,q}^Q$, $c\neq 3$, we find the
following upper limits for the dimensions of
singular vector spaces at level $m\in\bbbn$ and with charge $n\in\bbbz$
($\Delta=-m$ for no-label singular vectors).
\btab{|l|c|c|c|c|c|}
\hline \label{tab:dim2}
 & $n=-2$ & $n=-1$ & $n=0$ & $n=1$ & $n=2$ \\
\hline
$\Psi_{m,\ket{-m,q}^G}^{n}$ &
$0$ & $1$ & $1$ & $0$ & $0$ \\
\hline
$\Psi_{m,\ket{-m,q}^Q}^{n}$ &
$0$ & $0$ & $1$ & $1$ & $0$ \\
\hline
\etab{Maximal dimensions for spaces of no-label singular vectors
in $\vm_{\Delta,q}^G$ or in $\vm_{\Delta,q}^Q$.}
Singular vectors can only exist if they
contain in their normal form at least one non-trivial
term of the corresponding
ordering kernel of theorem \refoth{\ref{th:kernelG}} or
\refoth{\ref{th:kernelQ}}. Charges $n$ that are not given have
dimension $0$ and hence do not allow any singular vectors.
\eth

We now use the ordering kernels of section
\refoth{\ref{sec:aoGQ}} and section \refoth{\ref{sec:aonl}}
for chiral and no-label Verma modules. Again, simply by counting the number
of elements in the ordering kernels one obtains the corresponding
dimensions, given in the tables that follow.
\bth
For singular vectors in chiral Verma modules
$\vm_{0,q}^{GQ}$ or in no-label Verma modules
$\vm_{0,q}$, $c\neq 3$, we find the
following upper limits for the number of linearly independent
singular vectors at the same level $m\in\bbbn_0$ and with the
same charge $n\in\bbbz$.
\btab{|l|c|c|c|c|c|}
\hline \label{tab:dim3}
 & $n=-2$ & $n=-1$ & $n=0$ & $n=1$ & $n=2$ \\
\hline
$\Psi_{m,\ket{0,q}^{GQ}}^{G,n}$ &
$0$ & $0$ & $1$ & $1$ & $0$ \\
\hline
$\Psi_{m,\ket{0,q}^{GQ}}^{Q,n}$ &
$0$ & $1$ & $1$ & $0$ & $0$ \\
\hline
$\Psi_{0,\ket{0,q}^{GQ}}^{GQ,n}$ &
$0$ & $0$ & $0$ & $0$ & $0$ \\
\hline
$\Psi_{0,\ket{0,q}^{GQ}}^{n}$ &
$0$ & $0$ & $0$ & $0$ & $0$ \\
\hline
\etab{Maximal dimensions for singular vectors spaces
in $\vm_{0,q}^{GQ}$.}
\btab{|l|c|c|c|c|c|}
\hline \label{tab:dim4}
 & $n=-2$ & $n=-1$ & $n=0$ & $n=1$ & $n=2$ \\
\hline
$\Psi_{m,\ket{0,q}}^{G,n}$ &
$0$ & $1$ & $3$ & $3$ & $1$ \\
\hline
$\Psi_{m,\ket{0,q}}^{Q,n}$ &
$1$ & $3$ & $3$ & $1$ & $0$ \\
\hline
$\Psi_{0,\ket{0,q}}^{GQ,n}$ &
$0$ & $0$ & $1$ & $0$ & $0$ \\
\hline
$\Psi_{0,\ket{0,q}}^{n}$ &
$0$ & $0$ & $0$ & $0$ & $0$ \\
\hline
\etab{Maximal dimensions for singular vectors spaces
in $\vm_{0,q}$.}
Singular vectors can only exist if they
contain in their normal form at least one non-trivial
term of the corresponding
ordering kernel of theorem \refoth{\ref{th:kernelGQ}} or
\refoth{\ref{th:kernelnl}}.
Charges $n$ that are not given have dimension $0$ and hence do not allow
any singular vectors.
\eth

Tables \tab{\ref{tab:dim1}}, \tab{\ref{tab:dim2}}, \tab{\ref{tab:dim3}}
and \tab{\ref{tab:dim4}} prove the conjecture made in Ref.
\icite{beatriz2} about the possible existing types of topological
singular vectors. Namely, using the algebraic mechanism denoted
{\it the cascade effect} it was deduced (although not rigorously) the
existence of 4 types of singular vectors in chiral Verma modules
(the ones given in \tab{\ref{tab:dim3}}), and
29 types in complete Verma modules (the ones given in tables
\tab{\ref{tab:dim1}}, \tab{\ref{tab:dim2}} and \tab{\ref{tab:dim4}}).
In addition, low level examples were constructed for all these types
of singular vectors what proves that all these types do exist (already
at level 1, in fact,  except the type $\Psi_{0,\ket{0,q}}^{GQ,n}$ in
no-label Verma modules that only exists at level 0).

We ought to mention that the dimensions given in the previous three
theorems are consistent with the {\it spectral flow box diagrams} analysed
in Refs. \icite{beatriz4,beatriz1,beatriz2}. Namely, types of singular
vectors that are connected by the {\it topological} spectral flow
automorphism $\cal A$ always show the same singular vector space
dimensions\footnote{$\cal A$ is the universal odd spectral
flow\cite{BJI3,beatriz3,beatriz2}, discovered in Ref. \icite{BJI3},
which transform any topological singular vector into another topological
singular vector; in particular $\cal A$ transforms chiral singular
vectors into chiral singular vectors and no-label singular vectors into
no-label singular vectors.}.

Finally let us consider the results of \tab{\ref{tab:dim1}} in more
detail for the case when the conformal weight $\Delta$ is a negative
integer: $\Delta = -m \in -\bbbn_0$.
In this case, we easily find for each singular vector
$\Psi^{G,n}_{m,\ket{-m,q}^G}$ (which has zero conformal weight) a
companion $\cQ_0\Psi^{G,n}_{m,\ket{-m,q}^G}$ which is
of chiral type $\Psi^{GQ,n-1}_{m,\ket{-m,q}^G}$. Note that
$\cQ_0\Psi^{G,n}_{m,\ket{-m,q}^G}$ cannot be trivial. It is
rather a $secondary$ singular vector at level 0 with respect to the
singular vector $\Psi^{G,n}_{m,\ket{-m,q}^G}$.
Using the same arguments as for the no-label singular vectors
of \tab{\ref{tab:dim2}} we obtain that the dimension
for $\Psi^{G,n}_{m,\ket{-m,q}^G}$ is restricted by the
dimension for $\Psi^{GQ,n-1}_{m,\ket{-m,q}^G}$.
Similarly, we can act with $\cG_0$ on
$\Psi^{Q,n}_{m,\ket{-m,q}^G}$ in order to obtain a secondary
singular vector of chiral type $\Psi^{GQ,n+1}_{m,\ket{-m,q}^G}$.
The same statements are true for Verma modules of type
$\vm_{\Delta,q}^Q$. We hence obtain the following theorem.

\bth
For singular vectors at level $m\in\bbbn$ and with charge $n\in\bbbz$
in $\vm_{\Delta,q}^G$ or in
$\vm_{\Delta,q}^Q$, with $\Delta=-m $ and $c\neq 3$, we find the
following maximum dimensions for singular vector spaces.
\btab{|l|c|c|c|c|c|}
\hline \label{tab:dim5}
 & $n=-2$ & $n=-1$ & $n=0$ & $n=1$ & $n=2$ \\
\hline
$\Psi_{m,\ket{-m,q}^G}^{G,n}$ &
$0$ & $0$ & $1$ & $1$ & $0$ \\
\hline
$\Psi_{m,\ket{-m,q}^G}^{Q,n}$ &
$1$ & $1$ & $0$ & $0$ & $0$ \\
\hline
$\Psi_{m,\ket{-m,q}^Q}^{Q,n}$ &
$0$ & $1$ & $1$ & $0$ & $0$ \\
\hline
$\Psi_{m,\ket{-m,q}^Q}^{G,n}$ &
$0$ & $0$ & $0$ & $1$ & $1$ \\
\hline
\etab{Maximal dimensions for spaces of singular vectors at level $m$
in $\vm_{\Delta,q}^G$ or in $\vm_{\Delta,q}^Q$ with $\Delta = -m$.}
Charges $n$ that are not given have dimension $0$ and hence do not allow
any singular vectors.
\eth

The results of \tab{\ref{tab:dim5}} imply that if, for example, there
are two linearly independent singular vectors at level $m$ with charge $0$
in $\vm_{\Delta,q}^G$, with $\Delta = -m$, both annihilated
by $\cG_0$, then there exists a non-trivial linear combination of the two
singular vectors that turns out to be a chiral singular vector (annihilated
by $\cG_0$ $and$ $\cQ_0$). The space spanned by these two singular vectors
hence decomposes into a one-dimensional $\cG_0$-closed singular vector
space and a one-dimensional chiral singular vector space (see Ref.
\icite{beatriz2} for examples at level 3).

Some remarks are now in order concerning the existence of the considered
spaces of $N=2$ singular vectors. First of all observe that the
dimensions given by tables \tab{\ref{tab:dim1}} - \tab{\ref{tab:dim5}}
are the maximal possible dimensions for the spaces generated by
singular vectors of the corresponding types. That is, dimension 2 for a
given type of singular vector in \tab{\ref{tab:dim1}} does not
mean that all the spaces generated by singular vectors of such type are
two-dimensional. Rather, most of them are in fact one-dimensional and
only under certain conditions one finds two-dimensional spaces. The same
applies to the three-dimensional spaces in \tab{\ref{tab:dim4}}.
To be more precise, in Ref. \icite{paper2} it was
proved that for the Neveu-Schwarz $N=2$ algebra two-dimensional spaces
exist only for uncharged singular vectors and under certain conditions,
starting at level 2. For the topological $N=2$ algebra this
implies, as was shown in Ref. \icite{beatriz2},
that the four types of two-dimensional singular vector spaces of
\tab{\ref{tab:dim1}} must also exist starting at level 2, provided the
corresponding conditions are satisfied.
(To see this\cite{beatriz2,beatriz4} one only needs to apply the topological
twists to the singular vectors of the Neveu-Schwarz $N=2$ algebra and
then construct the box-diagrams using $\cG_0$, $\cQ_0$ and the odd spectral
flow automorphism $\cal A$). As a matter of fact, also in Ref.
\icite{beatriz2} several examples of these two-dimensional spaces were
constructed at level 3.

For the case of the three-dimensional singular
vector spaces in no-label Verma modules in \tab{\ref{tab:dim4}},
we do not know as yet of any conditions for them to exist. In
fact these are the only spaces, among all the spaces given in tables
\tab{\ref{tab:dim1}} - \tab{\ref{tab:dim5}}, which have not been observed so
far, although the corresponding types of singular vectors have been
constructed at level 1 generating one-dimensional\cite{beatriz2} as well as
two-dimensional spaces\footnote{In Ref. \icite{beatriz2} the existence of
these two-dimensional spaces of singular vectors at level 1 was overlooked.
We give examples of them here for the first time.}
(but not three-dimensional). The latter case is interesting, in addition,
because the corresponding two-dimensional spaces exist already at level 1
(in contrast with the two-dimensional spaces given by the
conditions of Ref. \icite{paper2}, which exist at levels 2 and higher).
Namely, for $c=9$ one can easily find
two-dimensional spaces of singular vectors of types
$\Psi_{1,\ket{0,-3}}^{G,0}$ and $\Psi_{1,\ket{0,-3}}^{Q,-1}$, in the
no-label Verma module $\vm_{0,-3}$, and two-dimensional spaces of singular
vectors of types $\Psi_{1,\ket{0,0}}^{G,1}$ and $\Psi_{1,\ket{0,0}}^{Q,0}$,
in the no-label Verma module $\vm_{0,0}$, all four types of singular
vectors belonging to the same box-diagram\cite{beatriz4,beatriz1,beatriz2}.
That is, the spectral flow automorphism $\cal A$, transforming the Verma
modules $\vm_{0,-3}$ and $\vm_{0,0}$ into each other, map
$\Psi_{1,\ket{0,-3}}^{G,0}$ to $\Psi_{1,\ket{0,0}}^{Q,0}$ and
$\Psi_{1,\ket{0,0}}^{G,1}$ to $\Psi_{1,\ket{0,-3}}^{Q,-1}$, and the other
way around, whereas $\cG_0$ and $\cQ_0$ transform the singular vectors
into each other inside a given Verma module. One of these two-dimensional
spaces is, for example, the space spanned by the singular vectors of type
$\Psi_{1,\ket{0,0}}^{Q,0}$:
\bea \Psi_{1,\ket{0,0,c}}^{Q,0} & = & \cL_{-1}\cQ_0\cG_0 \ket{0,0,c} \eea
\noi
and
\bea
\hat\Psi_{1,\ket{0,0,9}}^{Q,0} & = & [\cH_{-1} \cQ_0 \cG_0 + \cQ_{-1} \cG_0
    - 2 \cQ_0 \cG_{-1}] \ket{0,0,9} \,, \eea

\noi
the latter existing only for $c=9$. As one can see,
the canonical projections of the first singular vector into the generic
Verma modules $\vm_{0,0}^G$ and $\vm_{0,0}^Q$ vanish. On the contrary,
the canonical projections of the second singular vector into the generic
Verma modules $\vm_{0,0}^G$ and $\vm_{0,0}^Q$ are different from zero,
giving rise to the singular vectors
\bea
\Psi_{1,\ket{0,0,9}^G}^{Q,0} = \cQ_0 \cG_{-1}] \ket{0,0,9}^G \ \,, \ \ \ \
\Psi_{1,\ket{0,0,9}^Q}^{Q,0} = [\cQ_{-1}\cG_0 - 4 \cL_{-1}]\ket{0,0,9}^Q \,,
\eea

\noi
respectively. These are, in turn, the particular cases for
($\Delta=0,\, q=0,\, c=9$) of the
general expressions for $\Psi_{1,\ket{\Delta,q}^G}^{Q,0}$ and
$\Psi_{1,\ket{\Delta,q}^Q}^{Q,0}$, given in Ref. \icite{beatriz2}.


\section{ Dimensions for the Neveu-Schwarz and the Ramond $N=2$ algebras}
\label{sec:NSR}

Transferring the dimensions we have found for the topological $N=2$
algebra to the Neveu-Schwarz and to the Ramond $N=2$ algebras is
straightforward. The Neveu-Schwarz $N=2$ algebra is related to the
topological $N=2$ algebra through the topological twists ${\ }T_W^{\pm}$:
${\ }\cL_m=L_m\pm 1/2 H_m$, ${\ }\cH_m=\pm H_m$,
${\ }\cG_m=G^{\pm}_{m+1/2}$ and ${\ }\cQ_m=G^{\mp}_{m-1/2}$ . As a
consequence, the (non-chiral) Neveu-Schwarz highest weight vectors
correspond to $\cG_0$-closed topological highest weight vectors
(annihilated by $\cG_0$), whereas the chiral and antichiral Neveu-Schwarz
highest weight vectors (annihilated by $G^+_{-1/2}$ and by $G^-_{-1/2}$,
respectively), correspond to chiral topological highest weight vectors
(annihilated by $\cG_0$ and $\cQ_0$). This implies (see the details in
Refs. \icite{beatriz1,beatriz2}) that the standard
Neveu-Schwarz singular vectors (`normal', chiral and antichiral)
correspond to topological singular vectors of the types
$\Psi^{G,n}_{m,\ket{\Delta,q}^G}$ and $\Psi^{GQ,n}_{m,\ket{\Delta,q}^G}$,
whereas the Neveu-Schwarz singular vectors in chiral or antichiral
Verma modules correspond to topological singular vectors of only the type
$\Psi^{G,n}_{m,\ket{\Delta,q}^{GQ}}$ (as there are no chiral singular
vectors in chiral Verma modules). To be precise, for the singular vectors
of the Neveu-Schwarz $N=2$ algebra one finds the following results.

\bth \label{th:dimNS}
The spaces of Neveu-Schwarz singular vectors $\Psi_{m,\ket{\Delta,q}}^{n}$,
$\Psi_{m,\ket{\Delta,q}}^{ch,n}$ and $\Psi_{m,\ket{\Delta,q}}^{a,n}$, where
the supercripts $ch$ and $a$ stand for chiral and antichiral, respectively,
have the same maximal dimensions as the spaces of topological singular
vectors $\Psi_{m\pm n/2,\ket{\Delta\pm q/2,\pm q}^G}^{G,\pm n}$ and
$\Psi_{m\pm n/2,\ket{-m\mp n/2,\pm q}^G}^{GQ,\pm n}$, given in
\tab{\ref{tab:dim1}}.
Therefore for singular vectors in Neveu-Schwarz Verma modules
$\vm_{\Delta,q}^{NS}$ we find the following upper limits for the
number of linearly independent singular vectors at the same
level $m$ and with the same charge $n\in\bbbz$
($m \in\bbbn$ for $n=0$ while $m \in {\bf {N-1/2}}$ for $n=\pm 1$).

\btab{|l|c|c|c|}
\hline \label{tab:dim6}
 & $n=-1$ & $n=0$ & $n=1$  \\
\hline
$\Psi_{m,\ket{\Delta,q}}^{n}$ &
 $1$ & $2$ & $1$  \\
\hline
$\Psi_{m,\ket{\Delta,q}}^{ch,n}$ &
 $0$ & $1$ & $1$  \\
\hline
$\Psi_{m,\ket{-m,q}}^{a,n}$ &
$1$ & $1$ & $0$  \\
\hline
\etab{Maximal dimensions for singular vectors spaces in
$\vm_{\Delta,q}^{NS}$.}
(Chiral and antichiral singular vectors satisfy $\Delta + m=\pm 1/2(q+n)$,
respectively). Charges $n$ that are not given have dimension
$0$ and hence do not allow any singular vectors.
\eth

\bth
The spaces of Neveu-Schwarz singular vectors
$\Psi_{m,\ket{\Delta,q}^{ch}}^{n}$, $\Psi_{m,\ket{\Delta,q}^{a}}^{n}$
$\Psi_{m,\ket{\Delta,q}^{ch}}^{a,n}$ and
$\Psi_{m,\ket{\Delta,q}^{a}}^{ch,n}$ in chiral or antichiral
Verma modules, $\vm_{\Delta,q}^{NS,ch}$ or $\vm_{\Delta,q}^{NS,a}$
with $\Delta=\pm q/2$ respectively, have the same maximal dimensions
as the spaces of topological singular vectors
$\Psi_{m \pm n/2,\ket{0,\pm q}^{GQ}}^{G,\pm n}$ in chiral
topological Verma modules, given in \tab{\ref{tab:dim3}}.
Therefore for Neveu-Schwarz singular vectors in chiral or antichiral
Verma modules we find the following upper limits for the number of
linearly independent singular vectors at the same level
$m$ and with the same charge $n\in\bbbz$
($m \in\bbbn$ for $n=0$ while $m \in {\bf {N-1/2}}$ for $n=\pm 1$).
The supercripts $ch$ and $a$ stand for chiral and antichiral, respectively.

\btab{|l|c|c|c|}
\hline \label{tab:dim7}
 & $n=-1$ & $n=0$ & $n=1$  \\
\hline
$\Psi_{m,\ket{q/2,q}^{ch}}^{n}$ &
 $1$ & $1$ & $0$  \\
\hline
$\Psi_{m,\ket{q/2,q}^{ch}}^{a,n}$ &
 $1$ & $1$ & $0$  \\
\hline
$\Psi_{m,\ket{-q/2,q}^{a}}^{n}$ &
 $0$ & $1$ & $1$  \\
\hline
$\Psi_{m,\ket{-q/2,q}^{a}}^{ch,n}$ &
 $0$ & $1$ & $1$  \\
\hline

\etab{Maximal dimensions for singular vectors spaces in
$\vm_{q/2,q}^{NS,ch}$ and $\vm_{-q/2,q}^{NS,a}$ .}
($\Psi_{m,\ket{q/2,q}^{ch}}^{a,n}$ and $\Psi_{m,\ket{-q/2,q}^{a}}^{ch,n}$
satisfy in addition $q=\mp m-n/2$, respectively). Charges $n$ that are not
given have dimension $0$ and hence do not allow any singular vectors.
\eth

Observe that there are no chiral singular vectors in chiral Verma modules,
neither antichiral singular vectors in antichiral Verma modules; that is,
there are no Neveu-Schwarz singular vectors of types
$\Psi_{m,\ket{q/2,q}^{ch}}^{ch,n}$ and $\Psi_{m,\ket{-q/2,q}^{a}}^{a,n}$,
which would correspond to the non-existing chiral topological singular
vectors $\Psi_{m \pm n/2,\ket{0,\pm q}^{GQ}}^{GQ, \pm n}$ in chiral
topological Verma modules.

The first row of \tab{\ref{tab:dim6}} recovers the results
already proven\footnote{In Ref. \icite{paper2} it had not been explicitly
stated that the results do not hold for $c=3$. Also, the necessity
of the change of basis in the final consideration of the proof of theorem
\refoth{\ref{th:kernelG}} had been overlooked in Ref. \icite{paper2}.
Nevertheless, theorem \refoth{\ref{th:dimNS}} shows that all the results
of Ref. \icite{paper2} do hold.} in Refs. \icite{thesis,paper2}, using
adapted orderings in generalised (analytically continued) Verma modules.
That is, in complete Verma modules of the Neveu-Schwarz $N=2$ algebra
singular vectors can only exist with charges $n=0, \pm 1$ and, under certain
conditions, there exist two-dimensional spaces of (only) uncharged singular
vectors. \tab{\ref{tab:dim7}} proves the conjecture, made in Refs.
\icite{beatriz1,beatriz2}, that in chiral Neveu-Schwarz Verma modules
$\,\vm_{q/2,q}^{NS,ch}\,$ the charged singular vectors are always
negatively charged, with $n=-1$, whereas in antichiral Neveu-Schwarz
Verma modules $\,\vm_{-q/2,q}^{NS,a}\,$
 the charged singular vectors are always
positively charged, with $n=1$. In contrast to this, the chiral charged
singular vectors in the Verma modules $\,\vm_{\Delta,q}^{NS}\,$
and $\,\vm_{-q/2,q}^{NS,a}\,$ are always positively charged, with $n=1$,
whereas the antichiral charged singular vectors in the Verma modules
$\,\vm_{\Delta,q}^{NS}\,$ and $\,\vm_{q/2,q}^{NS,ch}$ are always negatively
charged, with $n=-1$. This fact was observed also in Ref. \icite{beatriz2}
and can be deduced from the results of Ref. \icite{paper2}.

\vskip .18in

As to the Ramond $N=2$ algebra, combining the topological twists
${\ }T_W^{\pm}$ and the spectral flows
it is possible to construct a one-to-one mapping between
every Ramond singular vector and every topological singular vector, at
the same levels and with the same charges\footnote{We define the charges
for the Ramond states in the same way as for the Neveu-Schwarz states,
see the details in Refs. \icite{beatriz1,DGR1}.}
(see the details in Ref. \icite{DGR3}). As a consequence, the
results of tables \tab{\ref{tab:dim1}} - \tab{\ref{tab:dim5}} can be
transferred to the Ramond singular vectors simply by exchanging the labels
$G \to (+), {\ }Q \to (-)$, where
the helicity $(+)$ denote the Ramond states annihilated by $G_0^+$ and
the helicity $(-)$ denote the Ramond states annihilated by $G_0^-$. The
{\it no-helicity} Ramond states, analogous to the {\it no-label} topological
states, have been overlooked
until recently in the literature (see Refs. \icite{DGR1,DGR3}).
They require conformal weight $\Delta+m=c/24 \,$ in the
same way that no-label states require zero conformal weight $\Delta+m=0$.


\section{Conclusions and prospects}
\label{sec:conclusions}

For the study of the highest weight representations of a Lie
algebra or a Lie super algebra, the determinant formula plays
a crucial r\^ole. However, the determinant formula does not give the
complete information about the submodules existing in a given Verma
module. Exactly which Verma modules contain proper submodules and at
which level can be found the lowest non-trivial grade space of the
biggest proper submodule is the information that may easily be obtained
from the determinant formula. But it does not give a proof that the
singular vectors obtained in that way are all the existing singular
vectors, i.e. generate the biggest proper submodules, neither does
it give the dimensions of the singular vector spaces. However, it has
been shown\cite{paper2,beatriz2} that singular vector spaces with more
than one dimension exist already for the $N=2$ superconformal algebra.

In this paper we have presented a method that can easily be applied to
many Lie algebras and Lie super algebras. This method is based on the
concept of adapted ordering, which
implies that any singular vector needs to contain
at least one non-trivial term included in the ordering kernel.
The size of the ordering kernel therefore limits the
dimension of the corresponding singular vector space.
Weights for which the ordering kernel is trivial do not allow any
singular vectors in the corresponding weight space. On the other
hand, non-trivial ordering kernels give us the maximal
dimension of a possible singular vector space.
The framework can easily be understood using the example of the Virasoro
algebra where the ordering kernel always has size one and therefore
Virasoro singular vectors at the same level in the same Verma module
are always proportional. In its original version, Kent\cite{adrian1}
used the idea of an ordering for generalised (analytically extended)
Virasoro Verma modules in order to show that all vectors
satisfying the highest weight conditions at level $0$
are proportional to the highest weight vector.

As an important application of this method, we have computed the maximal
dimensions of the singular vector spaces for Verma modules of the
topological $N=2$ algebra, obtaining maximal dimensions 0, 1, 2 or 3,
depending on the type of Verma module and the type of singular vector.
The results are consistent with the topological spectral flow
automorphisms and with all known examples of topological singular vectors.
On the one hand, singular vector spaces with maximal dimension bigger
than 1 agree with explicilty computed examples found before (and during)
this work, although in the case of the three-dimensional spaces
in no-label Verma modules, the singular vectors of the corresponding types
known so far generate only one and two-dimensional spaces. These exist
already at level 1, in contrast with the previously known two-dimensional
spaces, which exist at levels 2 and higher. On the other
hand, singular vector spaces with zero dimension imply that the `would-be'
singular vectors of the corresponding types do not exist. As a consequence,
our results provide a rigorous proof to the conjecture made in Ref.
\icite{beatriz2} about the possible existing types of topological singular
vectors: 4 types in chiral Verma modules and 29 types in complete Verma
modules.

Finally we have transferred the results found for the topological $N=2$
algebra to the Neveu-Schwarz and to the Ramond $N=2$ algebras. In the
first case we have recovered the results obtained in Ref. \icite{paper2}
for complete Verma modules: maximal dimensions 0, 1 or 2, the latter
only for uncharged singular vector spaces, and allowed charges only 0 and
$\pm 1$. In addition, we have proved the conjecture
made in Refs. \icite{beatriz1,beatriz2} on the possible existing types
of Neveu-Schwarz singular vectors in chiral and antichiral Verma modules.
In the case of the Ramond $N=2$ algebra we have found a one-to-one mapping
between the Ramond singular vectors and the topological singular vectors,
so that the corresponding results are essentially the same.

The only exception for which the adapted orderings presented in this
paper are not suitable is for central term $c=3$. This case needs a
separate consideration. The application of the
adapted ordering method to the twisted $N=2$ algebra will be the subject
of a forthcoming paper.

The example of the $N=2$ topological Verma modules
is only one out of many cases where the concept of adapted orderings
can be applied. For example, Bajnok\cite{zoltan2} showed that
the {\it analytic continuation} method of Kent\cite{adrian1}
can be extended to generalised Verma modules of the $WA_2$ algebra.
Not only will the concept of adapted orderings allow us to obtain
information about superconformal Verma modules with $N>2$,
it should also be easily applicable to any other Lie algebra whenever an
adapted ordering can be constructed with small ordering kernels.

\appendix

\acknowledgements
We are very grateful to Adrian Kent for numerous
discussions on the $N=2$ orderings and to Victor Kac
for many important comments with respect to the $N=2$
superconformal algebras.
M.D. is supported by a
DAAD fellowship and in part by NSF grant PHY-98-02709.

\noi



\begin{thebibliography}{10}

\bibitem{ademollo}
M.~Ademollo, L.~Brink, A.~d'Adda, R.~d'Auria, E.~Napolitano, S.~Sciuto, E.~del
  Giudice, P.~di~Vecchia, S.~Ferrara, F.~Gliozzi, R.~Musto, and R.~Pettorino.
\newblock Supersymmetric strings and colour confinement.
\newblock {\em Phys. Lett.} B62:105, 1976.

\bibitem{ast1}
A.B. Astashkevich.
\newblock On the structure of {Verma} modules over {Virasoro} and
  {Neveu-Schwarz} algebras.
\newblock {\em Commun. Math. Phys.} 186:531, 1997.

\bibitem{ast2}
A.B. Astashkevich and D.B. Fuchs.
\newblock Asymptotic of singular vectors in {Verma} modules over the
{Virasoro} {Lie} algebra.
\newblock {\em Pac. J. Math.} 177:No.2, 1997.

\bibitem{zoltan2}
Z.~Bajnok.
\newblock Singular vectors of the {$WA_{2}$} algebra.
\newblock {\em Phys. Lett.} B329:225, 1994.

\bibitem{adrian}
W.~Boucher, D.~Friedan, and A.~Kent.
\newblock Determinant formulae and unitarity for the {$N=2$} superconformal
  algebras in two dimensions or exact results on string compactification.
\newblock {\em Phys. Lett.} B172:316, 1986.

\bibitem{kac4}
S.-L. Cheng and V.G. Kac.
\newblock A new {$N=6$} superconformal algebra.
\newblock {\em Commun. Math. Phys.} 186:219, 1997.

\bibitem{DVV}
R.~Dijkgraaf, E.~Verlinde, and H.~Verlinde.
\newblock Topological strings in $d<1$.
\newblock {\em Nucl. Phys.} B352:59, 1991.

\bibitem{thesis}
M.~D{\"o}rrzapf.
\newblock {\em Superconformal field theories and their representations}.
\newblock PhD thesis, University of Cambridge, September 1995.

\bibitem{paper2}
M.~D{\"o}rrzapf.
\newblock Analytic expressions for singular vectors of the {$N=2$}
  superconformal algebra.
\newblock {\em Commun. Math. Phys.} 180:195, 1996.

\bibitem{paper5}
M.~D{\"o}rrzapf.
\newblock The embedding structure of unitary {$N=2$} minimal models.
\newblock {\em HUTP-97/A056 preprint, hep-th/9712169}, 1997.

\bibitem{DGR1}
M.~D{\"o}rrzapf and B. Gato-Rivera.
\newblock Transmutations between singular and subsingular vectors of the
 $N=2$ superconformal algebras.
\newblock {\em HUTP-97/A055, IMAFF-FM-97/04 preprint, hep-th/9712085},
1997, to be published in {\em Nucl. Phys.} B.

\bibitem{DGR3}
M.~D{\"o}rrzapf and B. Gato-Rivera.
\newblock Determinant formula for the topological $N=2$ superconformal
 algebra.
\newblock {\em HUTP-98/A055, IMAFF-98/07 preprint,} to appear in hep-th.

\bibitem{Dobrev}
V.K. Dobrev.
\newblock Characters of the unitarizable highest weight modules over
the {$N=2$} superconformal algebras.
\newblock {\em Phys. Lett.} B186 (1987) 43.

\bibitem{ff1}
B.L. Feigin and D.B. Fuchs.
\newblock {\em Representations of {Lie} groups and related topics}.
\newblock A.M. Vershik and A.D. Zhelobenko eds. , Gordon \& Breach, 1990.

\bibitem{beatriz4}
B.~Gato-Rivera.
\newblock Construction formulae for singular vectors of the topological
{$N=2$} superconformal algebra.
\newblock {\em IMAFF-FM-98/05, hep-th/9802204}, 1998.

\bibitem{beatriz3}
B.~Gato-Rivera.
\newblock The even and the odd spectral flows on the {$N=2$} superconformal
  algebras.
\newblock {\em Nucl. Phys.} B512:431, 1998.

\bibitem{BJI3}
B.~Gato-Rivera and J.I. Rosado.
\newblock Spectral flows and twisted topological theories.
\newblock {\em  Phys. Lett.} B369:7, 1996.

\bibitem{beatriz1}
B.~Gato-Rivera and J.I. Rosado.
\newblock Chiral determinant formulae and subsingular vectors for the {$N=2$}
  superconformal algebras.
\newblock {\em Nucl. Phys.} B503:447, 1997.

\bibitem{beatriz2}
B.~Gato-Rivera and J.I. Rosado.
\newblock Families of singular and subsingular vectors of the topological
  {$N=2$} superconformal algebra.
\newblock {\em Nucl. Phys.} B514:477, 1998.

\bibitem{kac3}
V.G. Kac.
\newblock Lie superalgebras.
\newblock {\em Adv. Math.} 26:8, 1977.

\bibitem{kac2}
V.G. Kac.
\newblock {\em Contravariant form for infinite-dimensional {Lie} algebras and
  superalgebras}, volume Lect. Notes in Phys. 94.
\newblock 1979.

\bibitem{kac5}
V.G. Kac.
\newblock Superconformal algebras and transitive groups actions on quadrics.
\newblock {\em Commun. Math. Phys.} 186:233, 1997.

\bibitem{kac6}
V.G. Kac and J.W. van~de Leuer.
\newblock {\em On classification of superconformal algebras}, volume Strings
  88.
\newblock World Scientific, 1988.

\bibitem{kato}
M.~Kato and S.~Matsuda.
\newblock Null field construction and {Kac} formulae of {$N=2$} superconformal
  algebras in two dimensions.
\newblock {\em Phys. Lett.} B184:184, 1987.

\bibitem{adrian1}
A.~Kent.
\newblock Singular vectors of the {Virasoro} algebra.
\newblock {\em Phys. Lett.} B273:56, 1991.

\bibitem{adrian3}
A.~Kent, M.~Mattis, and H.~Riggs.
\newblock Highest weight representations of the {$N=3$} superconformal algebras
  and their determinant formulae.
\newblock {\em Nucl. Phys.} B301:426, 1988.

\bibitem{adrian4}
A.~Kent and H.~Riggs.
\newblock Determinant formulae for the {$N=4$} superconformal algebras.
\newblock {\em Phys. Lett.} B198:491, 1987.

\bibitem{Kir}
E.~Kiritsis.
\newblock Character formula and the structure of the representations of the
$N=1$, $N=2$ superconformal algebras.
\newblock {\em Int. J. Mod. Phys.} A3:1871, 1988.

\bibitem{LVW}
W. Lerche, C. Vafa and N.P. Warner.
\newblock Chiral rings in $N=2$ superconformal theories.
\newblock {\em  Nucl. Phys.} B324:427, 1989.

\bibitem{malikov}
F.G. Malikov, B.L. Feigin, and D.B. Fuchs.
\newblock Singular vectors in {Verma} modules over {Kac-Moody} algebras.
\newblock {\em Funct. Anal. Appl.} 20:103, 1986.

\bibitem{matsuda}
S.~Matsuda.
\newblock Coulomb gas representations and screening operators of the {$N=4$}
  superconformal algebras.
\newblock {\em Phys. Lett.} B282:56, 1992.

\bibitem{meur}
A.~Meurman and A.~Rocha-Caridi.
\newblock Highest weight representations of the {Neveu-Schwarz} and {Ramond}
  algebras.
\newblock {\em Commun. Math. Phys.} 107:263, 1986.

\bibitem{nam}
S.~Nam.
\newblock The Kac formula for the {$N=1$} and {$N=2$} super-conformal
algebras.
\newblock {\em  Phys. Lett.} B172:323, 1986.


\end{thebibliography}

\end{document}